\documentclass[]{emulateapj}
\usepackage{natbib,amsmath}

\shorttitle{Simple Models of Cool Gas Outflows} 
\shortauthors{Prochaska, Kasen, \& Rubin}

\begin{document}

\newcommand{\nhn}{$n_{\rm H}^0$}
\newcommand{\mnhn}{n_{\rm H}^0}
\newcommand{\mnhf}{n_{\rm H}^{0,f}}
\newcommand{\msun}{M_\odot}
\newcommand{\tpk}{$\tau_{\rm pk}$}
\newcommand{\mtpk}{\tau_{\rm pk}}
\newcommand{\mdv}{\delta v}
\newcommand{\ewabs}{$W_{\rm a}$}
\newcommand{\ewe}{$W_{\rm e}$}
\newcommand{\rtt}{$r_{\tau_S = 0.2}$}
\newcommand{\mrtt}{r_{\tau_S = 0.2}}
\newcommand{\bturb}{$b_{\rm turb}$}
\newcommand{\mbturb}{b_{\rm turb}}
\newcommand{\taud}{$\tau_{\rm dust}$}
\newcommand{\mtaud}{\tau_{\rm dust}}
\newcommand{\maconfig}{2p$^6$3s}
\newcommand{\mbconfig}{2p$^6$3p}
\newcommand{\aconfig}{a~$^6$D$^0$}
\newcommand{\zconfig}{z~$^6$D$^0$}
\newcommand{\mvr}{v_{\rm r}}
\newcommand{\naid}{\ion{Na}{1}~$\lambda\lambda 5891, 5897$}
\newcommand{\mgiid}{\ion{Mg}{2}~$\lambda\lambda 2796, 2803$}
\newcommand{\mgiia}{\ion{Mg}{2}~$\lambda 2796$}
\newcommand{\mgiib}{\ion{Mg}{2}~$\lambda 2803$}
\newcommand{\feiia}{\ion{Fe}{2}~$\lambda 2586$}
\newcommand{\feiib}{\ion{Fe}{2}~$\lambda 2600$}
\newcommand{\feiic}{\ion{Fe}{2}*$ \; \lambda 2612$}
\newcommand{\feiie}{\ion{Fe}{2}*$ \; \lambda 2626$}
\newcommand{\feiid}{\ion{Fe}{2}~$\lambda\lambda 2586, 2600$}
\newcommand{\feiis}{\ion{Fe}{2}*}
\newcommand{\nmg}{$n_{\rm Mg^+}$}
\newcommand{\mnmg}{n_{\rm Mg^+}}
\newcommand{\nfe}{$n_{\rm Fe^+}$}
\newcommand{\mnfe}{n_{\rm Fe^+}}
\def\hub{h_{72}^{-1}}
\def\umfp{{\hub \, \rm Mpc}}
\def\mzq{z_q}
\def\zabs{$z_{\rm abs}$}
\def\mzabs{z_{\rm abs}}
\def\intl{\int\limits}
\def\cmma{\;\;\; ,}
\def\perd{\;\;\; .}
\def\ltk{\left [ \,}
\def\ltp{\left ( \,}
\def\ltb{\left \{ \,}
\def\rtk{\, \right  ] }
\def\rtp{\, \right  ) }
\def\rtb{\, \right \} }
\def\sci#1{{\; \times \; 10^{#1}}}
\def \rAA {\rm \AA}
\def \zem {$z_{\rm em}$}
\def \mzem {z_{\rm em}}
\def\smm{\sum\limits}
\def \cmm  {cm$^{-2}$}
\def \cmmm {cm$^{-3}$}
\def \kms  {km~s$^{-1}$}
\def \mkms  {\, {\rm km~s^{-1}}}
\def \lyaf {Ly$\alpha$ forest}
\def \Lya  {Ly$\alpha$}
\def \lya  {Ly$\alpha$}
\def \mlya  {Ly\alpha}
\def \Lyb  {Ly$\beta$}
\def \lyb  {Ly$\beta$}
\def \lyg  {Ly$\gamma$}
\def \ly5  {Ly-5}
\def \ly6  {Ly-6}
\def \ly7  {Ly-7}
\def \nhi  {$N_{\rm HI}$}
\def \mnhi  {N_{\rm HI}}
\def \lnhi {$\log N_{HI}$}
\def \mlnhi {\log N_{HI}}
\def \etal {\textit{et al.}}
\def \lyaf {Lyman--$\alpha$ forest}
\def \mnmin {\mnhi^{\rm min}}
\def \nmin {$\mnhi^{\rm min}$}
\def \O {${\mathcal O}(N,X)$}
\newcommand{\cm}[1]{\, {\rm cm^{#1}}}
\def \snrlim {SNR$_{lim}$}

\title{Simple Models of Metal-Line Absorption and Emission from
Cool Gas Outflows} 

\author{
J. Xavier Prochaska\altaffilmark{1}, 
Daniel Kasen\altaffilmark{2,3}, 
Kate Rubin\altaffilmark{1} 
}
\altaffiltext{1}{Department of Astronomy and Astrophysics, UCO/Lick Observatory, University of California, 1156 High Street, Santa Cruz, CA 95064}
\altaffiltext{2}{Departments of Physics and Astronomy, University of California, Berkeley, 366 LeConte Hall, Berkeley CA, 94720}
\altaffiltext{3}{Nuclear Science Division, Lawrence Berkeley National Laboratory, 1 Cyclotron Rd, Berkeley, CA, 94720}

\begin{abstract}
We analyze the absorption and emission-line profiles produced by a set
of simple, cool gas wind models motivated by galactic-scale outflow
observations.  
We implement monte carlo radiative
transfer techniques that track the propagation of scattered and
fluorescent photons to generate 1D spectra and 2D spectral images.
We focus on the \mgiid\ doublet and \ion{Fe}{2} UV1
multiplet at $\lambda \approx 2600$\AA, but the results are applicable
to other transitions that trace outflows (e.g.\ \ion{Na}{1},
\ion{H}{1} \lya, \ion{Si}{2}).   
By design, the resonance transitions show blue-shifted
absorption but one also predicts strong resonance and fine-structure line-emission at roughly
the systemic velocity.  This line-emission `fills-in' the absorption
reducing the equivalent width by up to $50\%$, shift the
absorption-lin centroid by tens of \kms, and 
reduce the effective
opacity near systemic.  Analysis of cool gas outflows that ignores this
line-emission may incorrectly infer that the gas is partially
covered, measure a significantly lower peak optical depth, and/or
conclude that gas at systemic velocity is absent   
(e.g.\ an interstellar or slowly infalling component). 
Because the \ion{Fe}{2} lines
are connected by optically-thin transitions to
fine-structure levels, their profiles
more closely reproduce the intrinsic opacity
of the wind.   Together these results naturally explain the
absorption and emission-line characteristics observed for star-forming
galaxies at $z<1$.
We also study a scenario promoted to describe the outflows
of $z \sim 3$ Lyman break galaxies and 
find prfiles inconsistent with the observations due to scattered
photon emission.
Although line-emission complicates the analysis of absorption-line
profiles, the 
surface brightness profiles 
offer a unique means of assessing the morphology and size
of galactic-scale winds.  Furthermore, the kinematics and line-ratios
offer powerful diagnostics of outflows,  motivating deep,
spatially-extended spectroscopic observations.

\end{abstract}

\keywords{galaxies: formation -- galaxies: starbust}

\section{Introduction}
\label{sec:intro}

Nearly all gaseous objects that shine are also
observed to generate gaseous flows.  This includes the jets of protostars, the
stellar winds of massive O and B stars, the gentle Solar wind
of our Sun, the associated absorption of bright quasars, and the
spectacular jets of radio-loud AGN.   These gaseous outflows 
regulate the metal and dust content and distribution within the 
objects and their surroundings, 
moderate the accretion of new material, and 
inject energy and momentum into gas on large scales. 
Developing a comprehensive model for these flows is critical to
understanding the evolution of the source and its impact on the
surrounding environment.

Starburst galaxies, whose luminosity is dominated by \ion{H}{2} regions
and massive stars, are also observed to drive gaseous outflows.  These
flows are generally expected (and sometimes observed) to have multiple
phases, for example a hot and diffuse phase traced by X-ray emission
together with a cool, denser phase traced by H$\alpha$ emission 
\citep[e.g.][]{ham90,martin99,shc+04,km10}. 
Several spectacular
examples in the local universe demonstrate that flows can extend to
up to $\sim 10$ kpc
from the galaxy \citep{lhw99,vsr+03,wsg08} carrying significant speed
to escape from the gravitational potential well of the galaxy's dark
matter halo \citep[e.g.][]{sh09}.

Galactic outflows are also
revealed by UV and optical absorption lines, e.g.\ \ion{Na}{1},
\ion{Mg}{2}, \ion{Si}{2} and \ion{C}{4} transitions.  With the galaxy
as a backlight, one observes gas that is predominantly
blue-shifted which indicates a flow toward
Earth and away from the galaxy.  These transitions are sensitive to
the cool (\ion{Mg}{2}; $T \sim 10^4$K) and warm (\ion{C}{4}; $T \sim
10^5$K) phases of the flow.  
The incidence of cool gas outflows is
nearly universal in vigorously
star-forming galaxies;  this includes systems at low $z$
which exhibit \ion{Na}{1} and \ion{Mg}{2} absorption
\citep{rvs05a,martin05,smn+09,mb09,cth+10}, 
$z \sim 1$ star-forming galaxies with winds traced by
\ion{Fe}{2} and \ion{Mg}{2} transitions \citep{wcp+09,rwk+10}, and
$z>2$ Lyman break galaxies (LBGs) that show blue-shifted \ion{Si}{2},
\ion{C}{2}, and \ion{O}{1} transitions \citep{sgp+96,lkg+97,shapley03}.

The observation of metal-line absorption is now a well-established
means of identifying outflows.  Furthermore, because the X-ray and
H$\alpha$ emission generated by winds is faint, absorption-line
analyses have traditionally been the only way to probe outflows in
distant galaxies. However, very little research
has been directed
toward comparing the observations against (even idealized) wind
models \citep[e.g.][]{fmm+09}.  
Instead, researchers have gleaned what limited information
is afforded by direct analysis of the absorption lines.  The data
diagnose the speed of the gas relative to the galaxy, 
yet they poorly constrain the
optical depth, covering fraction, density, temperature, and distance
of the flow from the galaxy.   In turn, constraints related to the
mass, energetics, and momentum of the flow suffer from
orders of magnitude uncertainty.  Both the origin and impact of
galactic-scale winds, therefore, remain open matters of debate
\citep{mqt05,sdr08,sh09,ssr02,od06,kkd+09}.

Recent studies of $z \lesssim 1$ star-forming galaxies have revealed that
the cool outflowing gas often exhibits significant resonant-line emission (e.g.\
\ion{Mg}{2}, \ion{Na}{1}) in
tandem with the nearly ubiquitous blue-shifted absorption
\citep{wcp+09,mb09,rwk+10,cth+10}.  The resultant spectra resemble the P-Cygni
profile characteristic of stellar winds.
This phenomenon was first reported by \citet{phillips93}, who observed 
blue-shifted absorption and red-shifted emission for the \ion{Na}{1}
transition in the spectrum of the local starburst galaxy NGC 1808.
More recently, \cite{wcp+09}, who
studied \ion{Mg}{2} absorption in $z \sim 1.4$ galaxies, reported
\ion{Mg}{2} emission in a small subset of the individual galaxy
spectra of their large sample.  These were excluded from the full
analysis on concerns that the emission was related to AGN activity.
The stacked spectra of the remaining galaxies, however, also indicated
\ion{Mg}{2} emission, both directly and when the authors modeled and
`removed' the $v \approx 0\mkms$ absorption component.  The
authors suggested the emission could be related to back-scattered
light in the wind, but presumed that it was related to weak
AGN activity.   Very similar \ion{Mg}{2} emission was observed by
\cite{rwk+10} who repeated the analysis of \cite{wcp+09} on a set
of lower redshift galaxies. 

Bright \ion{Mg}{2} line emission has also been
reported for individual galaxies at $z \sim 0.7$ by \cite[][see also
Rubin et al\ 2011, in prep.]{rubin+10c}.
In their analysis of a single galaxy spectrum, \cite{rubin+10c}
further demonstrated that the \ion{Mg}{2} line emission is spatially
extended, and used the size of the emission to infer that the wind 
extends to at least 7\,kpc from the galaxy.
These authors additionally detected line
emission from non-resonant \ion{Fe}{2}$^*$ transitions, and attributed
the emission to fluorescence powered by \ion{Fe}{2} resonant
absorption.  In other words, these photons are re-emitted by the wind
into our sightline, and are analogous to the emitted photons in a P-Cygni profile.
Line-emission that may be related to outflows is
also observed for $z \sim 3$ LBGs in the resonant \lya\ transition
and non-resonant \ion{Si}{2}$^*$ transitions\footnote{See also
  \cite{france10} for a new example at $z \sim 0$.} \citep{prs+02,shapley03}.
This emission likely arises from a completely different
physical process than those generating X-ray and H$\alpha$ emission
(e.g., shocks), and presumably probes both the cool gas at the base of
the wind and the outskirts of the flow (i.e., wherever a given
transition is optically thick).
A comprehensive analysis of the 
scattered and fluorescent emission
related to galactic-scale outflows
(e.g.\ via deep integral-field-unit [IFU] observations,
\citealt{ssb+05,wsg08}) 
may offer unique
diagnostics on the spatial extent, morphology, and density of the outflow 
from distant galaxies,
eventually
setting tighter constraints on the energetics of the flow.   


Although astronomers are rapidly producing a wealth of observational
datasets on galactic-scale winds, a key ingredient to a proper
analysis is absent.  Just as comparisons between observed supernova 
lightcurves and spectra and radiative transfer calculations of
of SNe models have provided crucial insight into the physics driving,
e.g., standard candle relationships for both Type Ia and IIP SNe 
\citep{kw07,kw09}, modeling of the observable signatures of galactic
outflows is necessary for understanding the physical
properties of the gas and ultimately the physics driving the flows. 
In this paper, we take the first steps toward modeling the absorption
and emission properties of cool gas outflows as observed in one
dimensional spectra and by IFUs.  Using Monte Carlo
radiative transfer techniques, we study the nature of \ion{Mg}{2} and
\ion{Fe}{2} absorption and emission for winds with a range of
properties, accounting for the effects of resonant scattering and
fluorescence. 
Although the winds are idealized, the results frequently
contradict our intuition and 
challenge the straightforward conversion of observables to (even crude) physical
constraints.  These findings have a direct bearing on
recent and upcoming surveys of galactic outflows, particularly those
which make use of \ion{Mg}{2} and \ion{Fe}{2} transitions to probe
outflow properties.

The paper is organized as follows.  In $\S$~\ref{sec:method}, we
describe the methodology of our radiative transfer algorithms.  These
are applied to a fiducial wind model in $\S$~\ref{sec:fiducial} and
variations of this model in $\S$~\ref{sec:variants}.  In
$\S$~\ref{sec:alternate}, we explore wind models with a broader range
of density and velocity laws. We discuss the principal results and
connect to observations in $\S$~\ref{sec:discuss}.  A brief summary is
given in $\S$~\ref{sec:summary}. 

\section{Methodology}
\label{sec:method}

This section describes our methodology for generating
emission/absorption profiles from simple wind models.

\begin{deluxetable}{lcccccc}
\tabletypesize{\footnotesize}
\tablecolumns{11}
\tablecaption{\ion{Mg}{2} and \ion{Fe}{2} Transitions Considered \label{tab:atomic}}
\tablewidth{0pt}
\tablehead{\colhead{} & \colhead{$\rm E_{high}$} & \colhead{$\rm E_{low}$} & \colhead{$J_{\rm high}$} & \colhead{$J_{\rm low}$} & \colhead{$\lambda$} & \colhead{$A$} \\
 & \colhead{($\rm cm^{-1}$)} & \colhead{($\rm cm^{-1}$)} &&& \colhead{(\AA)} & \colhead{($\rm s^{-1}$)} } 
\startdata
\ion{Fe}{2} UV1 & 38458.98 &     0.00 &   9/2 & 9/2 & 2600.173 &
2.36E+08  \\
           & 38458.98 &   384.79 &   9/2 & 7/2 & 2626.451 & 3.41E+07 \\
           & 38660.04 &     0.00 &   7/2 & 9/2 & 2586.650 & 8.61E+07 \\
           & 38660.04 &   384.79 &   7/2 & 7/2 & 2612.654 & 1.23E+08 \\
           & 38660.04 &   667.68 &   7/2 & 5/2 & 2632.108 & 6.21E+07 \\
           & 38858.96 &   667.68 &   5/2 & 5/2 & 2618.399 & 4.91E+07 \\
           & 38858.96 &   862.62 &   5/2 & 3/2 & 2631.832 & 8.39E+07 \\
\tableline \\ [-1.5ex]
\ion{Mg}{2}& 35760.89 &     0.00 &   3/2 &   0 & 2796.351 & 2.63E+08\\
           & 35669.34 &     0.00 &   1/2 &   0 & 2803.528 & 2.60E+08\\
\enddata
\tablecomments{Atomic data from \citet{morton03}.}
\end{deluxetable}

\begin{figure}
\includegraphics[width=3.5in]{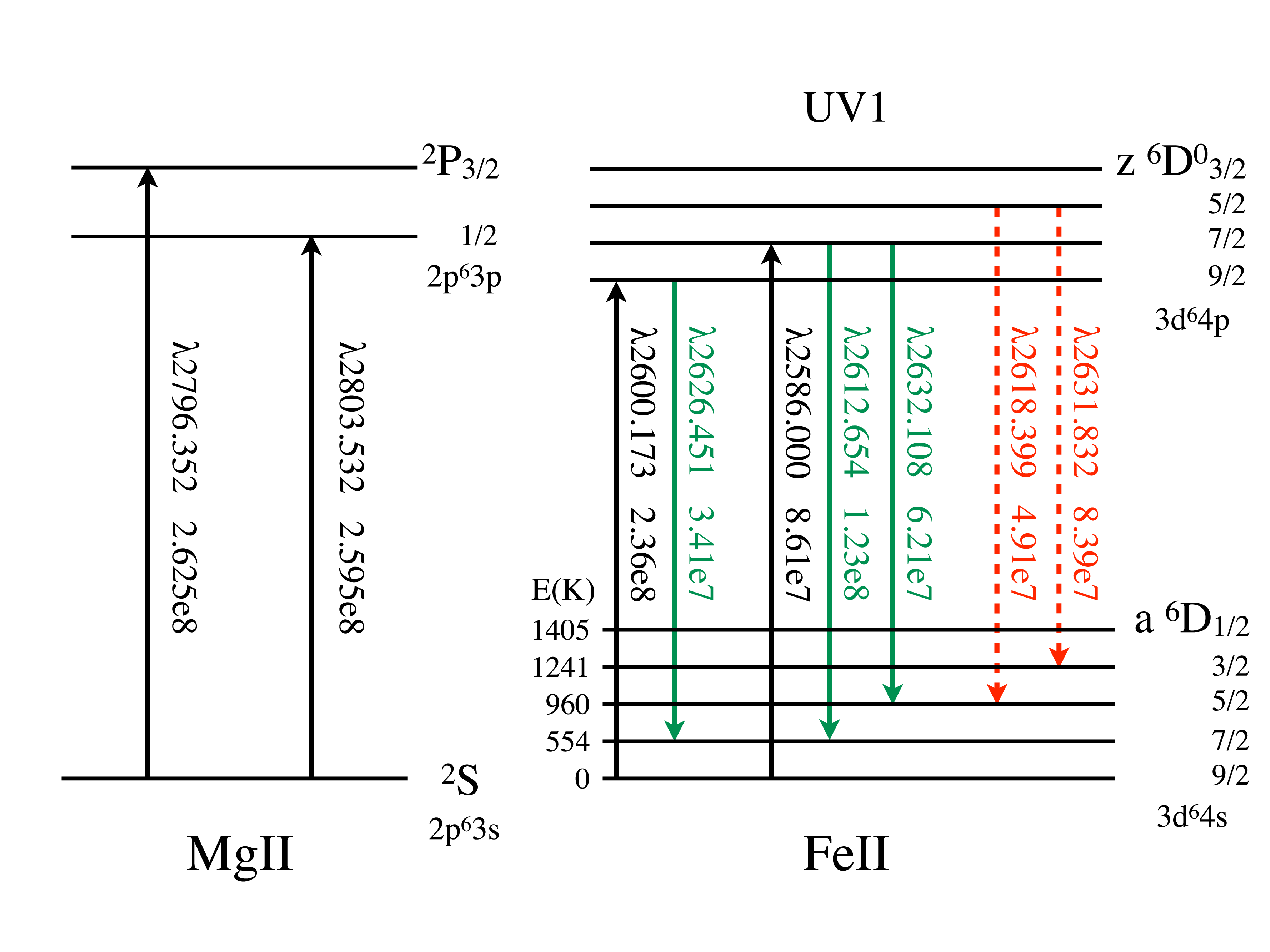}
\caption{
Energy level diagrams for the \mgiid\ doublet and the UV1
multiplet of \ion{Fe}{2} transitions   
(based on Figure~7 from \cite{hmt+99}).
Each transition shown is
labeled by its rest wavelength (\AA) and Einstein A-coefficient
(s$^{-1}$). Black upward arrows
indicate the resonance-line transitions, i.e.\ those connected to the ground
state.  The 2p$^6$3p configuration of Mg$^+$ is split into
two energy levels that give rise to the \mgiid\ doublet.  
Both the 3d$^6$4s ground state and 3d$^6$4p upper level of Fe$^+$
exhibit fine-structure splitting that gives rise to a series of
electric-dipole transitions. 
The downward (green) arrows show the \feiis\ transitions that are connected to the
resonance-line transitions (i.e.\ they share the same upper energy
levels).  We also show a pair of transitions (\ion{Fe}{2}$^* \lambda\lambda
2618,2631$; red and dashed lines) that arise from higher levels in the \zconfig\
configuration.  These transitions have not yet been observed in
galactic-scale outflows and are not considered in our analysis.
}
\label{fig:energy}
\end{figure}

\subsection{The Radiative Transitions}

In this paper, we focus on two sets of radiative transitions
arising from Fe$^+$ and Mg$^+$ ions
(Table~\ref{tab:atomic}, Figure~\ref{fig:energy}).
This is a necessarily limited
set, but the two ions and their transitions do have characteristics
shared by the majority of low-ion transitions
observed in cool-gas outflows. Therefore, many
of the results that follow may be generalized to observational studies that
consider other atoms and ions tracing cool gas.

The Mg$^+$ ion, with a single 3s electron in the ground-state,
exhibits an alkali doublet of transitions at $\lambda \approx
2800$\AA\ analogous to the
\lya\ doublet of neutral hydrogen.  Figure~\ref{fig:energy}
presents the energy level diagram for this 
\mgiid\ doublet.  In non-relativistic quantum
mechanics, the 2p$^6$3p energy level is said to be split by spin-orbit
coupling giving the observed line doublet.  These are the only
\ion{Mg}{2} electric-dipole transitions 
with wavelengths near 2800\AA\ and the transition connecting
the $\rm {}^2P_{3/2}$ and $\rm {}^2P_{1/2}$ states is forbidden by several
selection rules.  Therefore, an absorption from
\maconfig~$\to$~\mbconfig\
is followed $\approx 100\%$ of the time by a spontaneous decay
($t_{\rm decay} \approx 4\sci{-9}$s) to the
ground state. Our treatment will ignore any other possibilities
(e.g.\ absorption by a second photon when the electron is at the \mbconfig\ level).

In terms of radiative transfer, the 
\mgiid\ doublet is very similar to that for \ion{H}{1}
\lya, the \naid\ doublet, and many other doublets commonly
studied in the interstellar medium (ISM) of distant galaxies.  
Each of these has the ground-state connected to a pair of electric
dipole transitions with nearly identical energy.
The doublets differ only in 
their rest wavelengths and the energy of the doublet separation. 
For \ion{H}{1} \lya, the
separation is sufficiently small ($\Delta v = c \Delta E / E \approx
1.3 \mkms$) that most radiative transfer treatments actually ignore it
is a doublet.
This is generally justifiable for \lya\ because 
most astrophysical processes have turbulent motions that
significantly exceed the doublet's velocity separation and effectively mix the
two transitions.  For \ion{Mg}{2} ($\Delta v \approx 770 \mkms$),  
\ion{Na}{1} ($\Delta v \approx 304 \mkms$), and most of the other doublets
commonly observed, the separation is large and the transitions
must be treated separately.  

Iron exhibits the most complex set of energy levels for elements
frequently studied in astrophysics.  The Fe$^+$ ion alone has 
millions of energy levels recorded \citep{kurucz05}, and even this is an
incomplete list.  
One reason for iron's complexity is
that the majority of its configurations exhibit fine-structure splitting.
This includes the ground-state configuration (\aconfig) which is split
into 5 levels, 
labeled by the total angular momentum $J$, 
with excitation energies $T_{\rm ex} \equiv \Delta E / k$ ranging from
$T_{\rm ex} \approx 500-1500$\,K (Figure~\ref{fig:energy}).  
Transitions between these fine-structure levels are 
forbidden (magnetic-dipole) and have spontaneous decay times of several hours.  

In this paper, we examine transitions between the ground-state
configuration and the energy levels of the \zconfig\
configuration.  This set of transitions (named the
UV1 multiplet) have wavelengths near 2600\AA.
There are two resonance-line transitions\footnote{We adopt the
  standard convention that a ``resonance line'' is an electric-dipole
  transition connected to the ground-state.  We also label
  non-resonant transitions with an asterisk, e.g.\ \feiic.} 
associated with this multiplet (\feiid)
corresponding to $\Delta J = 0, -1$; these are indicated by upward (black) arrows
in Figure~\ref{fig:energy}. The solid (green) downward
arrows in Figure~\ref{fig:energy} mark the non-resonant \feiis\
transitions that are connected to
the upper energy levels of the resonance lines.  These transitions may
occur following the absorption of a single photon by Fe$^+$ in its
ground-state.   This process may also be referred to as fluorescence.
Note that two of these transitions (\feiis$\; \lambda\lambda 2626, 2632$) are
close enough in energy that their line profiles can overlap.

The Figure also shows (as dashed, downward arrows) two of the
\feiis\ transitions that connect to higher energy levels of the \zconfig\
configuration.  Ignoring collisions and recombinations, these
transitions may only occur after the absorption
of two photons: one to raise the electron from the ground-state to an
excited state and another to raise the electron from the excited state
to one of the \zconfig\ levels with $J \le 5/2$.  The excitation of
fine-structure levels by 
the absorption of UV photons is termed indirect UV pumping
\citep[e.g][]{silva02,pcb06} and requires the ion to lie
near an intense source of UV photons.  
Even a bright, star-forming galaxy emits too few photons at $\lambda
\approx 2500$\AA\ to UV-pump Fe$^+$ ions that are farther than $\sim
100$\,pc from the stars.
In the
following, we will assume that emission from this process is
negligible.

Our calculations also ignore collisional
processes\footnote{Recombination is also ignored.}, i.e.\ collisional
excitation and de-excitation of the various levels.  For the
fine-structure levels of the \aconfig\ configuration, the excitation
energies are modest ($T_{\rm ex} \sim 1000$\,K) but the critical
density $n_c$ is large.  For the \aconfig$_{9/2} \to
$\aconfig$_{7/2}$ transition, the critical density $n_e^C \approx 4
\sci{5} \cm{-3}$.  At these densities, one would predict  
detectable quantities of \ion{Fe}{1} which has not yet been observed
in galactic-scale outflows. 
If collisional excitation is insignificant 
then one may also neglect collisional de-excitation.  
Furthermore, observations rarely show
{\it absorption} from the
fine-structure levels of the \aconfig\ configuration and that material
is not significantly blue-shifted (Rubin et al., in prep). 
In the following, we assume that electrons only occupy the
ground-state, i.e.\ the gas has zero opacity to the non-resonant
lines.  
Regarding the \ion{Mg}{2} doublet, its
excitation energy is significantly higher implying 
negligible collisional processes at essentially any density.

\subsection{The Source}

Nearly all of the absorption studies of galactic-scale outflows 
have focused on intensely, star-forming galaxies.  The
intrinsic emission of these galaxies is a complex combination of
light from stars and \ion{H}{2} regions that is then modulated by dust and gas
within the ISM.  For the spectral regions studied
here, the hottest stars show a featureless continuum, but later spectral
types do show significant \ion{Mg}{2} and \ion{Fe}{2} absorption.
In addition, asymmetric and/or blue-shifted absorption is exhibited in these
transitions in A and F stars driving stellar winds \citep{sll+94}.
\ion{Mg}{2} P Cygni profiles are observed in a handful of F stars by \cite{sll+94}, who
attribute the emission to chromospheric activity rather than mass-loss effects.
\ion{H}{2} regions, meanwhile, are observed to emit weakly
at the \mgiid\ doublet, primarily due to recombinations in the outer
layers \citep{kbc+93}.
It is beyond the scope of this paper to
properly model the 
stellar absorption and \ion{H}{2} region emission, but the reader
should be aware that they can complicate the observed spectrum,
independently of any outflow, especially at velocities $v \approx 0 \mkms$.
In the following, we assume a simple flat continuum 
normalized to unit value.  The size of the emitting
region $r_{\rm source}$ is a free parameter, but we restrict its value
to be smaller than the minimum radial extent of any gaseous
component.   Lastly, the source does not absorb any scattered or
emitted photons.

\subsection{Monte Carlo Algorithm}
\label{sec:monte}

We calculated spectra using a 3D Monte Carlo radiation transport code
originally designed for supernova outflows  \citep{Kasen_2006} but
modified to treat resonant line transport  on a galactic scale
\citep{Kasen_lyman}.  Our methods are similar to those used in several
other codes which focus on \lya\
\citep[e.g.,][]{Zheng_2002,Dijkstra_2006,Verhame_2006,Laursen_2009},
except that here we include the effects of multiple line scattering
and fluorescence. 

For our 3-dimensional calculations, the  wind properties (density, temperature and velocity) were discretized on a $300^3$ Cartesian grid (for certain 1-D calculations, the wind properties were simply computed on the fly from analytic formulae).   
The radiation field was represented by $N$ photon packets (typically $N \sim 10^7$) which were initially emitted isotropically and uniformly throughout the source region ($r < r_{\rm source}$).  The wavelengths of the packets were sampled from a flat spectral energy distribution. 

Photon packets were tracked through randomized scattering and
absorption events until they escaped the computational domain.  The
distance to the next packet interaction event was determined by
Doppler shifting packets to the comoving frame and sampling the mean
free path to each resonance line (and dust, if present).  The resonant
line opacity followed a Voigt profile which was determined using the
analytic fits of  \cite{Tomi_2006}.      Non-resonant lines were
assumed to be completely optically thin, and the dust opacity was
assumed to be wavelength independent and completely absorbing
($\S$~\ref{sec:dust_method}).
Because of the relatively low optical depths encountered in
these models, no ``core-skipping'' scheme was applied to accelerate
the resonant line transport,  and recoil effects were ignored. 

In a resonant line interaction, a photon excites an atom from the
ground state to an excited level.  The end result is either scattering
(i.e., de-excitation back to the ground state) or fluorescence
(de-excitation to another excited  level).  The probability that the
atom de-excites to lower level $x$ is  $p_{ux} = A_{ux} / \sum_i A_{ui}$, where the $A_{ui}$ are the Einstein spontaneous emission coefficients and the sum runs over all levels accessible from the upper state $u$.  
For each interaction event, a random number was drawn to sample the
final state from this probability distribution.    If the result was a
scattering, the process was assumed to be coherent in the comoving
frame.  If the result was fluorescence, the packet was  reemitted at
the line center wavelength of the new transition.  Natural line
broadening in fluorescence was ignored given the high velocity
gradients in our models.  In all cases, the angular redistribution was
assumed to be isotropic. 

To generate multi-dimensional images and spectra of the system, an escape probability method was used.  At the initiation of a packet, and at every subsequent interaction event, we calculated the probability $P_x = e^{-\tau_{ux}} p_{ux}$,  that the packet de-excited into state $x$ and escaped the domain in some pre-specified direction. Here
$\tau_{ux}$ is the optical depth to infinity, which was  constructed
by integration along the path to escape, taking into account the
relevant Doppler shifts and possible absorption by lines or dust.  The
contribution of the packet to the final spectrum was then added in for
every possible final state, each shifted to the proper observer-frame
wavelength and weighted by the probability $P_x$. 

\begin{figure}
\includegraphics[width=3.5in]{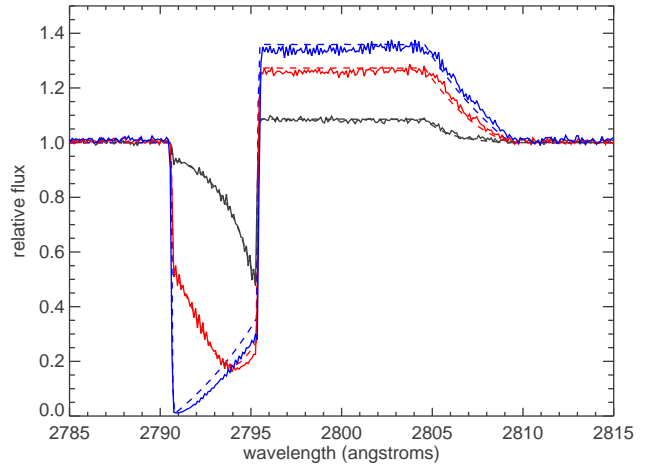}
\caption{
Line profiles of a test problem calculated using the monte carlo code (solid) and by direct numerical integration of the transport equation under the Sobolev approximation (dashed lines).  The configuration was  spherical and consisted of a homologously expanding wind  with density profile $n \propto r^{-2}$  lying above an opaque emitting surface at velocity coordinate 100~km~s$^{-1}$.  
The wind extended to 2,000 km~s$^{-1}$ and the Sobolev optical depth
had values of  $\tau = 1$ (black lines), $\tau = 10$ (red lines), and
$\tau = 100$ (blue lines).    Discrepancies in the $\tau = 100$ case
are likely due to the breakdown of the Sobolev approximation
($\S$~\ref{sec:Sobolev}) at high optical depth. 
}
\label{fig:oneline_test}
\end{figure}

To validate the monte carlo code, we calculated a series of test
problems consisting of a point source embedded in a spherical,
homogenous medium of varying resonant line optical depth.  The
resulting resonant line profiles  \citep[see][]{Kasen_lyman} were 
found to be in good agreement with the standard analytic solutions
\citep{Harrington_1973,Neufeld_1990,Dijkstra_2006}.  As a further
test, we calculated line profiles for the case of an extended
spherical source embedded in a homologously expanding wind with large
velocity gradient.  These model spectra
(Figure~\ref{fig:oneline_test}) are in good agreement with ones
determined by direct numerical integration of the radiation transport
equation under the Sobolev approximation
\citep[$\S$~\ref{sec:Sobolev}; e.g.,][]{Jeffery_Branch}.

\subsection{Dust}
\label{sec:dust_method}

For the majority of models studied in this paper, we assume the gas
contains no dust.
This is an invalid assumption, especially for material associated with
the ISM of a galaxy.  
Essentially all astrophysical environments that contain both cool gas
and metals also show signatures of dust depletion and extinction.  This includes the
ISM of star-forming and \ion{H}{1}-selected galaxies
\citep[e.g.][]{ss96,pw01,pcd+07}, strong \ion{Mg}{2} metal-line
absorption systems \citep{ykv+06,mnt+08}, and the galactic winds traced
by low-ion transitions \citep{prs+02,rvs05b}.  
Extraplanar material likely associated with a galactic-scale 
outflow has been observed to emit IR radiation characteristic of dust 
\citep[e.g.][]{hrg90,adb99,rkl01}.
In addition, \cite{msf+10} have argued from a statistical analysis
that dust is distributed to many tens of kpc from $z \sim 0.1$ galaxies
and have suggested it was transported from the galaxies by
galactic-scale winds.  
Although the galactic winds
traced specifically by \ion{Mg}{2} and \ion{Fe}{2} transitions have not (yet) been
demonstrated to contain dust, it is reasonable to consider its effects.

For analysis on normalized spectra (i.e.\ absorption lines), the effects of dust are largely minimized; 
dust has a nearly constant opacity over small spectral regions and all
features are simply scaled together.
For scattered and resonantly trapped photons, however, the relative effect of dust
extinction can be much greater.  
These photons travel a much longer distance to escape the medium and
may experience a much higher integrated opacity from dust.
Indeed, dust is frequently invoked to
explain the weak (or absent) \lya\ emission from star-forming galaxies
\citep[e.g.][]{shapley03}.  Although the transitions studied here have
much lower opacity than \lya, dust could still play an important role
in the predicted profiles.

In a few models, we include absorption by dust under the following
assumptions:
(i) the dust opacity scales with the density of the gas (i.e.\ we
adopt a fixed dust-to-gas ratio);
(ii) the opacity is independent of wavelength, a reasonable
approximation given the small spectral range analyzed;
(iii) dust absorbs but does not scatter photons;
(iv) the photons absorbed by dust are re-emitted at IR wavelengths
and are `lost' from the system.  The dust absorption is normalized by \taud, 
the integrated opacity of dust from the center of the system to
infinity.  The ambient ISM of a star-forming galaxy may be expected to exhibit
\taud\ values of one to a few at $\lambda \sim 3000$\AA\ \citep[e.g.][]{cf00}.

\subsection{The Sobolev Approximation}

A photon propagating through a differentially expanding medium
interacts with a line only when
its comoving frame wavelength is Doppler-shifted into near resonance with the line center rest wavelength. 
For a wind with a steep velocity law  (i.e.\ a large gradient $dv/dr$)
and/or a narrow intrinsic profile (i.e.\ a small Doppler parameter), the 
spatial extent of the region of resonance
may be much smaller than the length scale of the
wind itself.  The interaction can then 
be considered to occur at  a point.
In this case, \citet{sobolev60} introduced a formalism that gives the line optical depth $\tau(r)$ at  
a given point in terms of the density and
velocity gradient of the flow at that radius.  For a wind in homologous expansion,
this optical depth is independent of the direction of propagation and
is given by
\begin{equation}
\tau_S(r) = K_\ell   \lambda^0_\ell | dv/dr |^{-1} \cmma
\label{eqn:Sobolev}
\end{equation}
where 
\begin{equation}
K_\ell = \frac{\pi e^2}{m_e c} f_\ell n_\ell(r)
\end{equation}
is the integrated line opacity, $\lambda^0_\ell$ the line-center rest wavelength,
$f_\ell$ the oscillator strength and
$n_\ell$ the density in the lower level of the transition.   We have
neglected corrections for stimulated emission.   This optical
depth applies to a photon with wavelength $\lambda_S
\equiv \lambda_\ell (1- v(r)~\hat{r}  \cdot \hat{d} /c)$ where $\hat{d}$ is the 
direction of propagation.  The probability that such a photon
is scattered/absorbed at the point of resonance is simply $1 - \exp[-\tau_S(r)]$.
We find that the Sobolev approximation applies for nearly all of the
models presented in this paper, and therefore provides a convenient
approach to estimating the optical depth.

\section{The Fiducial Wind Model}
\label{sec:fiducial}

In this section, we study a simple yet illustrative wind model for
a galactic-scale outflow.  The properties of this wind were tuned, in
part, to yield a \ion{Mg}{2} absorption profile 
similar to those observed for $z \sim 1$, star-forming galaxies
\citep{wcp+09,rubin+10c}.  We emphasize, however, that we do not
favor this fiducial model over any other wind scenario nor do its
properties have special physical motivation.
Its role is to establish a baseline
for discussion.

The fiducial wind is isotropic, dust-free, and extends from an inner wind
radius $r_{\rm inner}$ to an outer wind radius $r_{\rm outer}$.  
It follows a density law,

\begin{equation}
n_{\rm H} (r) = n^0_{\rm H} \biggl( \frac{r_{\rm inner}}{r}  \biggr)^2\;\;\; , 
\label{eqn:density}
\end{equation}
and a velocity law with a purely radial flow

\begin{equation}
\vec v = v_r (r) \hat r = \frac{v_0 r}{r_{\rm outer}} \hat r  \perd
\label{eqn:vel}
\end{equation}
For the fiducial case,  the hydrogen density at
the inner radius is $n_{\rm H}^0 = 0.1~{\rm cm}^{-3}$ and the
velocity at the outer radius is $v_0 = 1,000$~\kms. Turbulent motions are
characterized by a Doppler parameter\footnote{
  The adopted Doppler parameter
 has a minor impact on the results.
  The absorption profiles are insensitive to its value and varying \bturb\
  only tends to modify the widths and modestly
  shift the centroids of emission lines.} 
$b_{\rm turb} = 15 \mkms$.  
We convert the hydrogen density $n_{\rm H}$ to the number densities of
Mg$^+$ and Fe$^+$ ions by assuming solar relative abundances with an
absolute metallicity of 1/2 solar and depletion factors of 1/10 and
1/20 for Mg and Fe respectively, i.e.\  $\mnmg = 10^{-5.47} n_{\rm H}$ 
and \nfe=\nmg/2. At the center of the wind is a homogeneous source of
continuum photons with size $r_{\rm source}$. The parameters for the
fiducial wind model are summarized in Table~\ref{tab:fiducial}.   

\begin{deluxetable}{ccl}
\tablewidth{0pc}
\tablecaption{Wind Parameters: Fiducial Model\label{tab:fiducial}}
\tabletypesize{\footnotesize}
\tablehead{\colhead{Property} & \colhead{Parameter} & \colhead{Value} } 
\startdata
Density law  & $n(r)$ & $\propto r^{-2}$ \\
Velocity law  & $v_r$ & $ \propto r$ \\
Inner Radius & $r_{\rm inner}$ & 1\,kpc \\
Outer Radius & $r_{\rm outer}$ & 20\,kpc \\
Source size  & $r_{\rm source}$ & 0.5\,kpc \\
Density Normalization & $n^0_{\rm H}$ & $0.1\cm{-3}$ at $r_{\rm inner}$ \\
Velocity Normalization & $v_0$ & 1000\kms at $r_{\rm outer}$ \\
Turbulence   & $b_{\rm turb}$  & 15 \kms \\
Mg$^+$ Normalization & \nmg\ & $10^{-5.47} n_{\rm H}$ \\
Fe$^+$ Normalization & \nfe\ & \nmg/2 \\
\enddata

\end{deluxetable}

\begin{figure}
\includegraphics[height=3.5in,angle=90]{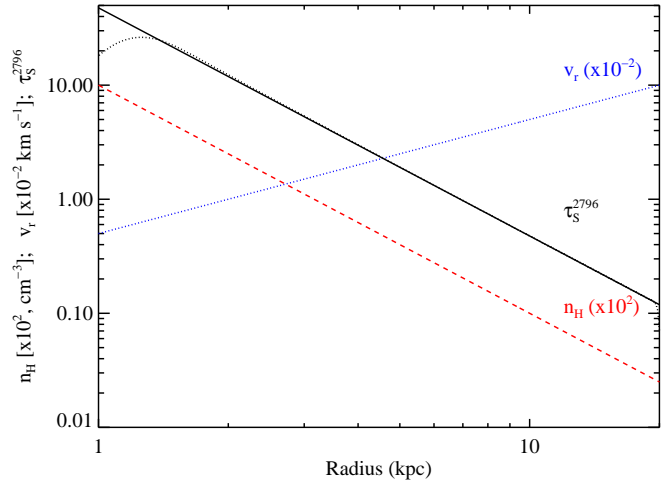}
\caption{
Density (dashed; red), radial velocity (dotted; blue), and
\mgiia\ optical depth profiles (solid and dotted; black) for the fiducial
wind model (see Table~\ref{tab:fiducial} for details).
The density and velocity laws are simple $r^{-2}$ and $r^1$
power-laws; these curves have been scaled for plotting
convenience.  
The optical depth profile was calculated two ways: (i) using the
Sobolev approximation (Equation~\ref{eqn:Sobolev}; solid black curve)
and (ii) summing
the opacity at small and discrete radial intervals in velocity space
and then converting to radius with the velocity law (doted black
curve).  These give
very similar results. 
The wind parameters were set to give an optically thick medium at the
inner radius ($r_{\rm inner} = 1$\,kpc) that becomes optically thin at
the outer radius ($r_{\rm outer} = 20$\,kpc). 
}
\label{fig:fiducial_nvt}
\end{figure}

In Figure~\ref{fig:fiducial_nvt} we plot the density and velocity
laws against radius;  
their simple power-law expressions are evident.  The figure also
shows the optical depth profile for the \mgiia\ transition
($\tau_{2796}$), estimated from the Sobolev approximation\footnote{We
  have verified this approximation holds by calculating 
  $\tau_{2796}$ first
  as a function of velocity by summing the opacity for a series of
  discrete and small radial intervals
  between $r_{\rm inner}$  and $r_{\rm outer}$.   We then mapped
  $\tau_{2796}$ onto radius using the velocity law
  (Equation~\ref{eqn:vel}). This optical depth profile is shown as the
black dotted line in Figure~\ref{fig:fiducial_nvt}.}
(Equation~\ref{eqn:Sobolev}). 
The $\tau_{2796}$ profile peaks with $\tau_{2796}^{\rm max} \approx 30$
at a velocity $\mvr \approx 65 \mkms$ corresponding to $r \approx 1.3
r_{\rm inner}$.  The optical depth profile for the \mgiib\ transition
(not plotted) is scaled down by the $f\lambda$ ratio but is otherwise identical.  Similarly,
the optical depth profiles for the \feiid\ transitions are
scaled down by $f \lambda$ and the \nfe/\nmg\ ratio.  

For the spatially integrated (i.e., 1D) spectra, the actual dimensions and density of the wind are
unimportant provided they scale together to give nearly the same
optical depth. Therefore, one may consider the choices for
$r_{\rm inner}, r_{\rm outer}$, and $n_{\rm H}^0$ as largely arbitrary.
Nevertheless, we adopted values for this fiducial model with
some astrophysical motivation,  e.g., values that correspond to
galactic dimensions and a normalization that gives $\tau_{2796}^{\rm
  max} \sim 10$.

\begin{deluxetable}{ccr}
\tablewidth{0pc}
\tablecaption{Equivalent Widths for the Fiducial Model\label{tab:fiducial_EW}}
\tabletypesize{\footnotesize}
\tablehead{\colhead{Transition} & \colhead{$v_{\rm int}^a$} & \colhead{$W_{\lambda}^b$} \\ 
& (\kms) & (\AA) }
\startdata
  MgII 2796  &[$-1030,-65$]& 2.96\\
&[$-65,311$]&$-1.75$\\
  MgII 2803  &[$-458,-57$]& 1.29\\
&[$-57,692$]&$-2.34$\\
  FeII 2586  &[$-365,-46$]& 0.59\\
&[$-46,186$]&$-0.10$\\
  FeII 2600  &[$-626,-78$]& 1.16\\
&[$-49,470$]&$-0.94$\\
  FeII* 2612 &[$-219,183$]&$-0.33$\\
  FeII* 2626 &[$-194,177$]&$-0.27$\\
  FeII* 2632 &[$-155,130$]&$-0.16$\\
\enddata
\tablenotetext{a}{Velocity interval over which the equivalent width is calculated.}
\tablenotetext{b}{Negative values indicate line-emission.}
\end{deluxetable}
 
\begin{figure*}
\includegraphics[height=7.0in,angle=90]{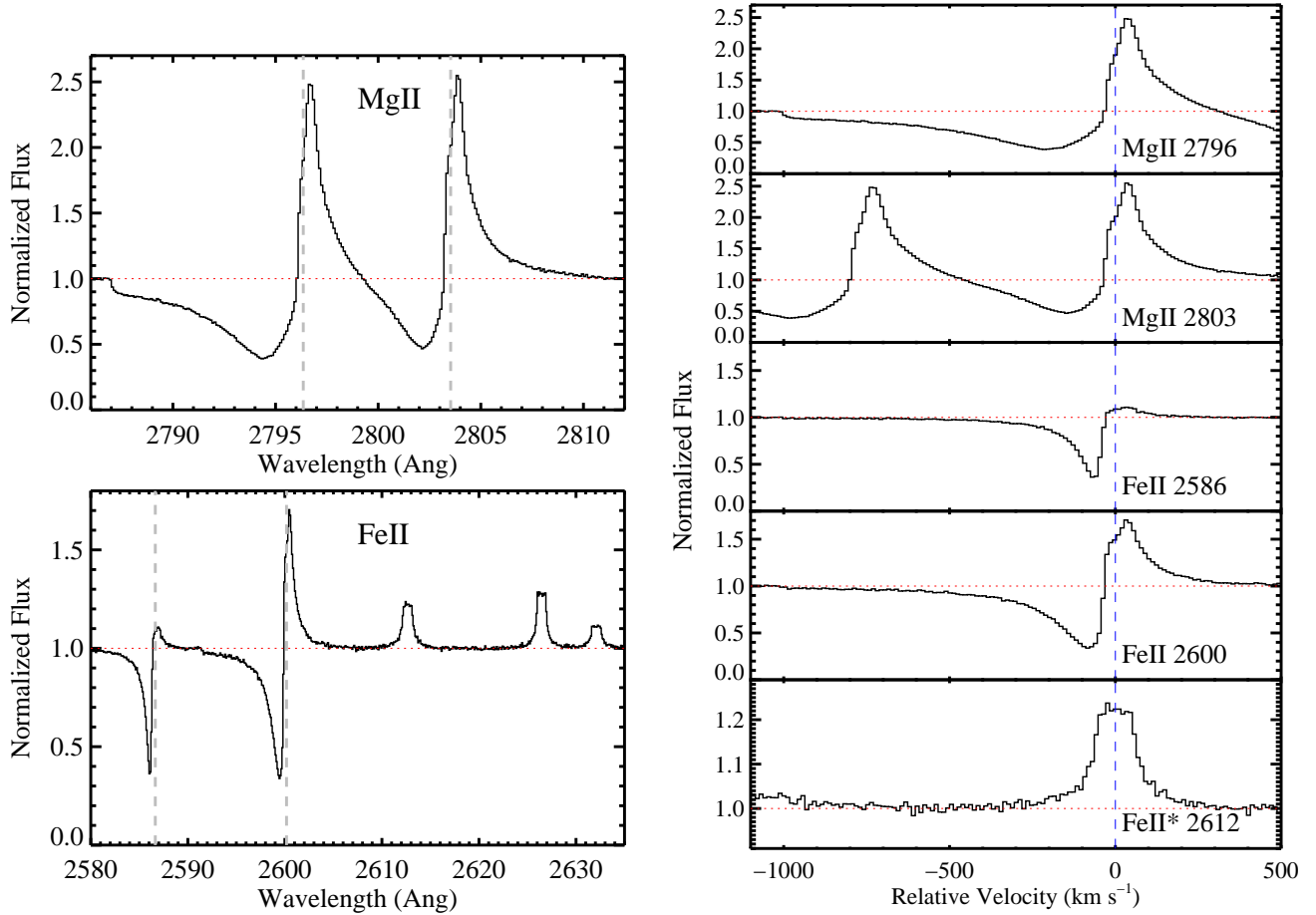}
\caption{
{\it Left} -- (Upper) \mgiid\ profiles for the fiducial wind model
described in Table~\ref{tab:fiducial} and
Figure~\ref{fig:fiducial_nvt}.  The doublet shows the P-Cygni profiles
characteristic of an outflow with significant absorption blueward of
line-center (dashed vertical lines) extending to $v = -1000\mkms$
and significant emission redward of line-center.  Note
that even though the peak optical depth of the \ion{Mg}{2} transitions
is $\tau_{2796}^{\rm max} \approx 30$ at $v \approx -70 \mkms$,
photons scattered off the outflow significantly fill-in the absorption.
(Lower) \ion{Fe}{2} absorption and emission profiles for the UV1
multiplet at $\lambda \approx 2600$\AA.  The \feiid\ resonance lines 
show weaker absorption due to the smaller Fe$^+$ number density and
lower $f\lambda$ values.  Each also shows a P-Cygni profile, although
the emission for \feiia\ is much weaker than that for the
\feiib\ and \mgiid\ transitions.  This is because a majority of the
absorbed \feiia\ photons fluoresce into
\ion{Fe}{2}$^*~\lambda\lambda 2612, 2632$ photons.
{\it Right} -- A subset of the transitions displayed in a velocity
plot.
}
\label{fig:fiducial_1d}
\end{figure*}

Using the methodology described in $\S$~\ref{sec:method}, we
propagated photons from the source and through the outflow to an
`observer' at $r \gg r_{\rm outer}$ who views the entire wind+source
complex.  Figure~\ref{fig:fiducial_1d} presents the 1D spectrum
that this observer would record, with the unattenuated flux
normalized to unit value.   The \ion{Mg}{2} doublet
shows the canonical `P-Cygni profile' that characterizes a continuum
source embedded within an outflow.  Strong absorption is evident at
$\delta v  \equiv (\lambda/\lambda_0 - 1) < -50 \mkms$ in both
transitions (with equivalent widths $W^{2796}_a =
3.0$\AA\ and $W^{2803}_a = 1.3$\AA) and each shows emission at
positive velocities.  For an isotropic and dust-free model, the
total equivalent width of the doublet must be zero,
i.e.\ every photon
absorbed eventually escapes the system, typically at lower
energy.  The wind simply shuffles the photons in frequency space.
A simple summation of the absorption and emission equivalent widths
(Table~\ref{tab:fiducial_EW}) confirms this expectation.

\begin{figure}
\includegraphics[width=3.5in]{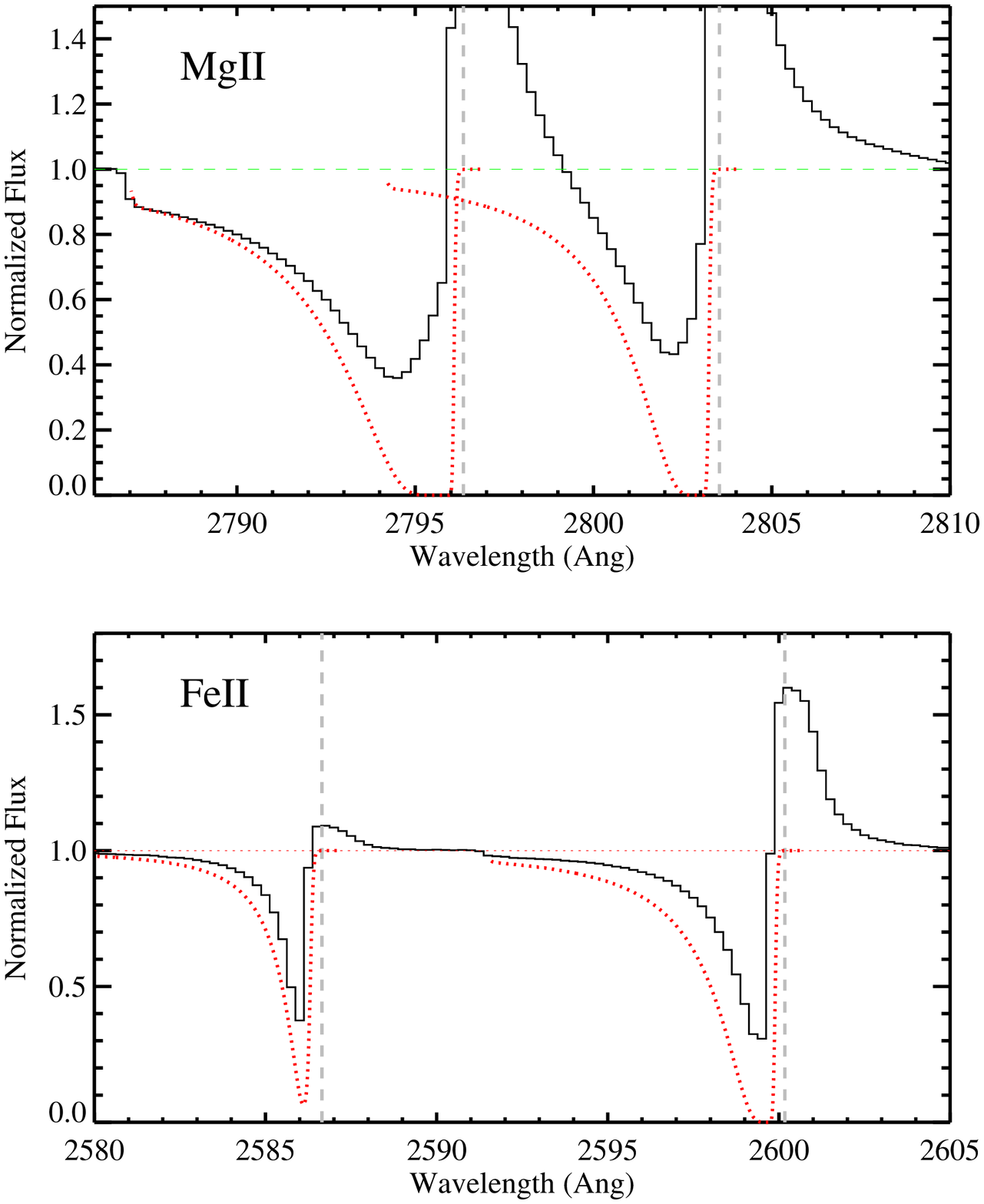}
\caption{
The solid curves show the line profiles (absorption and emission) of
the \ion{Mg}{2} and \ion{Fe}{2} resonance lines for the fiducial wind
model.  These include the effects of scattered photons and show the
canonical P-Cygni profiles of a source 
embedded within an outflow.  Overplotted on each transition (dotted
red line) is
the predicted absorption profile under the constraint that every
absorbed photon is lost from the system (referred to as the
intrinsic profile), i.e.\ no scattering or
re-emission occurs.   The intrinsic profiles have been `filled-in' 
significantly with photons scattered by the wind into our sightline.  Ignoring this
process, one would model the absorption lines with a systematically lower
optical depth and/or (incorrectly) conclude that the source is partly covered by the
gas.  
}
\label{fig:noemiss}
\end{figure}

Focusing further on the \ion{Mg}{2} absorption, one notes that the profiles lie
well above zero intensity and have similar depth even though their $f\lambda$
values differ by a factor of two.  In standard absorption-line
analysis, this is 
the tell-tale signature of a `cloud' that has a high optical depth (i.e.\
is saturated) which only partially covers the emitting source
\citep[e.g.][]{rvs05a,hkp+10}.  Our fiducial wind model, however, 
{\it entirely covers the source}; the apparent partial covering must
be related to a different effect.
Figure~\ref{fig:noemiss} further emphasizes this point by comparing the 
absorption profiles from Figure~\ref{fig:fiducial_1d} against an
artificial model where no absorbed photons are 
re-emitted.   As expected from the
$\tau_{2796}$ profile (Figure~\ref{fig:fiducial_nvt}), this
`intrinsic' model
produces a strong \mgiid\ doublet that absorbs all photons at
$\delta v \approx -100 \mkms$, i.e.\ $I_{2796}^{\rm min} = \exp(-\tau^{\rm
  max}_{2796}) \approx 0$.
The true model, in contrast, has been `filled in' at $\delta v \approx -100
\mkms$ by photons scattered in the wind.  An
absorption-line analysis that ignores these effects
would (i) systematically underestimate the true optical
depth and/or (ii) falsely conclude that the wind partially covers the
source.  We will find that these are
generic results, even for wind models that include
dust and are not fully isotropic.

Turning to the emission profiles of the \mgiid\ doublet, one notes
that they are quite similar with comparable equivalent widths.
This is because the gas is optically
thick yielding comparable total absorption. The
flux of the \mgiib\ transition even exceeds that for \mgiia\ 
because the wind speed is greater than the velocity separation
of the doublet, $|\mvr|_{\rm max} > (\Delta v)_{\rm MgII}$.
Therefore, the red wing of the
\mgiia\ emission profile is partially absorbed by \mgiib\ and
re-emitted at lower frequency.
This yields a
line ratio that is far below the $2:1$ ratio that one may have naively
expected (e.g.\ if the line-emission resulted from recombinations),
and leads us to conclude that the relative 
strengths of the emission lines are sensitive to both the opacity and
velocity extent of the wind. 

Now consider the \ion{Fe}{2} transitions:
the bottom left panel of Figure~\ref{fig:fiducial_1d} covers the
majority of the \ion{Fe}{2} UV1 transitions and several are
shown in the velocity plot.  The line
profile for \feiib\ is very similar to the \mgiid\ doublet;
one observes strong absorption to negative velocities and strong
emission at $\mdv > 0 \mkms$ producing a characteristic P-Cygni profile. 
Splitting
the profile at $\mdv = -50 \mkms$, we measure an equivalent width
$W^{2600}_{\rm a} = 1.16$\AA\ in absorption and $W^{2600}_{\rm e} =
-0.94$\AA\ in emission (Table~\ref{tab:fiducial_EW}) for a total
equivalent width of $W^{2600}_{\rm TOT} \approx 0.22$\AA.  
In contrast, the \feiia\ resonance line shows much weaker emission and
a much higher total equivalent width ($W^{2586}_{\rm TOT} = 0.49$\AA),
even though the line has a $2 \times$ lower $f\lambda$ value.
These differences between the \ion{Fe}{2} resonance lines (and between
\ion{Fe}{2} and \ion{Mg}{2}) occur because of the complex of non-resonant
\feiis\ transitions that are coupled to the resonance lines
(Figure~\ref{fig:energy}).  Specifically, 
a resonance photon absorbed at \feiid\ has a finite probability of
being re-emitted as a non-resonant photon which then escapes the system
without further interaction.  The principal effects are to reduce the
line emission of \feiid\ and to produce non-resonant line-emission (e.g.\
\feiic).

The reduced \feiia\ emission relative to \feiib\ is related to
two factors:
(i) there is an additional downward transition from the
\zconfig$_{7/2}$ level and 
(ii) the Einstein A
coefficients of the non-resonant lines coupled to \feiia\ are comparable to and even
exceed the Einstein A coefficient of
the resonant transition.  In contrast, 
the \feiie\ transition (associated with \feiib)
has an approximately  $4\times$ smaller A coefficient than the
corresponding
resonance line.  Therefore, the majority of photons absorbed at
$\lambda \approx 2600$\AA\ are re-emitted as \feiib\ photons, whereas 
the majority of photons absorbed at $\lambda \approx 2586$\AA\ are re-emitted 
at longer wavelengths (\feiic\ or $\lambda 2632$).
If we 
increase $\tau_{2600}$ (and especially if we include
gas with $\mvr \approx 0 \mkms$) then the \feiib\ emission is
significantly suppressed (e.g.\ $\S$~\ref{sec:ISM}).
The total equivalent width, however, of the three lines connected to the
\zconfig$_{7/2}$ upper level must still vanish (photons are conserved
in this isotropic, dust-free model).

\begin{figure}
\includegraphics[width=3.5in]{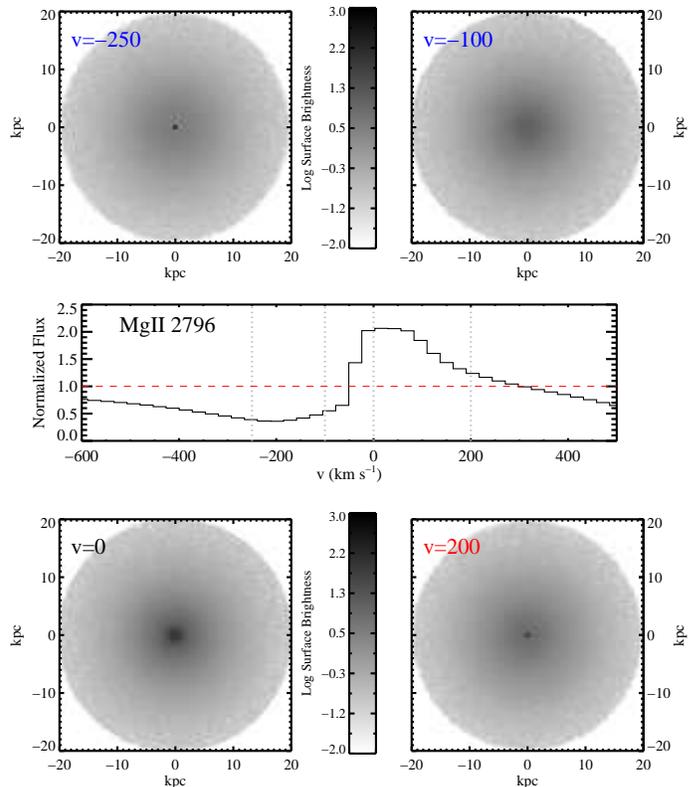}
\caption{
Surface-brightness emission maps around the \mgiia\ transition for the
source+wind complex of the fiducial model.  The middle panel shows the
1D spectrum with $v=0 
\mkms$ corresponding to $\lambda = 2796.35$\AA\ and the dotted vertical
curves indicate the velocity slices for the emission maps.  The
source has a size $r_{\rm source} = 0.5$\,kpc, traced by a few
pixels at the center of each map.   At $v=-250 \mkms$, the wind has an optical
depth of $\tau_{2796} \approx 1$ and the source contributes
roughly half of the observed flux.  At $v=-100 \mkms$ the
wind absorbs all photons from the source and the observed emission is
entirely due to photons scattered by the wind.  Amazingly, this
emission exceeds the integrated flux at $v = -250\mkms$ such that the
\mgiia\ line center is offset from the (intrinsic) peak in optical depth.  At $v \ge 0
\mkms$,  both the source and wind contribute to the observed emission.
At all velocities, the majority of emission comes from the inner
$\approx 5$\,kpc.
}
\label{fig:fiducial_ifu_mgii}
\end{figure}

The preceding discussion emphasizes the filling-in of resonance
absorption at $\mdv
\lesssim -50 \mkms$ and the generation of emission lines at $\mdv \approx
0 \mkms$ by photons scattered in the wind.  We also
mapped the emission of the fiducial
model to sudy its spatial extent (see $\S$~\ref{sec:monte} for a description of the algorithm).
The output is a set of surface-brightness maps in a series of 
frequency channels yielding a
dataset analogous to integral-field-unit (IFU) observations.  In
Figure~\ref{fig:fiducial_ifu_mgii}, we present the output 
at several velocities relative to the \mgiia\
transition. At $\mdv = -250 \mkms$, where the wind has an optical
depth $\tau_{2796} < 1$ (Figure~\ref{fig:fiducial_nvt}),
the source contributes roughly half of the observed flux.  
At $\mdv=-100 \mkms$, however, the
wind absorbs all photons from the source and the observed emission is
entirely from photons scattered by the wind.  This scattered emission
actually exceeds the source+wind emission at 
$\mdv = -250 \mkms$ such that the absorption profile is
negatively offset from the velocity where $\tau_{2796}$ is maximal
(Figure~\ref{fig:fiducial_nvt}).
The net result is
weaker \ion{Mg}{2} absorption that peaks blueward of the actual peak in the
optical depth profile.  
Table~\ref{tab:line_diag} reports several kinematic measurements of
the absorption and emission features.
Clearly, these effects complicate estimates for the
speed, covering fraction, and total column density of the wind.  At
$\mdv = 0 \mkms$, the wind and source have comparable total flux with the
latter dominating at higher velocities.  

\begin{figure}
\includegraphics[width=3.5in]{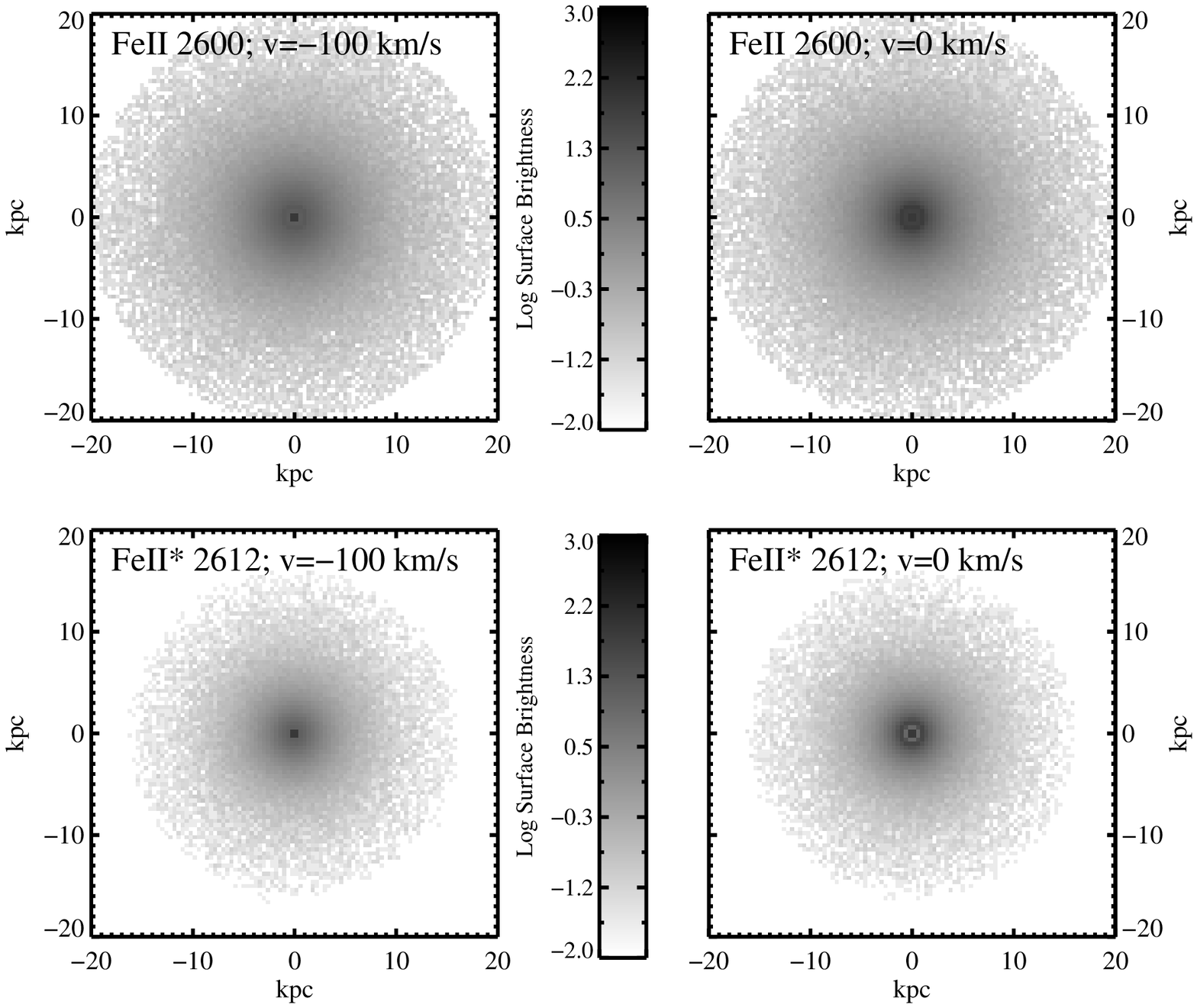}
\caption{
(Upper) Surface-brightness emission maps around the \feiib\ transition for the
source+wind complex.  
The results are very similar to those observed for the \mgiid\ doublet
(Figure~\ref{fig:fiducial_ifu_mgii}).
(Lower) Surface-brightness emission maps around the \ion{Fe}{2}~$\lambda
2612$ transition for the 
source+wind complex.  
In this case, the source is unattenuated yet scattered photons from
the wind also make a significant contribution. 
}
\label{fig:fiducial_ifu_feii}
\end{figure}

Similar results are observed for the \ion{Fe}{2} resonance
transitions (Figure~\ref{fig:fiducial_ifu_feii}).
For transitions to fine-structure levels of the \aconfig\
configuration, the source
is unattenuated but there is a significant contribution from photons
generated in the wind. 

At all velocities, the majority of light comes from the inner regions
of the wind. 
The majority of
\ion{Mg}{2} emission occurs within the inner few kpc, e.g.\ $50-60\%$
of the light at $\mdv=-100$ to $+100 \mkms$
comes from $|r| < 3$\,kpc.
The emission
is even more centrally concentrated for the \ion{Fe}{2} transitions.
A proper treatment of these
distributions is critical for interpreting observations
acquired through a slit, i.e.\ where the aperture has a limited extent
in one or more dimensions.  A standard longslit on 10m-class
telescopes, for example, subtends $\approx 1''$ corresponding to
$5-10$\,kpc for $z \sim 1$.    We return to this issue in
$\S$~\ref{sec:discuss}. 

The results presented in Figures~\ref{fig:fiducial_ifu_mgii} and
\ref{fig:fiducial_ifu_feii} are sensitive to the radial extent,
morphology, density and velocity profiles of this galactic-scale
wind.  Consequently, IFU observations of line emission from low-ion
transitions may offer the most direct constraints on galactic-scale
wind properties. 

\begin{figure}
\includegraphics[width=3.5in]{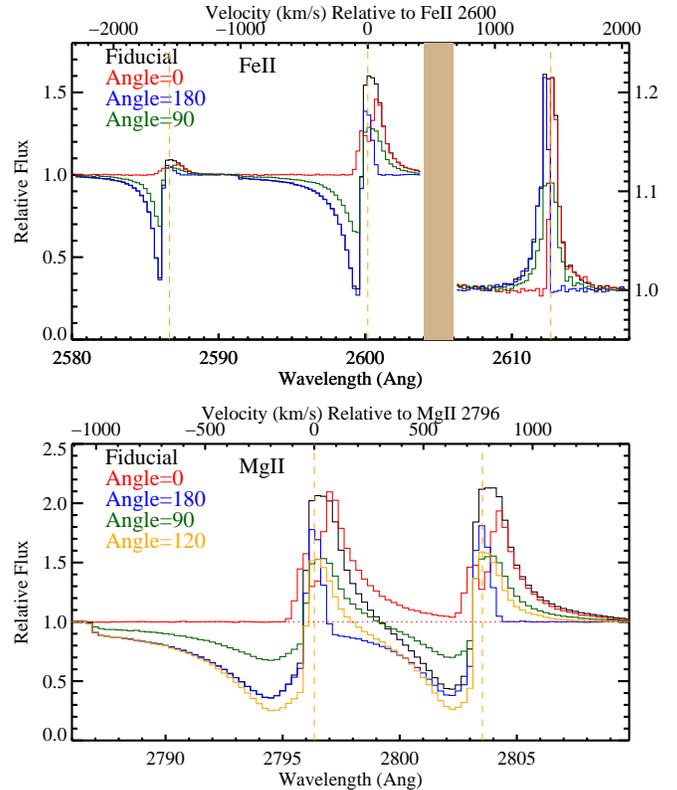}
\caption{
Profiles of the \ion{Fe}{2} and \ion{Mg}{2} profiles for the fiducial
case (black lines) compared against an anisotropic wind blowing into
only $2\pi$ steradians as viewed from $\phi = 0^\circ$ (source
uncovered) to $\phi = 180^\circ$ (source covered).  One detects
significant emission for all orientations but significant
absorption only for $\phi \ge 90^\circ$.
The velocity centroid of the 
emission shifts from positive to negative velocities as $\phi$
increases and one transitions from viewing the wind as lying behind the
source to in front of it.  The velocity centroid of emission, therefore,
diagnoses the degree of anisotropy for the wind.
}
\label{fig:anisotropic}
\end{figure}

\section{Variations to the Fiducial Model}
\label{sec:variants}

In this section, we investigate a series of more complex wind
scenarios
through modifications to the fiducial model.  These include relaxing
the assumption of isotropy, introducing dust, adding an ISM
component within $r_{\rm inner}$, and varying the normalization of the
optical depth profiles.

\subsection{Anisotropic Winds}
\label{sec:anisotropic}

The fiducial model assumes an
isotropic wind with only radial variations in velocity and density. 
Angular isotropy is obviously an idealized case, but
it is frequently assumed in studies of galactic-scale outflows
\citep[e.g.][]{steidel+10}.   There are several reasons, however, to
consider anisotropic winds.  Firstly, galaxies are not spherically
symmetric;  the sources driving the
wind (e.g.\ supernovae, AGN) are very unlikely to be spherically distributed
within the galaxy.  
Secondly, the galactic ISM frequently has a disk-like morphology
which will suppress the wind preferentially at low galactic latitudes,
perhaps yielding a bi-conic morphology \citep[e.g.][]{ham90,wws02}.
Lastly, the galaxy could be surrounded by an
aspherical gaseous halo whose interaction would produce an irregular 
outflow.

With these considerations in mind, we reanalyzed the fiducial model
with the 3D algorithm after departing from isotropy.  It is beyond the
scope of this paper to explore a full suite of anisotropic profiles.
We consider two simple examples: (i) half the fiducial model, where 
the wind density is set to zero for $2\pi$ steradians (i.e.\ a
hemispherical wind).  This model is
viewed from $\phi = 0^\circ$ (source uncovered) 
to $\phi = 180^\circ$ (source covered); and 
(ii) a bi-conical wind which fills 
$|\theta| < \theta_b$ both into and out of the plane of the sky and that is
viewed along the axis of rotational symmetry ($\theta = 0^\circ$ is
defined along this axis).

\begin{figure}
\includegraphics[width=3.5in]{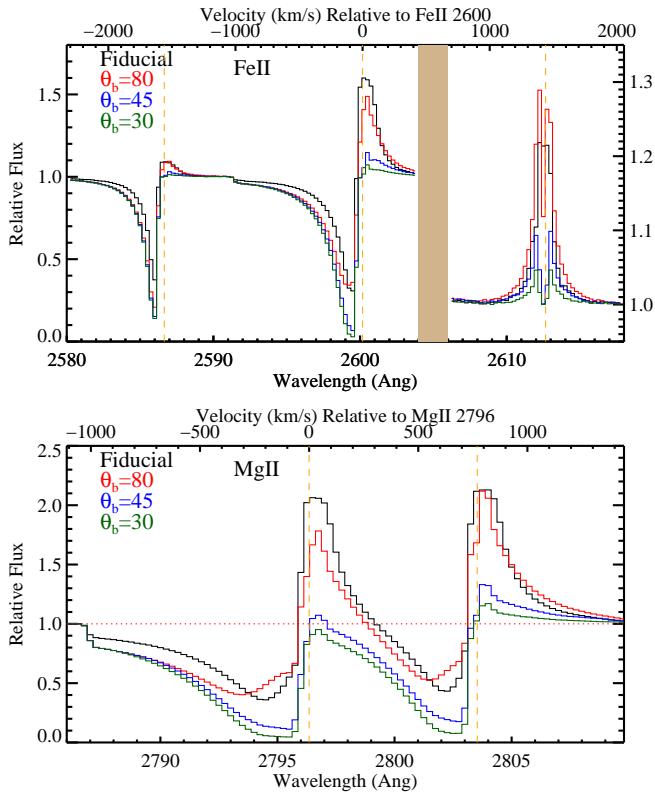}
\caption{
Profiles of the \ion{Fe}{2} and \ion{Mg}{2} profiles for the fiducial
case (black lines) compared against a bi-conical wind with varying
opening angle ($\theta_b = 0^\circ$ is a full wind).  For modest angles, one
observes similar results to the fiducial model but larger angles
($\theta_b > 45^\circ$) the emission is significantly suppressed at
all velocities.  
}
\label{fig:biconical}
\end{figure}

The resulting \ion{Mg}{2} and
\ion{Fe}{2} profiles for the half wind are compared against the fiducial model
(isotropic wind) in Figure~\ref{fig:anisotropic}.  
Examining the \mgiid\ doublet, 
the $\phi = 0^\circ$ model only shows
line-emission from photons scattered
off the back side.  These photons, by definition, have $\mdv \gtrsim 0 \mkms$
relative to line-center (a subset have $\mdv \lesssim 0 \mkms$ because
of turbulent motions in the wind). 
When viewed from the opposite direction ($\phi = 180^\circ$), the
absorption lines dominate, but there is still significant
line-emission at $\mdv \approx 0 \mkms$ and at $\mdv < 0 \mkms$ 
from photons that scatter through the wind
which fills in the absorption. 
The key difference between this and the isotropic wind is the absence of photons
scattered to $\mdv > 100 \mkms$;  this also implies deeper 
\mgiib\ absorption at $\mdv \approx -100 \mkms$. The 
shifts in velocity centroid and asymmetry of the emission lines
serve to diagnose the degree of wind isotropy, especially in
conjunction with analysis of the absorption profiles. 
The results are similar for the \feiid\ resonance lines.  The
\ion{Fe}{2}$^* \; \lambda 2612$ line, meanwhile, shows most clearly the
offset in velocity between the source unobscured ($\phi = 0^\circ$)
and source covered ($\phi = 180^\circ$) cases.  The offset of the
\feiis\ lines is the most significant 
difference from the fully isotropic wind.

We have also analyzed the predicted profiles for a series of bi-conical
winds with $\theta_b = 30-80^\circ$, each viewed along the axis of
rotational asymmetry with the wind covering the
source. Figure~\ref{fig:biconical} and
Table~\ref{tab:line_diag} summarizes the results for several cases.  For
$\theta_b = 80^\circ$, the results are
very similar to the fiducial model;  the profiles show
similar equivalent widths for absorption and emission.  The key
quantitative difference is that the \feiis\ line-emission is
modestly suppressed at $\delta v \approx 0 \mkms$ because the portion of the
wind with that projected velocity has been removed.  For
more highly collimated winds
$\theta_b<45^\circ$, however, the
line-emission is several times weaker than the fiducial model.
Similarly, the absorption-line profiles more closely track the
intrinsic optical depth of the outflow.   Of course, these same models
when viewed `sideways' would show no absorption but strong
line-emission.  Such events are sufficiently rare (Rubin et
al., in prep) that we consider this level of anisotropy to be
uncommon.  Nevertheless, the results for a bi-conic clearly have
important implications for the nature of absorption and line-emission
in our fiducial model.

\begin{figure}
\includegraphics[width=3.5in]{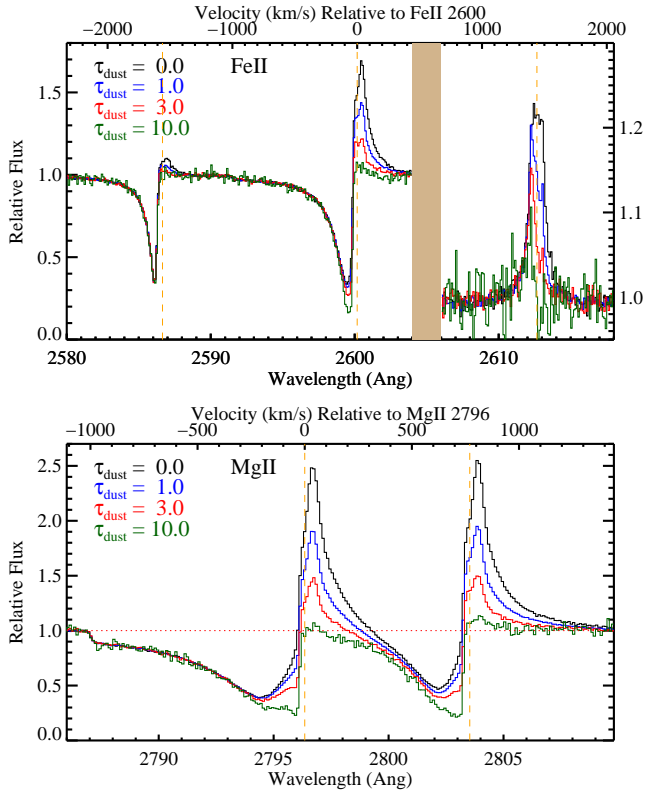}
\caption{
Profiles of the \ion{Fe}{2} and \ion{Mg}{2} profiles for the fiducial
model (black) compared against a series of models that include 
dust extinction, parameterize by the integrated dust opacity from
source to infinity (\taud).  The primary effect of dust is to suppress 
the line-emission relative to the continuum. 
A more subtle but important effect is that the redder photons in the
emission lines (corresponding to positive velocity offsets relative to
line-center) suffer greater extinction.  
This occurs because the `redder' photons that we view have travelled 
farther to scatter off the backside of the wind.  Note that
the absorption lines are nearly unmodified until $\mtaud = 10$, a
level of extinction that would preclude observing the source
altogether.
}
\label{fig:dust}
\end{figure}

\subsection{Dust}
\label{sec:dust}

As described in $\S$~\ref{sec:dust_method}, 
one generally expects dust in astrophysical environments that contain
cool gas and metals.  This dust 
modifies the observed wind profiles in two ways. 
First, it is a source of opacity for all of 
the photons.  This suppresses the flux at all
wavelengths by $\approx \exp(-\mtaud)$ but because we re-normalize the
profiles, this effect is essentially ignored.  Second, photons that are
scattered by the wind must travel a greater
distance to escape and therefore suffer from greater extinction.  A photon that is
trapped for many scatterings 
has an increased 
probability of being absorbed
by dust.  
Section~\ref{sec:dust_method} describes the details of our treatment of dust; we 
remind the reader here that we assume a constant dust-to-gas ratio 
that is normalized by the total optical
depth \taud\ photons would experience if they traveled from the
source to infinity without scattering. 

\begin{figure}
\includegraphics[height=3.5in,angle=90]{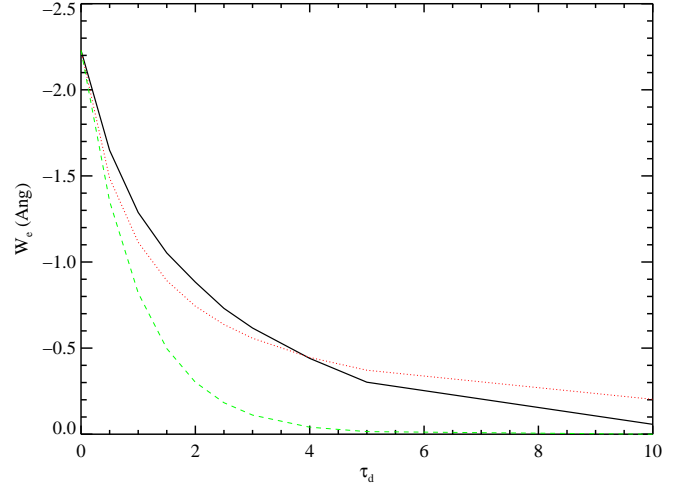}
\caption{
Predicted equivalent widths for \mgiib\ line-emission from the
fiducial wind model, as attenuated by dust with a range of \taud\
values.  
The black (solid) curve shows the model results.
The dashed (green) curve shows
$W_{\rm e}(\tau_d) = W_{\rm e}(\tau_d = 0) \cdot \exp(-\tau_d)$.
It is obvious that the equivalent widths (i.e., the flux relative to
the continuum) do not follow this scaling.
The dotted (red) line, meanwhile, plots $W_{\rm e}(\tau_d) = W_{\rm e}(\tau_d = 0)
\cdot (1+\tau_d)^{-1}$. 
This simple approximation is a good
representation of the results for our radiative transfer calculations.   
}
\label{fig:dust_tau}
\end{figure}

In Figure~\ref{fig:dust}, we show the \ion{Mg}{2} and \ion{Fe}{2}
profiles of the fiducial model ($\mtaud = 0$) against a series of
models with $\mtaud > 0$.  For the \ion{Mg}{2} transitions, the
dominant effect is the suppression of line emission at $\mdv \ge 0
\mkms$.  These `red' photons have scattered off the
backside of the wind and must travel a longer path than other
photons.  Dust leads to a differential reddening that increases with 
velocity relative to line-center. This is a natural consequence of dust
extinction and is most evident in the \feiic\ 
emission profile which is symmetrically distributed around
$\mdv= 0 \mkms$ in the $\mtaud=0$ model.   
The degree of suppression of the line-emission is relatively modest,
however.  Specifically, we find that the flux is reduced by a factor 
of the order of (1+\taud)$^{-1}$ (Figure~\ref{fig:dust_tau}),
instead of the factor
exp(--\taud) that one may have naively predicted. 
In terms of absorption, the profiles are
nearly identical for $\mtaud \le 3$.  One requires very high
extinction to 
produce a deepening of the profiles at 
$\mdv \approx -100 \mkms$.

We conclude that dust has only a modest influence on this fiducial model and,
by inference, on models with moderate peak optical depths and
significant velocity gradients with radius (i.e.\ scenarios in which the
photons scatter only one to a few times before exiting).
For qualitative changes, one requires an extreme level of
extinction ($\mtaud = 10$).  In this case, the source would be
extinguished by 15\,magnitudes and could never be observed. 
Even $\mtaud = 3$ is larger than typically inferred for the
star-forming galaxies that drive outflows \citep[e.g.][]{cf00}.
For the emission lines,
the dominant effect is a reduction in the flux 
with a greater extinction at higher velocities relative to line-center.
In these respects, dust extinction crudely mimics the behavior of the
anisotropic
wind described in Section~\ref{sec:anisotropic}. 

\begin{figure}
\includegraphics[width=3.5in]{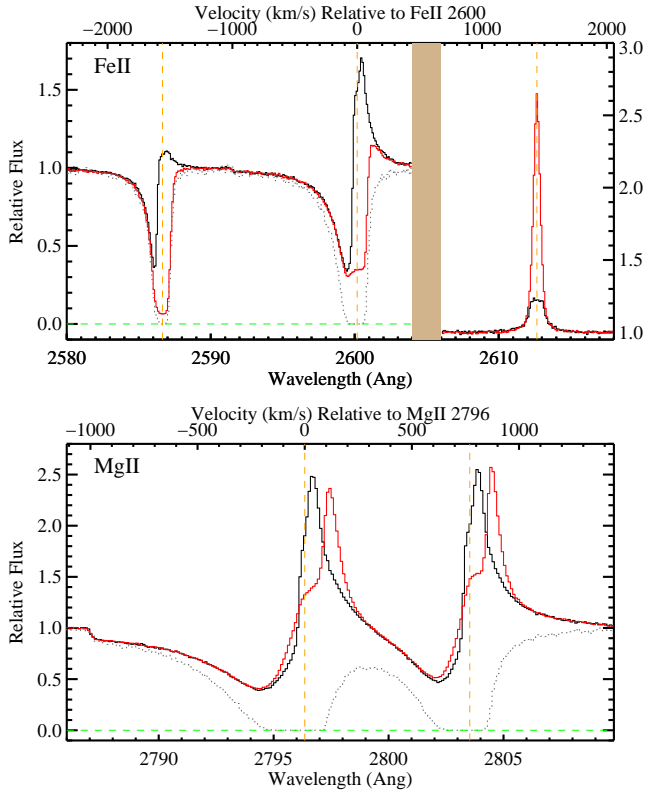}
\caption{
Profiles of the \ion{Fe}{2} and \ion{Mg}{2} profiles for the ISM+wind
model (red) compared against the fiducial wind model (black). 
The dotted line traces the predicted absorption profile in the absence
of re-emission and scattering.
Regarding the \ion{Mg}{2}~doublet, the primary difference between the
ISM+wind and the fiducial models is the shift of $\approx 100 \mkms$
in the emission lines from $v \approx 0 \mkms$ for the fiducial
model to $v \approx +100\mkms$ for the ISM+wind model. 
The \ion{Fe}{2} profiles, meanwhile, show several qualitative differences. 
The \feiid\ resonance transitions each exhibit much
greater absorption at $v \approx 0 \mkms$ than the fiducial model.
The resonant line-emission is also substantially
reduced, implying much higher fluxes for the non-resonant lines (e.g.\
\feiic,2626).  
Lastly, we note that the \feiia\ absorption profile provides a good
(albeit imperfect) representation of the wind opacity and therefore
offers the best characterization of an ISM component.
}
\label{fig:ISM_spec}
\end{figure}

\subsection{ISM}
\label{sec:ISM}

The fiducial model does not include gas associated with
the ISM 
of the galaxy, i.e.\ material at $r \sim
0$\,kpc with $v_{\rm r} \approx 0 \mkms$.  This allowed us to focus on
results related solely to a wind component.  The decision to ignore the ISM
was also motivated by the general absence of significant absorption at
$\mdv \approx 0 \mkms$ in galaxies that exhibit outflows 
\citep[e.g.][]{wcp+09,rwk+10,steidel+10}.
On the other hand, the stars that comprise the
source are very likely embedded within and fueled by gas
from the ISM.  
Consider, then, a modification to the fiducial model that 
includes an ISM. 
Specifically, we assume the ISM component has density $n_{\rm H} = 1 \cm{-3}$ for
$r_{\rm ISM} \le r < r_{\rm inner}$ with $r_{\rm ISM} = 0.5$\,kpc, 
an average velocity of $\mvr = 0 \mkms$, and a larger turbulent velocity $b_{\rm ISM} = 40 \mkms$.
The resultant optical depth profile $\tau_{2796}$ is identical to the
fiducial model for $r > 2$\,kpc, has a slightly higher opacity at
$r=1-2$\,kpc, and has a very large opacity at $r = 0.5-1$\,kpc.

In Figure~\ref{fig:ISM_spec}, the solid curves show the \ion{Mg}{2} and
\ion{Fe}{2} profiles for the ISM+wind and fiducial 
models. In comparison, the
dotted curve shows the intrinsic absorption profile for the ISM+wind
model corresponding to the (unphysical) case
where none of the absorbed photons are scattered or re-emitted.   Focus first on the
\ion{Mg}{2} doublet.  As expected, the dotted curve shows strong
absorption at $\mdv \approx 0 \mkms$ and blueward.  The full models,
in contrast, show non-zero flux at these velocities and even a
normalized flux exceeding unity at $\mdv \approx 0 \mkms$.  In
fact, the ISM+wind model is nearly identical to the fiducial model;
the only quantitative difference is that the velocity centroids of
the emission lines are shifted redward by $\approx +100 \mkms$.
We have also examined the spatial distribution of emission from 
this model and find results
qualitatively similar to the fiducial wind model
(Figures~\ref{fig:fiducial_ifu_mgii} and \ref{fig:fiducial_ifu_feii}).

There are, however, several qualitative differences 
for the \ion{Fe}{2} transitions. 
First, the \feiia\ transition in the ISM+wind model
shows much stronger absorption at $\mdv \approx 0$
to $-100 \mkms$.  In contrast to the \ion{Mg}{2} doublet,
the profile is not filled in by scattered photons. Instead, 
the majority of \feiia\ photons that are absorbed are re-emitted as
\feiis$\lambda\lambda 2612, 2632$ photons.  In fact, the
\feiia\ profile very nearly matches the profile without re-emission 
(compare to the dotted lines); this transition provides a
very good description of the intrinsic ISM+wind optical depth profile.  
We conclude that resonant transitions that are coupled to (multiple)
non-resonant, electric dipole transitions offer the best
diagnosis of ISM absorption.

The differences in the \ion{Fe}{2} absorption profiles are reflected
in the much higher strengths ($3-10\times$) of emission from
transitions to the excited states of the \aconfig\ configuration.   This occurs because:
(1) there is greater absorption by the \feiid\
resonance lines; and (2) the high opacity of the ISM component leads to
an enhanced conversion of resonance photons with $\mdv \approx 0\mkms$
into \feiis\ photons.  This is especially notable for the
\feiib\ transition whose coupled \feiis\ transition shows an equivalent width nearly
$10\times$ stronger than for the fiducial model.  The relative
strengths of the \feiib\ and \feiie\ lines provide a direct
diagnostic of the degree to which the resonance line photons are
trapped, i.e.\ the peak optical depth of \ion{Fe}{2}~2600 and
the velocity gradient of the wind.

Because of the high degree of photon trapping within the ISM
component, this model does suffer more from dust extinction than the
fiducial model.  We have studied the ISM+wind model including dust
with $\mtaud=1$ (normalized to include the ISM gas).  All of the 
emission lines are significantly reduced.  The \ion{Mg}{2} emission is
affected most because these lines are resonantly trapped.  The \feiis\
emission is also reduced relative to the dust-free model, but the
absolute flux still exceeds the fiducial model
(Table~\ref{tab:line_diag}).  At large negative velocity offsets from
systemic, the two profiles are nearly identical.
We conclude that a dusty ISM model could show significant absorption
at $\delta v \approx 0 \mkms$ in resonance lines (including
\ion{Mg}{2}) with strong line-emission in the \feiis\
transitions.

\begin{figure}[ht]
\includegraphics[width=3.5in]{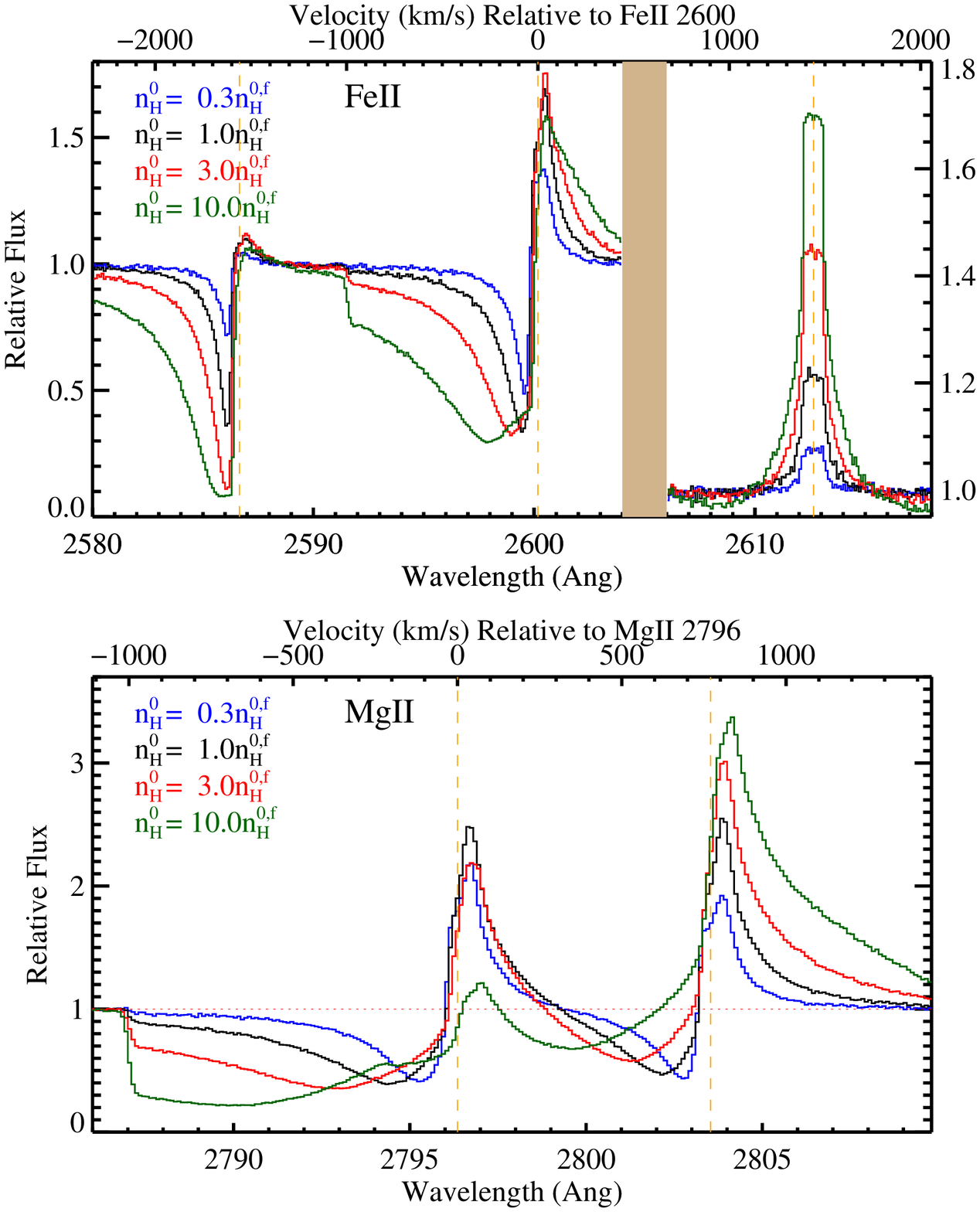}
\caption{
\ion{Mg}{2} and \ion{Fe}{2} profiles for the fiducial model with
varying normalization, parameterized by \nhn.  As expected, the
strength of absorption increases with increasing \nhn; this also
results in stronger line-emission.  Note that the \feiia\ emission is
always weak. Only its absorption increases with \nhn\ and actually exceeds
the depth of \feiib\ for $\mnhn > \mnhf$.  The depth of the
\ion{Mg}{2} doublet, meanwhile, always falls below a relative flux of
0.3 while the \mgiib\ emission rises steadily with \nhn.
}
\label{fig:norm}
\end{figure}

\begin{deluxetable*}{ccrccccccccccc}
\tablewidth{0pc}
\tablecaption{Line Diagnostics for the Fiducial Model and Variants
\label{tab:line_diag}}
\tabletypesize{\scriptsize}
\tablehead{\colhead{Transition} & \colhead{Model} & \colhead{$v_{\rm int}^a$} & \colhead{$W_{\rm i}$} & \colhead{$W_{\rm a}$} & \colhead{$\tau_{\rm pk}$} & \colhead{$v_\tau$} 
& \colhead{$v_{\bar \tau}$}
& \colhead{$v_{\rm int}^e$} & \colhead{$W_{\rm e}$} & \colhead{$f_{\rm pk}$} & \colhead{$v_f$} 
& \colhead{$v_{\bar f}$} & \colhead{$\Delta v_{\rm e}$}
\\
&& (\kms) & (\AA) & (\AA) && (\kms) & (\kms) & (\kms) & (\AA) & & (\kms) & (\kms) & (\kms)}
\startdata
  MgII 2796  \\
&Fiducial&[$-1009,-43$]& 4.78& 2.83&0.94&$ -215$&$ -372$&[$-32,311$]&$-1.77$& 2.48&$   32$&$  117$& 215\\
&$\phi=0^\circ$&$\dots$&$\dots$&$\dots$&$\dots$&$\dots$&$\dots$&[$-145,606$]&$-2.34$& 2.10&$   70$&$  208$& 483\\
&$\phi=180^\circ$&[$-1030,-65$]& 4.78& 2.98&1.03&$ -199$&$ -369$&[$-65,70$]&$-0.51$& 1.78&$  -11$&$    6$& 107\\
&$\theta_b=80^\circ$&[$-1030,-65$]& 4.78& 3.55&0.91&$ -306$&$ -442$&[$-65,257$]&$-1.03$& 1.78&$   43$&$   93$& 215\\
&$\theta_b=45^\circ$&[$-1003,-11$]& 4.78& 4.79&2.19&$  -91$&$ -332$&[$16,96$]&$-0.04$& 1.07&$   43$&$   55$&  54\\
&\taud=1&[$-1009,-32$]& 4.77& 2.94&0.95&$ -215$&$ -360$&[$-32,257$]&$-0.95$& 1.90&$   32$&$  101$& 193\\
&\taud=3&[$-998,-32$]& 4.78& 3.07&1.03&$ -193$&$ -342$&[$-22,182$]&$-0.40$& 1.48&$   43$&$   76$& 150\\
&ISM&[$-1009,-65$]& 6.36& 2.67&0.90&$ -204$&$ -390$&[$-54,311$]&$-1.60$& 2.36&$  118$&$  125$& 236\\
&ISM+dust&[$-998,150$]& 6.42& 4.06&1.06&$ -193$&$ -270$&$\dots$&$\dots$&$\dots$&$\dots$&$\dots$&$\dots$\\
  MgII 2803  \\
&Fiducial&[$-437,-41$]& 3.29& 1.19&0.76&$ -148$&$ -193$&[$-41,676$]&$-2.24$& 2.55&$   34$&$  269$& 449\\
&$\phi=0^\circ$&$\dots$&$\dots$&$\dots$&$\dots$&$\dots$&$\dots$&[$-84,665$]&$-1.86$& 1.93&$   77$&$  257$& 482\\
&$\phi=180^\circ$&[$-538,-57$]& 3.29& 1.82&0.97&$ -137$&$ -232$&[$-57,104$]&$-0.59$& 1.81&$   -3$&$   22$& 107\\
&$\theta_b=80^\circ$&[$-485,-57$]& 3.29& 1.33&0.65&$ -217$&$ -237$&[$-57,746$]&$-2.30$& 2.12&$   23$&$  303$& 535\\
&$\theta_b=45^\circ$&[$-538,-57$]& 3.29& 2.60&1.74&$  -84$&$ -217$&[$-30,692$]&$-0.72$& 1.33&$   23$&$  320$& 562\\
&\taud=1&[$-479,-41$]& 3.28& 1.41&0.83&$ -137$&$ -195$&[$-41,591$]&$-1.30$& 1.95&$   34$&$  247$& 417\\
&\taud=3&[$-533,-30$]& 3.26& 1.67&0.94&$ -116$&$ -195$&[$-30,484$]&$-0.66$& 1.50&$   34$&$  213$& 375\\
&ISM&[$-426,-62$]& 6.49& 1.03&0.67&$ -169$&$ -206$&[$-51,655$]&$-2.09$& 2.57&$   98$&$  265$& 439\\
&ISM+dust&[$-554,130$]& 6.51& 2.59&1.02&$ -137$&$ -143$&[$173,195$]&$-0.01$& 1.03&$  184$&$  184$&  21\\
  FeII 2586  \\
&Fiducial&[$-348,-35$]& 0.82& 0.61&1.01&$  -70$&$ -119$&[$-35,128$]&$-0.10$& 1.11&$   35$&$   47$& 128\\
&$\phi=0^\circ$&$\dots$&$\dots$&$\dots$&$\dots$&$\dots$&$\dots$&[$-104,186$]&$-0.10$& 1.06&$   70$&$   41$& 232\\
&$\phi=180^\circ$&[$-365,-46$]& 0.82& 0.60&1.01&$  -75$&$ -133$&[$-46,70$]&$-0.03$& 1.06&$  -17$&$   12$&  87\\
&$\theta_b=80^\circ$&[$-481,-46$]& 0.82& 0.96&1.57&$  -75$&$ -148$&[$-46,186$]&$-0.10$& 1.09&$   41$&$   71$& 203\\
&$\theta_b=45^\circ$&[$-481,-46$]& 0.82& 1.03&1.88&$  -75$&$ -143$&[$41,99$]&$-0.02$& 1.03&$   41$&$   69$&  58\\
&\taud=1&[$-348,-35$]& 0.82& 0.61&1.04&$  -70$&$ -118$&[$-35,116$]&$-0.05$& 1.06&$   23$&$   41$& 116\\
&\taud=3&[$-313,-35$]& 0.94& 0.60&1.06&$  -70$&$ -113$&[$-23,46$]&$-0.02$& 1.04&$  -12$&$   12$&  58\\
&ISM&[$-348,104$]& 1.90& 1.60&2.75&$    0$&$  -29$&$\dots$&$\dots$&$\dots$&$\dots$&$\dots$&$\dots$\\
&ISM+dust&[$-313,116$]& 1.84& 1.63&2.96&$   23$&$  -25$&[$278,290$]&$-0.00$& 1.03&$  278$&$  284$&  12\\
  FeII 2600  \\
&Fiducial&[$-580,-37$]& 1.87& 1.18&1.08&$  -83$&$ -181$&[$-37,459$]&$-0.83$& 1.70&$   32$&$  191$& 312\\
&$\phi=0^\circ$&$\dots$&$\dots$&$\dots$&$\dots$&$\dots$&$\dots$&[$-135,442$]&$-0.78$& 1.46&$   67$&$  144$& 346\\
&$\phi=180^\circ$&[$-597,-49$]& 1.87& 1.18&1.31&$  -78$&$ -188$&[$-49,95$]&$-0.25$& 1.39&$  -20$&$   22$&  87\\
&$\theta_b=80^\circ$&[$-856,-49$]& 1.87& 1.72&1.07&$ -106$&$ -249$&[$-49,557$]&$-0.77$& 1.49&$   38$&$  241$& 404\\
&$\theta_b=45^\circ$&[$-856,-49$]& 1.87& 2.12&2.71&$  -78$&$ -204$&[$-20,413$]&$-0.24$& 1.15&$   38$&$  194$& 346\\
&\taud=1&[$-591,-37$]& 1.87& 1.20&1.14&$  -83$&$ -176$&[$-37,332$]&$-0.50$& 1.45&$   32$&$  138$& 265\\
&\taud=3&[$-580,-37$]& 1.95& 1.25&1.31&$  -72$&$ -169$&[$-37,228$]&$-0.23$& 1.22&$   44$&$   93$& 196\\
&ISM&[$-568,90$]& 3.12& 1.88&1.19&$  -72$&$  -99$&[$101,378$]&$-0.19$& 1.15&$  113$&$  237$& 242\\
&ISM+dust&[$-603,101$]& 3.06& 2.12&1.40&$   44$&$  -93$&[$182,194$]&$-0.00$& 1.02&$  182$&$  188$&  12\\
  FeII* 2612 \\
&Fiducial&&&&&&&[$-173,183$]&$-0.34$& 1.24&$  -23$&$    5$& 241\\
&$\phi=0^\circ$&&&&&&&[$-46,183$]&$-0.16$& 1.22&$   11$&$   67$& 172\\
&$\phi=180^\circ$&&&&&&&[$-190,68$]&$-0.16$& 1.21&$  -46$&$  -61$& 172\\
&$\theta_b=80^\circ$&&&&&&&[$-276,241$]&$-0.46$& 1.29&$  -46$&$  -17$& 316\\
&$\theta_b=45^\circ$&&&&&&&[$-18,155$]&$-0.07$& 1.10&$   40$&$   68$& 144\\
&\taud=1&&&&&&&[$-173,114$]&$-0.23$& 1.21&$  -46$&$  -28$& 207\\
&\taud=3&&&&&&&[$-150,91$]&$-0.14$& 1.16&$  -46$&$  -30$& 184\\
&ISM&&&&&&&[$-184,160$]&$-1.10$& 2.66&$    0$&$   -9$& 138\\
&ISM+dust&&&&&&&[$-138,137$]&$-0.45$& 1.53&$    0$&$   -1$& 161\\
  FeII* 2626 \\
&Fiducial&&&&&&&[$-171,183$]&$-0.39$& 1.29&$  -46$&$    5$& 217\\
&$\phi=0^\circ$&&&&&&&[$-52,177$]&$-0.13$& 1.16&$   34$&$   62$& 171\\
&$\phi=180^\circ$&&&&&&&[$-194,63$]&$-0.13$& 1.16&$  -52$&$  -66$& 143\\
&$\theta_b=80^\circ$&&&&&&&[$-280,234$]&$-0.58$& 1.41&$  -52$&$  -22$& 314\\
&$\theta_b=45^\circ$&&&&&&&[$6,148$]&$-0.08$& 1.14&$   34$&$   76$& 114\\
&\taud=1&&&&&&&[$-160,126$]&$-0.25$& 1.23&$  -46$&$  -17$& 194\\
&\taud=3&&&&&&&[$-171,68$]&$-0.14$& 1.17&$  -46$&$  -51$& 194\\
&ISM&&&&&&&[$-194,194$]&$-1.76$& 4.10&$    0$&$    0$& 103\\
&ISM+dust&&&&&&&[$-160,114$]&$-0.47$& 1.70&$    0$&$  -21$& 137\\
  FeII* 2632 \\
&Fiducial&&&&&&&[$-132,119$]&$-0.16$& 1.12&$   16$&$   -6$& 194\\
&$\phi=0^\circ$&&&&&&&[$-41,130$]&$-0.08$& 1.09&$   45$&$   44$& 142\\
&$\phi=180^\circ$&&&&&&&[$-155,16$]&$-0.08$& 1.11&$  -41$&$  -69$& 142\\
&$\theta_b=80^\circ$&&&&&&&[$-212,159$]&$-0.23$& 1.14&$   16$&$  -26$& 285\\
&$\theta_b=45^\circ$&&&&&&&[$-639,102$]&$-0.18$& 1.14&$ -611$&$ -271$& 684\\
&\taud=1&&&&&&&[$-121,85$]&$-0.11$& 1.10&$  -52$&$  -18$& 160\\
&\taud=3&&&&&&&[$-98,16$]&$-0.05$& 1.08&$  -41$&$  -41$& 103\\
&ISM&&&&&&&[$-121,130$]&$-0.55$& 1.81&$    5$&$    4$& 137\\
&ISM+dust&&&&&&&[$-121,107$]&$-0.22$& 1.30&$    5$&$   -6$& 148\\
\enddata
\tablecomments{{L}isted are the equivalent widths (intrinsic, absorption, and emission), the peak optical depth for the absorption
$\tau_{\rm pk} \equiv -\ln(I_{\rm min})$, the velocity where the optical depth peaks $v_\tau$, the optical depth-weighted velocity centroid 
$v_{\bar \tau} \equiv \int dv \, v \ln[I(v)] / \int dv \ln[I(v)]$, the peak flux $f_{\rm pk}$ in emission, the velocity where the flux peaks 
$v_f$, the flux-weighted velocity centroid of the emission line $v_{\bar f}$ (occasionally affected by blends with neighboring emission lines), and the $90\%$ width $\Delta v_{\rm e}$.
The $v^a_{\rm int}$ and $v^e_{\rm int}$ columns give the velocity range used to calculate the absorption and emission characteristics, respecitvely.  These were defined by the velcoities where the profile crossed 0.95 in the normalized flux.}
\end{deluxetable*}

\subsection{Varying $n_{\rm H}^0$}

The final modification to the fiducial model considered was 
a uniform variation of
the normalization of the optical depth profiles.
Specifically, we ran a series of additional models with $n_{\rm H}^0$ at
1/3, 3, and 10 times the fiducial value of $n_{\rm H}^{0,f} = 0.1 \cm{-3}$.
The resulting \ion{Mg}{2} and \ion{Fe}{2} profiles are compared
against the fiducial model in Figure~\ref{fig:norm}.  Inspecting
\ion{Mg}{2}, one notes that the models with 1/3 and $3 n_{\rm H}^{0,f}$
behave as expected.  Higher/lower optical depths lead to
greater/weaker \mgiia\ absorption and stronger/weaker \mgiib\ emission.
In the extreme case of
$n_{\rm H}^0 = 10 n_{\rm H}^{0,f}$, the \mgiib\ absorption and \mgiia\
emission have nearly disappeared and one primarily observes very
strong \mgiia\ absorption and \mgiib\ emission.
In essence, this wind has converted all of the photons absorbed by the
\ion{Mg}{2} doublet into \mgiib\ emission.

Similar behavior is observed for the \ion{Fe}{2} transitions.
Interestingly, the \feiia\ transitions never show significant emission,
only stronger absorption with increasing $n_{\rm H}^0$.  In fact, for $n_{\rm H}^0
> 2 n_{\rm H}^{0,f}$ the peak optical depth of \feiia\ actually exceeds that
for \feiib\ because the latter remains significantly filled-in by
scattered photons.  By the same token, the flux of the \feiic\
emission increases with \nhn.

\subsection{Summary Table}

Table~\ref{tab:line_diag} presents a series of quantitative measures
of the \ion{Mg}{2} and \ion{Fe}{2} absorption and emission lines for
the fiducial model ($\S$~\ref{sec:fiducial}) and a subset of the models
presented in this section.  Listed are the absorption and emission
equivalent widths ($W_{\rm a}, W_{\rm e}$), the peak optical depth
for the absorption $\tau_{\rm pk}$, the
velocity where the optical depth peaks $v_\tau$, the optical
depth-weighted velocity centroid $v_{\bar \tau} \equiv \int dv \, v
\ln[I(v)] / \int dv \ln[I(v)]$, the peak flux $f_{\rm pk}$ in
emission, the velocity where the flux peaks $v_f$, and the
flux-weighted velocity centroid of the emission line $v_{\bar f}$.
We discuss several of these measures in $\S$~\ref{sec:discuss}.

\section{Alternate Wind Models}
\label{sec:alternate}

This section presents several additional wind scenarios that differ
significantly from the fiducial model explored in the previous
sections.  In each case, we maintain the simple assumption of
isotropy.

\subsection{The Lyman Break Galaxy Model}
\label{sec:lbg}

The Lyman break galaxies (LBGs), UV color-selected galaxies at $z \sim 3$,
exhibit cool gas outflows in \ion{Si}{2}, \ion{C}{2},
etc.\ transitions with speeds up to 1000\,\kms\
\citep[e.g.][]{lkg+97,pks+98}.
Researchers have invoked these winds to explain enrichment of
the intergalactic medium \citep[e.g.][]{ahs+01,spa+06}, the origin of the
damped \lya\ systems \citep{nbf98,schaye01a}, and the formation of
`red and dead' galaxies \citep[e.g.][]{spf01}.  Although the
presence of these outflows was established over a decade ago,
the processes that drive them remain
unidentified.  Similarly,  current estimates of the mass and energetics of the
outflows suffer from orders of magnitude uncertainty.

\begin{figure}
\includegraphics[width=3.5in]{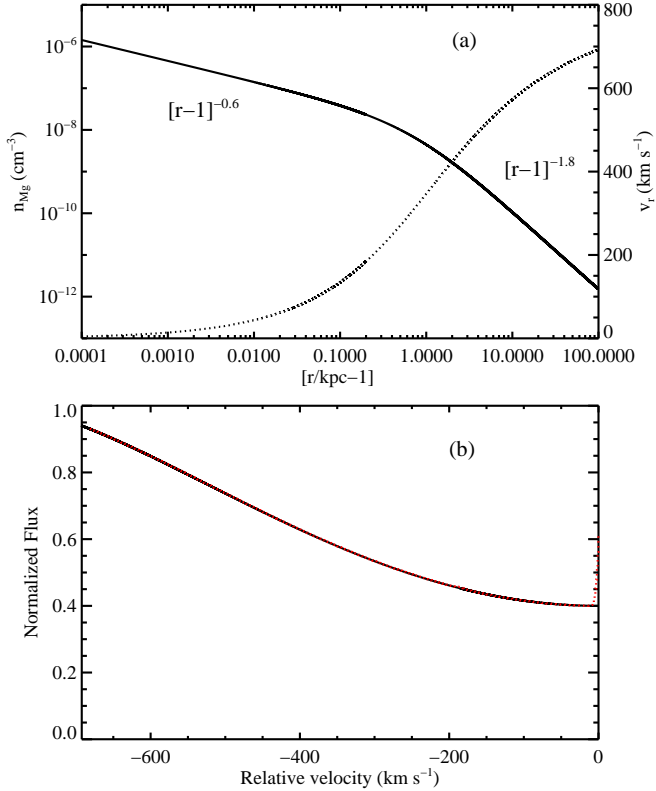}
\caption{
(a) The dotted curve shows the velocity law (labeled on the
right axis) for the LBG model
of S10 plotted against $(r/{\rm kpc} -1)$ to highlight the evolution
in the quantities at small radii.  For example,  note how rapidly the velocity rises from $r =
1$ to 2\,kpc.  The solid curve shows the Sobolev solution for the
Mg$^+$ gas
density derived from 
the average absorption profile of LBG galaxies (see below).
The density falls off initially as $(r/{\rm kpc}-1)^{-0.6}$ and then
steepens to $(r/{\rm kpc}-1)^{-1.8}$.
(b) The average profile for LBG cool gas absorption as
measured and defined by S10
(black solid curve).  Overplotted on this curve is a (red) dotted line that
shows the \mgiia\ absorption profile derived from the density (and velocity)
law shown in the upper panel.  
}
\label{fig:LBG_Sobolev}
\end{figure}

Recently, \citet[][hereafter S10]{steidel+10} introduced a model to
explain jointly the average absorption they observed
along the sightlines to several hundred LBGs and the average absorption in gas
observed transverse to these galaxies.  
Their wind model is defined by two
expressions: (i) a radial velocity law $\mvr(r)$; and (ii) the covering
fraction of optically thick `clouds' $f_c(r)$.  For the latter, S10
envision an 
ensemble of small, optically thick $(\tau \gg 1)$ clouds that only
partially cover the galaxies.
For the velocity law, they adopted the following functional
form:

\begin{equation}
\mvr = \ltp \frac{A_{\rm LBG}}{1-\alpha} \rtp^{1/2} \ltk r_{\rm
  inner}^{1-\alpha} - r^{1-\alpha} \rtk^{1/2}
\label{eqn:LBG_vlaw}
\end{equation}
with $A_{\rm LBG}$ the constant that sets the terminal speed,
$r_{\rm inner}$ the inner radius of the wind (taken to be 1\,kpc), and
$\alpha$ the power-law exponent that describes how steeply the velocity curve rises.  Their
analysis of the LBG absorption profiles implied
a very steeply rising curve with $\alpha \approx 1.3$.
This velocity expression is shown as a dotted line in 
Figure~\ref{fig:LBG_Sobolev}a.  

The covering fraction of optically thick clouds, meanwhile, was assumed to have
the functional form

\begin{equation}
f_c(r) = f_{c,max} \ltp \frac{r}{r_{\rm inner}} \rtp^{-\gamma} \cmma
\label{eqn:covering}
\end{equation}
with $\gamma \approx 0.5$ and $f_{c,max}$ the maximum covering
fraction.  From this expression and the velocity law, one can recover
an absorption profile $I_{\rm LBG}(v) = 1 - f_c[r(v)]$, written
explicitly as

\begin{equation}
I_{\rm LBG}(v) = 1 - f_{c,max} \ltk r_{\rm inner}^{1-\alpha} - \ltp
\frac{1-\alpha}{A_{\rm LBG}} \rtp v^2 \rtk^{\gamma/(\alpha-1)}
\perd
\label{eqn:LBG_I}
\end{equation}
The resulting profile for $f_{c,max} = 0.6$, $\gamma=0.5$,
$\alpha=1.3$, and $A_{\rm LBG} = -192,000 \, \rm km^2 \, s^{-2} \, kpc^{-2}$ 
is displayed in Figure~\ref{fig:LBG_Sobolev}b.  

In the following, we consider two methods for analyzing the LBG wind.
Both approaches assume an isotropic wind and adopt the velocity law given by
Equation~\ref{eqn:LBG_vlaw}.  In the first model, we treat the cool gas as a
diffuse medium with unit covering fraction and a radial density
profile determined from the Sobolev approximation.  We then apply 
the Monte Carlo methodology used for the other wind models to predict
\ion{Mg}{2} and \ion{Fe}{2} line profiles.  In the second LBG model,
we modify our algorithms to more precisely mimic the concept of an
ensemble of optically thick clouds with a partial covering fraction on
galactic scales.

\subsubsection{Sobolev Inversion}
\label{sec:Sobolev}

As demonstrated in Figure~\ref{fig:LBG_Sobolev}, the wind velocity for
the LBG model rises very steeply with increasing radius before
flattening at large radii.  Under these conditions, the Sobolev
approximation provides an accurate description of the optical depth
for the flowing medium. 
The Sobolev line optical depth profile can be derived from
the absorption profile (Equation~\ref{eqn:LBG_I}), 
\begin{equation}
\tau_{\rm LBG}(v) = -\ln \ltk I_{\rm LBG}(v) \rtk \perd
\label{eqn:tauLBG}
\end{equation}
This simple expression assumes that the source size is small compared the wind dimensions
and neglects the effects of light scattered in the wind (for more general inversion
formulae that do not make these assumptions, see \citet{Kasen_2002}).
The Sobolev approximation (Equation~\ref{eqn:Sobolev}) provides a simple expression for the optical
depth at each radius in terms of the local density and velocity gradient.  In general, the velocity 
 gradient must be taken along the direction of propagation of a photon, however
in this model the central source is relatively small  and the radiation
field is therefore primarily radially directed.  In this 
case the velocity gradient in the radial direction, $d\mvr/dr$, is the relevant quantity, and 
we may use Equation~\ref{eqn:Sobolev}  to invert
the optical depth and determine the density profile,
\begin{equation}
n_{\rm LBG}(r) =  \tau_{\rm LBG}(r)  \frac{ |d\mvr/dr |}{f_\ell  \lambda^0_\ell}
\biggl( \frac{ m_e c}{\pi e^2} \biggl)
 \perd
\label{eqn:nSobolev}
\end{equation}

The solid curve in Figure~\ref{fig:LBG_Sobolev}a shows the
resultant density profile for Mg$^+$ assuming that the \mgiia\ line
follows the intensity profile drawn in Figure~\ref{fig:LBG_Sobolev}b. 
This is a relatively extreme density profile.  From the inner radius
of 1\,kpc to 2\,kpc, the density drops by over 2 orders of
magnitude including nearly one order of magnitude over the first
10\,pc.  Beyond 2\,kpc, the density drops even more rapidly, falling
orders of magnitude from 2 to 100\,kpc.

We verified that the Sobolev-derived density profile shown in
Figure~\ref{fig:LBG_Sobolev}a
reproduces the proper
absorption profile by discretizing the wind into a series of layers
and calculating the integrated absorption profile.  This
calculation is shown as a dotted red curve in
Figure~\ref{fig:LBG_Sobolev}b; it is an excellent
match to the desired profile (black curve).
To calculate the optical depth profiles for the other \ion{Mg}{2} and
\ion{Fe}{2} transitions, we
assume $\mnfe = \mnmg/2$ and also scale $\tau$ by the $f\lambda$ product.

\begin{figure}
\includegraphics[width=3.5in]{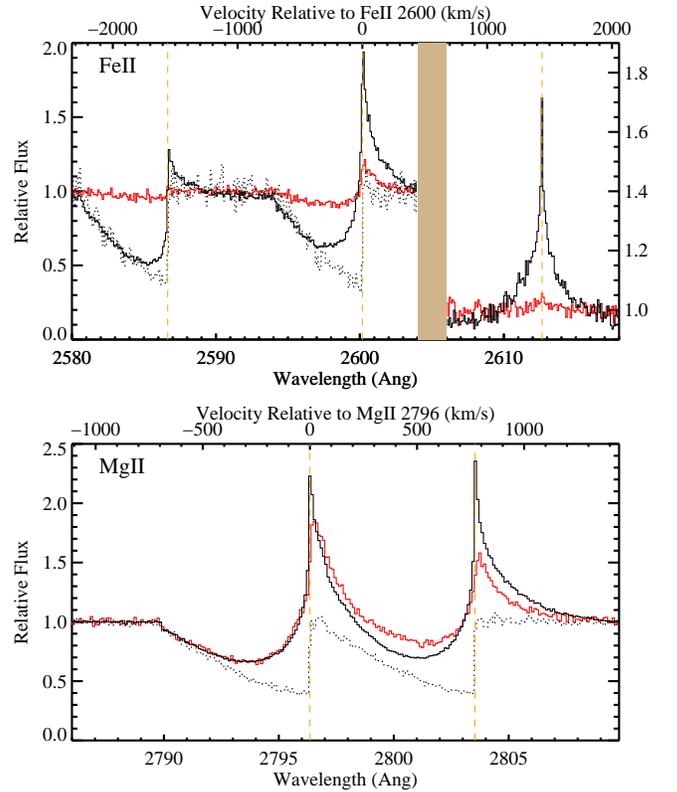}
\caption{
Profiles of the \ion{Fe}{2} and \ion{Mg}{2} profiles for the two
LBG models considered: (red) the profiles predicted using the 
Sobolev approximation 
and (black) a model constructed to faithfully
represent the wind model advocated by S10.  The dotted line
shows the absorption profiles of the latter model if one were
to neglect scattered and re-emitted photons.  Similar to the fiducial
model (Figure~\ref{fig:fiducial_1d}), scattered and re-emitted photons
significantly modify the absorption profiles (especially \ion{Mg}{2}
and \feiib)
and produce strong emission lines.  
The result, for \ion{Mg}{2} especially, is a set of profiles that do
not match the average observed LBG absorption profile.
}
\label{fig:LBG_spec}
\end{figure}

We generated \ion{Mg}{2} and \ion{Fe}{2} profiles for this LBG-Sobolev
wind
using the 1D algorithm with no dust extinction; these are shown as
red curves in Figure~\ref{fig:LBG_spec}.   For this analysis, one
should focus on the \mgiia\ transition.  The dotted line in the Figure
shows the intensity profile when one ignores re-emission
of absorbed photons.  By construction, it is the same profile\footnote{
  The other transitions in
  the LBG-Sobolev model are significantly weaker than the input model
  because these are scaled down by $f\lambda$ and 
  for \ion{Fe}{2} the reduced Fe$^+$ abundance.} 
described by Equation~\ref{eqn:LBG_I} and plotted in
Figure~\ref{fig:LBG_Sobolev}b.   In comparison, the full model (solid,
red curve)
shows much weaker absorption, especially at $v = 0$ to $-300 \mkms$
because scattered photons fill in the absorption profile.
In this respect, our LBG-Sobolev model is an
inaccurate description of the observations which show more uniform
equivalent widths among differing transitions (S10). The model
also predicts significant emission in the \ion{Mg}{2} lines and several of
the \ion{Fe}{2}$^*$ transitions.   Emission associated with cool gas
has been observed for \ion{Si}{2}$^*$
transitions in LBGs \citep{prs+02,shapley03}, but
the \ion{Fe}{2}$^*$ transitions
modeled here lie in the near-IR and have not yet been investigated.
On the other hand, there have been no 
reported detections of significant line-emission related to low-ion resonance
transitions (e.g.\ \ion {Mg}{2}) in LBGs, only $z \le 1$ star-forming galaxies
\citep{wcp+09,rwk+10}.  
The principal result of the LBG-Sobolev model is that the scattering and re-emission of
absorbed photons significantly alters the predicted absorption profiles
for the input 
model.  This is, of course, an unavoidable
consequence of an isotropic, dust-free model with unit covering
fraction.

\begin{figure}
\includegraphics[height=3.5in,angle=90]{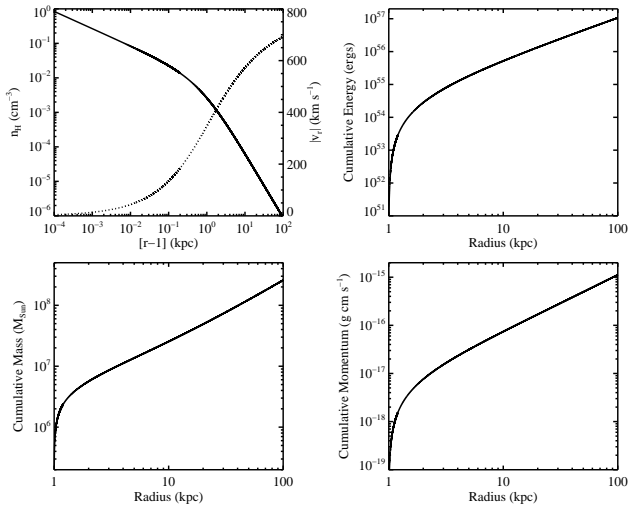}
\caption{
The upper left panel shows the density (solid) and velocity (dotted)
laws for the LBG-Sobolev wind model (Figure~\ref{fig:LBG_Sobolev}).
The remaining panels show the cumulative mass, energy, and momentum of
this flow.  All of the curves rise very steeply at small radii and
then rise steadily to large radii.  The principal result is that the
majority of mass, energy, and momentum in the wind are contained at
large radii.  We derive similar results for the LBG-partial covering model
if we assume the wind is composed of identical clouds optically thick
to strong metal-line transitions (e.g.\ \ion{Si}{2}~1526).  The
energy and momentum in the outer regions of the outflow may be very
difficult to generate with standard galactic-scale wind scenarios.
}
\label{fig:LBG_cumul}
\end{figure}

Because we have explicit velocity and density profiles for the LBG-Sobolev
model, it is straightforward to calculate the radial
distributions of mass, energy, and momentum of
this wind.  These are shown in cumulative form in
Figure~\ref{fig:LBG_cumul}.  Before discussing the results, we offer
two cautionary comments: (i) the conversion of \nmg\ to $n_{\rm H}$
assumes a very poorly constrained scalar factor of $10^{5.47}$.  One
should give minimal weight to the absolute values for any of the
quantities;
(ii) the Sobolev approximation is not a proper description of the S10
LBG wind model (see the following sub-section).
These issues aside, we may inspect Figure~\ref{fig:LBG_cumul} to reveal
global properties of this LBG-Sobolev wind.
One obvious result is that all of the curves rise very steeply at
small radii.  We find, for example, that $\approx 10\%$ of the mass is
contained within the inner 2\,kpc.  All of the curves continue to
rise, however, such that the majority of energy, mass, and momentum in
the wind is transported by its outermost layers (i.e., $r > 20$\,kpc).
This is somewhat surprising given that the density is $\approx
5$~orders of magnitude lower at these radii than at $r = 1$\,pc.
More importantly, we question whether any physical process could
produce a wind with such extreme profiles.

\subsubsection{Partial Covering Fraction}
\label{sec:Covering}

In the previous sub-section, we described a Sobolev inversion that
reproduces the average absorption profile of LBGs in cool gas
transitions when scattered photons are ignored.  A proper analysis
that includes scattered photons,
however, predicts line-profiles that are qualitatively
different from the observations because scattered photons fill-in
absorption and generate significant line-emission (similar to the
fiducial wind model; $\S$~\ref{sec:fiducial}).
We also argued that the implied mass, energy, and momentum
distributions of this wind (Figure~\ref{fig:LBG_cumul}) were too
extreme. 
This LBG-Sobolev model, however, is not precisely
the one introduced by S10;  those authors proposed an ensemble of optically
thick clouds with a partial covering fraction described by
Equation~\ref{eqn:covering}.  In contrast, the LBG-Sobolev model assumes
a diffuse medium with a declining density profile but a unit covering
fraction.  One may question, therefore,  whether these differences
lead to the failures of the LBG-Sobolev model. 

To more properly model the LBG wind described in S10, we 
performed the following Monte Carlo calculation.  First, we propagate
a photon from the source until its velocity relative to line-center
resonates with the wind (the photon escapes if this never occurs).
The photon then has a probability $P = f_c(r)$ of scattering.  If it
scatters, we track the photon until it comes into
resonance again\footnote{Because this wind has a monotonically
  increasing velocity law, the photon may only scatter again if a
  longer wavelength transition is available, e.g.\ \mgiib\ for a
  scattered \mgiia\ photon.} or escapes the system.  In this model, all of the
resonance transitions are assumed to have identical (infinite) optical
depth at line center. 

The results of the full calculation (absorption plus scattering) are
shown as the black curve in Figure~\ref{fig:LBG_spec}.  Remarkably, the results
are essentially identical to the LBG-Sobolev calculation for the
\mgiia\ transition; scattered photons fill-in
the absorption profiles at $v \approx 0 \mkms$ and yield significant
emission lines at $v \gtrsim 0 \mkms$. 
Similar results are observed for the other resonance lines and  
significant emission is observed 
for the \feiis\ transitions, centered at $\mdv \approx 0 \mkms$.  
In contrast to the \ion{Mg}{2} profiles, the 
\feiia\ line much more closely resembles its intrinsic
profile (dotted curves).  This occurs because most of these
absorbed photons fluoresce into \feiis\ emission.
The equivalent width of \feiia\ even exceeds that for \feiib, an
inversion that, if observed, would strongly support this model.

A robust conclusion of our
analysis is that these simple models cannot reproduce the
observed profiles of resonantly trapped lines (like the \ion{Mg}{2}
doublet) because of the emission from scattered photons.  
In particular, the observations show much greater opacity at
velocities within $\approx 100 \mkms$ of systemic (S10).
In order to
achieve a significant opacity at $\mdv \approx 0 \mkms$, one must
either suppress the line-emission (e.g.\ with severe dust extinction
or anisotropic winds) 
or include a substantial ISM component that absorbs light at $\mdv
\gtrsim 0 \mkms$ (e.g.\ $\S$~\ref{sec:ISM}).  
Absent such corrections (which may be insufficient),
we conclude that the LBG model
introduced by S10 is not a valid description of the data.  Also,
similar to our LBG-Sobolev calculation (Figure~\ref{fig:LBG_Sobolev}),
this clump model predicts the majority of mass, energy, and momentum are
carried in the outer regions of the wind. 
In fact, for clumps that are optically thick to strong metal-line
transitions (e.g.\ \ion{Si}{2}~1526) we estimate\footnote{Note that
  this estimate differs from the LBG-Sobolev calculation shown in
  Figure~\ref{fig:LBG_cumul} because the latter assumes a diffuse
  medium with unit filling factor.}
that the wind would
carry nearly a total mass of $\approx 10^{10} \msun$, with the
majority at large radii.

The resonant line-emission predicted for \ion{Mg}{2} should also arise
in other transitions.  Indeed, we caution that the \ion{C}{4}
line-emission observed in LBG spectra
\citep[e.g.][]{prs+02} may result from the galactic-scale outflow, not
only the stellar winds of massive stars.
A proper treatment of these radiative transfer effects is required to
quantitatively analyze these features.

\subsection{Radiation Pressure}
\label{sec:radiative}

As discussed in the Introduction, the importance of various physical mechanisms
in driving galactic outflows remains an open
question.  Several studies advocate a primarily energy-driven
wind scenario, in which the hot wind produced by the thermalized
energy from SNe ejecta entrains cool clouds via ram pressure \citep[e.g.][]{cc85,ham90,sh09}.
Alternatively, momentum deposition by photons emitted by luminous star clusters onto
dust grains in the ISM may also play an integral role in driving
large-scale outflows \citep[e.g.][]{mqt05,mmt10}.  Both of these
mechanisms likely contribute more or less significantly in a given
galaxy; however, the latter (radiation-pressure driving) has the added
advantage that it does not destroy cool gas clouds as it acts.  Here
we explore the line-profiles produced by a radiation-pressure driven
wind model as described by \cite{mqt05}.

\begin{figure}
\includegraphics[width=3.5in]{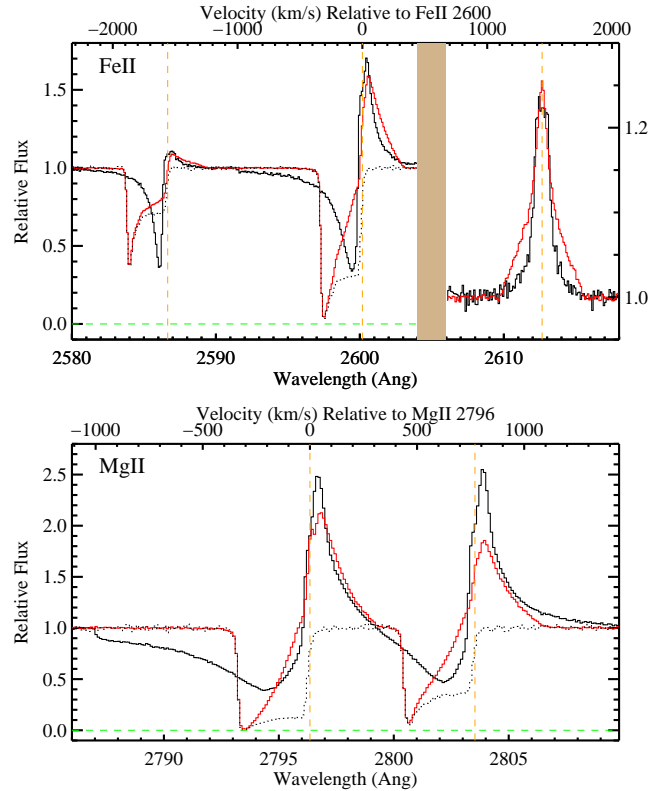}
\caption{
\ion{Mg}{2} and \ion{Fe}{2}
profiles for a radiation-driven wind model (red) compared against the
fiducial wind model (black).  In contrast to the fiducial model, the
optical depth peaks at large velocity (here $\mdv
\approx -350 \mkms$).  At these velocities, the absorption is not
filled-in by scattered photons and one recovers absorption lines
that closely resemble the intrinsic optical depth profile (dotted
curve).  Although the \ion{Mg}{2} absorption profiles are very
different from the fiducial model, the \ion{Mg}{2} line-emission has
similar peak flux and velocity centroids 
The \feiis\ emission also has a similar
equivalent width, yet the lines are much broader in the
radiation-driven model. 
}
\label{fig:rad_spec}
\end{figure}

These authors start by assuming the galaxy has luminosity $L$ and is an isothermal sphere
with mass profile $M_{gal}(r) = 2\sigma^2r/G$, where $\sigma$ is the
velocity dispersion.  The gas mass profile is
simply $M_g(r) = f_g M_{gal}(r)$, with $f_g$ a constant gas fraction.  
The dust has opacity $\kappa$ and is optically thin, such that the force per unit mass of wind
material due to radiation is $\frac{\kappa L}{4\pi c r^2}$.
Ignoring gas pressure, and assuming a steady state for the flow (i.e.,
the mass outflow rate, $dM_{\rm wind}/dt$, is constant), the momentum
equation for the wind is 
\begin{equation}
\mvr \frac{d\mvr}{dr} = -\frac{GM_{gal}(r)}{r^2} + \frac{\kappa L}{4\pi c r^2}.
\end{equation}
Substituting $2\sigma^2 r/G$ for $M_{gal}(r)$ and solving for $\mvr$,
\citet{mqt05} find
\begin{equation}
\mvr(r) = 2\sigma \sqrt{R_g \ltp \frac{1}{R_0} - \frac{1}{r} \rtp
   + \ln\ltp R_0/r \rtp }
\end{equation}
(their Equ. 26),
where $\mvr = 0$ at $R_0$ and $R_g = \frac{\kappa L}{8\pi c \sigma^2}$.
The corresponding density profile for the gas is given by
\begin{equation}
n_{\rm H}(r) = \frac{dM_{\rm wind}/dt}{m_p r^2 v(r)}, 
\end{equation}
with $m_p$ the mass of the hydrogen atom.
Similar to the fiducial wind model, this wind has a decreasing density
law with $n_{\rm H}$ roughly proportional to $r^{-2}$. 
In contrast to the fiducial model, the velocity law is nearly
constant with radius before decreasing sharply at large radii.
To produce a wind whose \mgiia\ optical depth profile peaks at 
$\tau_{2796}^{\rm max} \gtrsim 10$, we set $R_0 = r_{\rm inner} =
1\,\rm kpc$, $R_g=4\,\rm kpc$, and $\sigma = 125 \mkms$.  The density
peaks at $r=r_{\rm inner}$ where $n_{\rm H}^0 = 0.03
\cm{-3}$.  The velocity, meanwhile, peaks at $r=R_g$ with $\mvr
\approx 400 \mkms$.  This corresponds to a mass flow of
$dM_{\rm wind}/dt \approx 0.2 \msun\; \rm yr^{-1}$.  

Figure~\ref{fig:rad_spec} presents the \ion{Mg}{2} and \ion{Fe}{2}
profiles for this radiation-driven wind model compared against the
fiducial wind model.
In contrast to the fiducial model, 
the optical depth peaks at much larger velocity (here $\mdv
\approx -400 \mkms$).  At these velocities, the absorption is not
filled-in by scattered photons and one recovers absorption lines
that more closely follow the intrinsic optical depth profile (dotted
curve). The \ion{Mg}{2} line-emission, meanwhile, has
similar peak flux and velocity centroids to the fiducial model (see also
Table~\ref{tab:plaw_diag}).  The \feiis\ emission also has a similar
equivalent width, yet the lines of the radiation-driven model are much
broader. This reflects the fact that the majority of absorption
occurs at $\mdv \ll 0 \mkms$.
The kinematics of \feiis\ line-emission, therefore, independently
diagnose the intrinsic optical depth profile of the flow.
We conclude that the radiation-driven model has characteristics that
are qualitatively similar to the fiducial model (e.g.\ strong
blue-shifted absorption and significant line-emission) with a few
quantitative differences that, in principle, could distinguish them
with high fidelity observations.

\subsection{Power-Law Models}
\label{sec:power}

\begin{figure}
\includegraphics[width=3.5in]{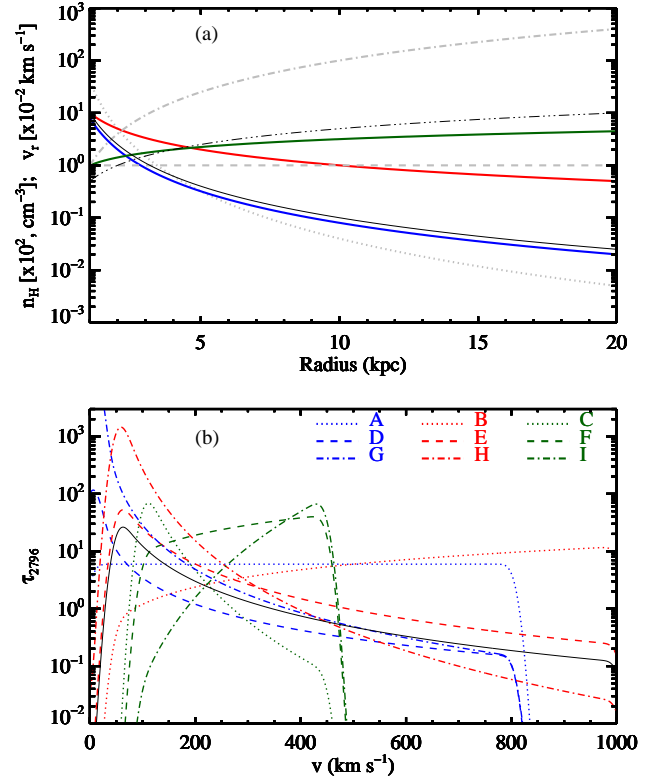}
\caption{
In panel (a) we illustrate the density (gray curves; dashed -- $\mnhn
\propto r^0$, dotted -- $\mnhn \propto r^{-3}$, dot-dash -- $\mnhn
\propto r^{2}$) and velocity laws (colored; green -- $v \propto
r^{0.5}$, red -- $v \propto r^{-1}$, blue -- $v \propto r^{-2}$) for a
series of power-law models (Table~\ref{tab:plaw_parm}).  For
comparison we also plot the velocity and density laws for the fiducial
model (thin, black curves).  
Panel (b) plots the optical depth profiles of the \mgiia\ line for
the nine models built from the density and velocity laws of panel (a).
Each of these were normalized to have peak optical depths of 10 to
1000 and to be optically thin at small or large velocity.
}
\label{fig:plaws}
\end{figure}

The fiducial model assumed power-law descriptions for both the density and velocity
laws (Equations~\ref{eqn:density},\ref{eqn:vel}).
The power-law exponents were arbitrarily chosen, i.e.\ with little physical
motivation.  In this sub-section, we cursorily explore the results for a series
of other power-law expressions, also arbitrarily defined.
Our intention is to illustrate the diversity of \ion{Mg}{2} and
\ion{Fe}{2} profiles that may result from modifications to the density
and velocity laws.
We consider three different density laws ($n_{\rm H} \propto
r^{-3}, r^{0}, r^2$) and three different velocity laws ($\mvr
\propto r^{-2}, r^{-1}, r^{0.5}$) for nine wind models
(Table~\ref{tab:plaw_parm}, Figure~\ref{fig:plaws}a).  
Each of these winds extends over the same inner and outer radii as the
fiducial model.
The density and velocity normalizations $n_{\rm H}^0, v^0$ have been
modified to yield a \mgiia\ optical depth profile that peaks at
$\tau_{2796}^{\rm max} \approx 10-10^3$ and then usually decreases to
$\tau_{2796} < 1$ (Figure~\ref{fig:plaws}b). 
This choice of normalization was observationally driven to yield
significant absorption lines.
All models assume a Doppler parameter
$b=15\mkms$, full isotropy, and a dust-free environment.
Lastly, the abundances of Mg$^+$ and Fe$^+$ scale with hydrogen as
in the fiducial model (Table~\ref{tab:fiducial}).

\begin{deluxetable}{cccccc}
\tablewidth{0pc}
\tablecaption{Wind Parameters: Power-law Models\label{tab:plaw_parm}}
\tabletypesize{\footnotesize}
\tablehead{\colhead{Label} & \colhead{$n(r)$} & \colhead{$n_{\rm
      H}^0$} & \colhead{$v(r)$} & \colhead{$v^0$} &
  \colhead{$\log \tau_{2796}^{\rm max}$} }
\startdata
A & $r^{-3}$& 0.4000 & $r^{-2.0}$&   2.0 & 0.8 \\ 
B & $r^{-3}$& 0.5000 & $r^{-1.0}$&  50.0 & 1.1 \\ 
C & $r^{-3}$& 0.3000 & $r^{ 0.5}$& 100.0 & 1.8 \\ 
D & $r^{ 0}$& 0.0100 & $r^{-2.0}$&   2.0 & 2.1 \\ 
E & $r^{ 0}$& 0.0100 & $r^{-1.0}$&  50.0 & 1.7 \\ 
F & $r^{ 0}$& 0.0200 & $r^{ 0.5}$& 100.0 & 1.6 \\ 
G & $r^{ 2}$& 0.0100 & $r^{-2.0}$&   2.0 & 4.4 \\ 
H & $r^{ 2}$& 0.0010 & $r^{-1.0}$&  50.0 & 3.2 \\ 
I & $r^{ 2}$& 0.0001 & $r^{ 0.5}$& 100.0 & 1.8 \\ 
\enddata
\tablecomments{For velocity laws that decrease with radius, the
normalization $v^0$ refers to the velocity at $r=r_{\rm outer}$
instead of $r=r_{\rm inner}$.  All models assume a Doppler parameter
$b=15\mkms$, full isotropy, and a dust-free environment.}

\end{deluxetable}

\begin{figure*}
\includegraphics[height=7.0in,angle=90]{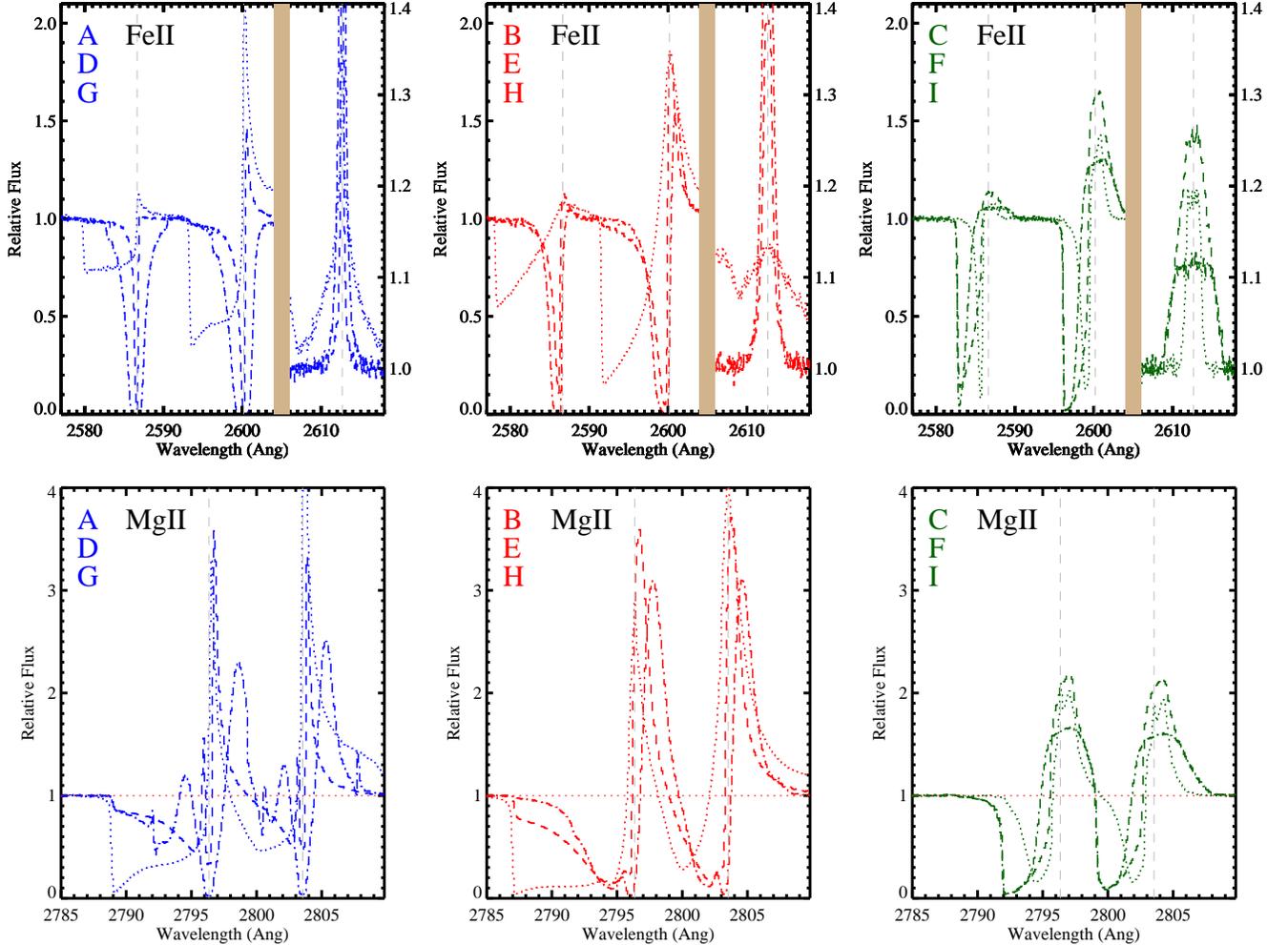}
\caption{
\ion{Mg}{2} and \ion{Fe}{2} profiles for a series of power-law wind
models (see Table~\ref{tab:plaw_parm} for details).  These panels reveal
the diversity of absorption and emission that results from a set of
intrinsic optical depth profiles that have relatively modest
differences (Figure~\ref{fig:plaws}b).  As designed, all of the models
show significant absorption blueward of systemic velocity.  As
important, each shows strong line-emission at $\mdv \sim 0 \mkms$,
with the flux proportional to the degree of absorption.
We also note that the width of the \feiic\ emission scales with the
velocity offset of peak optical depth for the \feiia\ absorption.
Therefore, the emission kinematics trace the intrinsic
optical depth profile.
}
\label{fig:plaws_spec}
\end{figure*}

Figure~\ref{fig:plaws_spec} presents the \ion{Mg}{2} and \ion{Fe}{2}
profiles for the full suite of power-law models (see also
Table~\ref{tab:plaw_diag}).  Although these
models differ qualitatively from the fiducial model in their density
and velocity laws, the resultant profiles share many of the same
characteristics.  Each shows significant absorption at $\mdv < 0
\mkms$, extending to the velocity where $\tau_{2796}$ drops below 0.1
(Figure~\ref{fig:plaws}b).  All of the models also exhibit strong
line-emission, primarily at velocities $\mdv \sim 0 \mkms$.  
This emission fills-in the \ion{Mg}{2} absorption at $\mdv \gtrsim
-100 \mkms$ such that the profiles rarely achieve a relative flux less
than $\approx 0.1$ at these velocities. 
Similar to the radiation-driven wind ($\S$~\ref{sec:radiative}),
the power-law models that have significant opacity at $\mdv <
-300 \mkms$ tend to have larger peak optical depths.  Another commonality is
the weak or absent line-emission at \feiia; one instead notes strong
\feiic\ emission that generally exceeds the equivalent width of the
fiducial model.  One also notes that the width of the \feiis\ emission
is systematically higher for models where the optical depth in
absorption peaks at large velocity from the systemic.
One of the few obvious distinctions between these models and the fiducial wind is
the higher peak optical depth of absorption in the \feiib\
transition.  This occurs primarily because these models have 
higher intrinsic optical depths (i.e.\ higher $n_{\rm Fe^+}$ values).

\begin{deluxetable*}{ccrccccccccccc}
\tablewidth{0pc}
\tablecaption{Line Diagnostics for the Fiducial Model and Variants
\label{tab:plaw_diag}}
\tabletypesize{\scriptsize}
\tablehead{\colhead{Transition} & \colhead{Model} & \colhead{$v_{\rm int}^a$} & \colhead{$W_{\rm i}$} & \colhead{$W_{\rm a}$} & \colhead{$\tau_{\rm pk}$} & \colhead{$v_\tau$} 
& \colhead{$v_{\bar \tau}$}
& \colhead{$v_{\rm int}^e$} & \colhead{$W_{\rm e}$} & \colhead{$f_{\rm pk}$} & \colhead{$v_f$} 
& \colhead{$v_{\bar f}$} & \colhead{$\Delta v_{\rm e}$}
\\
&& (\kms) & (\AA) & (\AA) && (km/s) & (km/s) & (km/s) & (\AA) & & (km/s) & (km/s) & (km/s)}
\startdata
  MgII 2796  \\
&Fiducial&[$-1009,-43$]& 4.78& 2.83&0.94&$ -215$&$ -372$&[$-32,311$]&$-1.77$& 2.48&$   32$&$  117$& 215\\
&LBG Sob.&[$-687,-86$]& 2.62& 1.35&0.45&$ -247$&$ -347$&[$-65,257$]&$-1.02$& 1.84&$   21$&$   90$& 193\\
&LBG Cov.&[$-698,-65$]& 2.67& 1.40&0.41&$ -258$&$ -344$&[$-54,214$]&$-0.93$& 2.23&$    0$&$   72$& 172\\
&Radiation&[$-365,-65$]& 2.86& 1.72&3.00&$ -301$&$ -250$&[$-54,322$]&$-1.74$& 2.13&$   53$&$  119$& 236\\
&A&[$-848,-22$]& 9.29& 5.02&2.92&$ -773$&$ -499$&[$-22,171$]&$-1.34$& 3.30&$   21$&$   64$& 118\\
&B&[$-1084,-65$]&11.08& 7.54&3.00&$ -976$&$ -589$&[$-65,193$]&$-1.83$& 2.89&$    0$&$   50$& 172\\
&C&[$-451,-65$]& 2.54& 1.82&1.86&$ -204$&$ -200$&[$-65,332$]&$-1.79$& 2.09&$    0$&$  116$& 247\\
&D&[$-794,21$]& 4.21& 2.14&1.11&$  -11$&$ -276$&[$32,311$]&$-1.57$& 3.58&$   43$&$  141$& 172\\
  MgII 2803  \\
&Fiducial&[$-437,-41$]& 3.29& 1.19&0.76&$ -148$&$ -193$&[$-41,676$]&$-2.24$& 2.55&$   34$&$  269$& 449\\
&LBG Sob.&[$-426,-73$]& 1.53& 0.46&0.24&$ -244$&$ -244$&[$-62,387$]&$-0.80$& 1.58&$   23$&$  154$& 332\\
&LBG Cov.&[$-533,-73$]& 2.56& 0.98&0.37&$ -233$&$ -285$&[$-51,558$]&$-1.48$& 2.35&$    2$&$  227$& 396\\
&Radiation&[$-362,-84$]& 2.32& 1.29&2.65&$ -308$&$ -256$&[$-73,323$]&$-1.31$& 1.86&$   45$&$  115$& 246\\
&A&[$-554,-41$]& 7.06& 1.98&0.78&$ -330$&$ -293$&[$-41,805$]&$-5.72$& 5.03&$   13$&$  304$& 610\\
&B&[$-554,-169$]& 5.74& 1.80&1.27&$ -372$&$ -359$&[$-84,1008$]&$-7.34$& 4.20&$    2$&$  338$& 749\\
&C&[$-405,-73$]& 2.17& 1.49&1.85&$ -169$&$ -179$&[$-73,355$]&$-1.52$& 1.94&$   77$&$  122$& 246\\
&D&[$-30,13$]& 3.69& 0.19&0.77&$   -9$&$   -1$&[$23,558$]&$-1.48$& 3.33&$   34$&$  251$& 332\\
  FeII 2586  \\
&Fiducial&[$-348,-35$]& 0.82& 0.61&1.01&$  -70$&$ -119$&[$-35,128$]&$-0.10$& 1.11&$   35$&$   47$& 128\\
&LBG Sob.&[$-151,-46$]& 0.18& 0.04&0.07&$ -139$&$  -99$&[$-12,35$]&$-0.01$& 1.05&$    0$&$   12$&  35\\
&LBG Cov.&[$-684,-12$]& 2.38& 1.93&0.65&$ -162$&$ -276$&[$-12,360$]&$-0.34$& 1.33&$   12$&$  170$& 313\\
&Radiation&[$0,0$]& 1.00& 0.00&0.02&$    0$&$    0$&[$12,255$]&$-0.11$& 1.09&$   35$&$  132$& 220\\
&A&[$-812,-12$]& 1.88& 1.66&0.33&$ -742$&$ -422$&[$-12,232$]&$-0.12$& 1.12&$   23$&$  109$& 197\\
&B&[$-1009,-186$]& 2.20& 1.97&0.60&$ -951$&$ -694$&[$-93,418$]&$-0.24$& 1.09&$   12$&$  161$& 429\\
&C&[$-278,-81$]& 0.79& 0.71&2.51&$ -116$&$ -132$&[$-70,162$]&$-0.10$& 1.08&$   93$&$   46$& 174\\
&D&[$-209,23$]& 0.88& 0.68&3.00&$  -12$&$  -26$&[$35,70$]&$-0.01$& 1.03&$   46$&$   52$&  35\\
  FeII 2600  \\
&Fiducial&[$-580,-37$]& 1.87& 1.18&1.08&$  -83$&$ -181$&[$-37,459$]&$-0.83$& 1.70&$   32$&$  191$& 312\\
&LBG Sob.&[$-453,-72$]& 0.55& 0.24&0.11&$ -187$&$ -263$&[$-26,228$]&$-0.21$& 1.23&$   20$&$   99$& 208\\
&LBG Cov.&[$-695,-60$]& 2.43& 1.38&0.44&$ -360$&$ -336$&[$-49,493$]&$-1.12$& 2.02&$    9$&$  203$& 381\\
&Radiation&[$-360,-26$]& 2.21& 1.44&3.00&$ -303$&$ -247$&[$-26,320$]&$-0.84$& 1.59&$   44$&$  138$& 242\\
&A&[$-822,-26$]& 4.61& 3.51&1.05&$ -753$&$ -455$&[$-26,793$]&$-1.78$& 2.34&$    9$&$  347$& 612\\
&B&[$-1018,-199$]& 5.10& 4.20&1.88&$ -960$&$ -701$&[$-176,1001$]&$-2.84$& 1.86&$   -3$&$  371$& 935\\
&C&[$-360,-72$]& 1.34& 1.09&2.04&$ -141$&$ -154$&[$-72,228$]&$-0.62$& 1.43&$   78$&$   75$& 219\\
&D&[$-430,20$]& 1.68& 1.13&3.00&$  -14$&$  -58$&[$32,309$]&$-0.31$& 1.47&$   44$&$  163$& 219\\
  FeII* 2612 \\
&Fiducial&&&&&&&[$-173,183$]&$-0.34$& 1.24&$  -23$&$    5$& 241\\
&LBG Sob.&&&&&&&[$-69,57$]&$-0.04$& 1.07&$    0$&$   -6$& 115\\
&LBG Cov.&&&&&&&[$-356,436$]&$-1.04$& 1.79&$    0$&$   35$& 563\\
&Radiation&&&&&&&[$-299,298$]&$-0.50$& 1.26&$    0$&$   -1$& 425\\
&A&&&&&&&[$-402,574$]&$-0.88$& 1.35&$    0$&$   80$& 735\\
&B&&&&&&&[$-402,597$]&$-0.88$& 1.14&$   34$&$   94$& 873\\
&C&&&&&&&[$-184,172$]&$-0.40$& 1.20&$  -81$&$   -6$& 241\\
&D&&&&&&&[$-115,114$]&$-0.47$& 2.37&$    0$&$   -1$& 103\\
  FeII* 2626 \\
&Fiducial&&&&&&&[$-171,183$]&$-0.39$& 1.29&$  -46$&$    5$& 217\\
&LBG Sob.&&&&&&&[$-23,46$]&$-0.03$& 1.06&$   23$&$   11$&  57\\
&LBG Cov.&&&&&&&[$-251,274$]&$-0.28$& 1.22&$    0$&$   11$& 434\\
&Radiation&&&&&&&[$-286,263$]&$-0.60$& 1.42&$    0$&$  -11$& 343\\
&A&&&&&&&[$-663,320$]&$-1.60$& 1.71&$    0$&$ -152$& 720\\
&B&&&&&&&[$-891,320$]&$-1.45$& 1.22&$    0$&$ -270$&1039\\
&C&&&&&&&[$-171,160$]&$-0.46$& 1.24&$   57$&$   -5$& 217\\
&D&&&&&&&[$-160,148$]&$-0.87$& 3.53&$    0$&$   -5$& 103\\
  FeII* 2632 \\
&Fiducial&&&&&&&[$-132,119$]&$-0.16$& 1.12&$   16$&$   -6$& 194\\
&LBG Sob.&&&&&&&[$-29,16$]&$-0.01$& 1.07&$   -7$&$   -7$&  46\\
&LBG Cov.&&&&&&&[$-292,301$]&$-0.53$& 1.36&$    5$&$    4$& 479\\
&Radiation&&&&&&&[$-235,210$]&$-0.23$& 1.12&$    5$&$  -12$& 353\\
&A&&&&&&&[$-326,370$]&$-0.59$& 1.19&$   -7$&$   15$& 581\\
&B&&&&&&&[$-326,563$]&$-0.78$& 1.17&$ -314$&$  108$& 741\\
&C&&&&&&&[$-143,153$]&$-0.19$& 1.10&$   62$&$    5$& 228\\
&D&&&&&&&[$-86,96$]&$-0.23$& 1.63&$    5$&$    4$&  91\\
\enddata
\tablecomments{{L}isted are the equivalent widths (intrinsic, absorption, and emission), the peak optical depth for the absorption
$\tau_{\rm pk} \equiv -\ln(I_{\rm min})$, the velocity where the optical depth peaks $v_\tau$, the optical depth-weighted velocity centroid 
$v_{\bar \tau} \equiv \int dv \, v \ln[I(v)] / \int dv \ln[I(v)]$, the peak flux $f_{\rm pk}$ in emission, the velocity where the flux peaks 
$v_f$, the flux-weighted velocity centroid of the emission line $v_{\bar f}$ (occasionally affected by blends with neighboring emission lines), and the $90\%$ width $\Delta v_{\rm e}$.
The $v^a_{\rm int}$ and $v^e_{\rm int}$ columns give the velocity
range used to calculate the absorption and emission characteristics,
respecitvely.  These were defined by the velcoities where the profile
crossed 0.95 in the normalized flux. [See the electronic version for
the complete Table].}
\end{deluxetable*}

While there is commonality between models, the specific
characteristics of the absorption/emission profiles do exhibit
significant diversity.
Table~\ref{tab:plaw_diag} compares measures of the absorption and
emission profiles against the fiducial model.
In detail, the line profiles differ in their peak optical depths,
the velocity centroids of absorption/emission, and their equivalent widths.
We find that many of the differences are driven by differences in
the velocity laws, i.e.\ the kinematics of the outflow.
At the same time, models with very different density/velocity laws can
produce rather similar results.  For example, profiles that have declining
opacity with increasing velocity offset from systemic can be obtained
with a radial velocity law that increases (e.g.\ the fiducial model)
or decreases (model~D).  The implication is that absorption profiles
alone are likely insufficient to fully characterize the physical
characteristics of the outflow.

\section{Discussion}
\label{sec:discuss}

We now discuss the principal results of our analysis and comment on
the observational consequences and implications. 

The previous sections presented idealized wind models for
cool gas outflows, and explored the absorption/emission profiles of
the \mgiid\ doublet and the \ion{Fe}{2} UV1 multiplet.  In addition to
the ubiquitous presence of blue-shifted absorption, the wind models
also predict strong line-emission in both resonance
and non-resonance transitions.  For isotropic and dust-free scenarios,
this is 
due to the 
simple conservation of photons: every photon emitted by
the source eventually escapes the system to maintain
zero total equivalent width. A principal result of this paper,
therefore, is that galaxies with observed cool gas outflows should
also generate detectable line-emission.
 
Indeed, line-emission from low-ion transitions has been reported from 
star-forming galaxies that exhibit cool gas outflows.  
This includes emission related to \ion{Na}{1} \citep{phillips93,cth+10},
\ion{Mg}{2} emission \citep{wcp+09,rwk+10}, 
\ion{Si}{2}* emission from LBGs \citep{shapley03},
and, most recently, significant \feiis\ emission \citep{rubin+10c}. 
A variety of origins have been proposed for this
line-emission including AGN activity, recombination in \ion{H}{2}
regions, and back-scattering off the galactic-scale wind.
Our results suggest that the majority of the observed line-emission is 
from scattered photons in the cool gas outflows of these star-forming
galaxies.  Indeed, line-emission should be generated by all 
galaxies driving a cool gas outflow.

On the other hand,
there are many examples of galaxies where blue-shifted absorption is
detected yet the authors report no significant line emission.  This
includes \ion{Na}{1} flows \citep{rvs05a,martin05,smn+09}, 
\ion{Mg}{2} and \ion{Fe}{2} absorption in
ULIRGS \citep{mb09}, and the extreme \ion{Mg}{2} outflows identified
by \cite{tmd07}.  Similarly, there have been no reports of
resonance line-emission from low-ion transitions in the
LBGs\footnote{\citet{psa+00}
  reported and analyzed \ion{C}{4} line-emission from the lensed LBG
  cb58, attributing this flux to stellar winds.  We caution, however,
  that the galactic-scale wind proposed for cb58 should also
  have contributed to the \ion{C}{4} line-emission.}, and many
$z \lesssim 1$ galaxies show no detectable \ion{Mg}{2} or \ion{Fe}{2}*
emission despite significant blue-shifted absorption (Rubin et al., in prep). 
These non-detections appear to contradict a primary conclusion of 
this paper.   We are motivated, therefore, to
reassess several of the effects that can reduce the line-emission
and consider whether these prevent its detection in many
star-forming galaxies.  

Dust is frequently invoked to explain the suppression of line-emission
for resonantly trapped transitions (e.g.\ \lya).  Indeed, a photon
that is trapped for many scatterings within a dusty medium will 
be preferentially extincted relative to a non-resonant photon.  In
$\S$~\ref{sec:dust}, we examined the effects of dust extinction on 
\ion{Mg}{2} and \ion{Fe}{2} emission.  The general result
(Figure~\ref{fig:dust_tau}) is a modest
reduction in flux that scales as (1+\taud)$^{-1}$ instead of
exp(--\taud). Although the \ion{Mg}{2} photons are resonantly
trapped, they require only one to a few scatterings
to escape the wind thereby limiting the effects of
dust.  This reflects the moderate opacity of the \ion{Mg}{2}
doublet (e.g.\ relative to \lya) and also the velocity law of the
fiducial wind model.
In scenarios where the \ion{Mg}{2} photons are
more effectively trapped, dust does suppress
the emission (e.g.\ in the ISM+wind model from $\S$~\ref{sec:ISM}).
In contrast to \ion{Mg}{2}, 
the \ion{Fe}{2} resonance photons may be
converted to non-resonant photons that freely escape the wind.
Therefore, a wind model that traps \ion{Mg}{2}
photons for many scatterings does not similarly trap 
\feiid\ photons.  The effects of dust are much 
reduced and, by inference, the same holds for any other set of
transitions that are coupled to a fine-structure level (e.g.\
\ion{Si}{2}/\ion{Si}{2}*).  In summary,
dust does reduce the line flux relative to the continuum, but 
it generally has only a modest effect on the predicted \ion{Mg}{2}
emission and a minor effect on \feiis\ emission.

Another factor that may reduce line-emission is 
wind 
anisotropy. The flux is lower, for example, if one
eliminates the backside to the wind (e.g.\ the source itself could
shadow a significant fraction of the backside).  
Similarly, a bi-conical
wind can have significantly lower line-emission, at 
least for the fiducial model ($\S$~\ref{sec:anisotropic}). 
For the anisotropic winds explored in this paper,
the line-emission scales 
roughly as $\Omega/4\pi$ where $\Omega$ is the angular extent of the
wind\footnote{There is also a first order dependence on the actual
  orientation of the wind, with maximal emission when the backside is
  fully viewed.}.  Because we require that the wind 
points toward us, it is difficult to reduce $\Omega$ much
below $2 \pi$ and, therefore, anisotropy reduces the
emission by a factor of order unity.  
Similarly, an anisotropically emitting source would modify the predicted
line-emission.  For example, if the backside of the galaxy were
brighter/fainter then would one predict brighter/fainter emission
relative to the observed continuum.
To reduce the emission, the brightest regions
of the galaxy would need to be oriented toward Earth.  Although this
would not be a
generic orientation, most spectroscopic samples are magnitude-limited
and biased towards detecting galaxies when observed at their
brightest.  In principle, this might imply a further
reduction of order unity.

Together, dust and anisotropic models may reduce the line-emission
of \ion{Mg}{2} and \ion{Fe}{2}
by one to a few factors of order unity.  For some galaxies, these
effects may explain the absence of significant line-emission,
but they may not be sufficient to preclude its detection. 

There is another (more subtle) effect that could
greatly reduce the observed line-emission: slit-loss.  As
described in Figure~\ref{fig:fiducial_ifu_mgii}, 
emission from the wind is spatially extended with
a non-zero surface brightness predicted at large radii. This implies 
a non-negligible luminosity emitted beyond the angular extent of the galaxy.  
The majority of observations of star-forming galaxies to date have
been taken through a spectroscopic slit designed to cover
the brightest continuum regions. Slits with angular
extents of $1-1.5''$ subtend roughly $8-15$\,kpc for galaxies at $z>0.5$.
Therefore, a $1''$ slit covering a galaxy with our fiducial wind would cover
less than half the wind.  The result is reduced line-emission
when compared to the galaxy continuum.

\begin{figure}
\includegraphics[height=3.5in,angle=90]{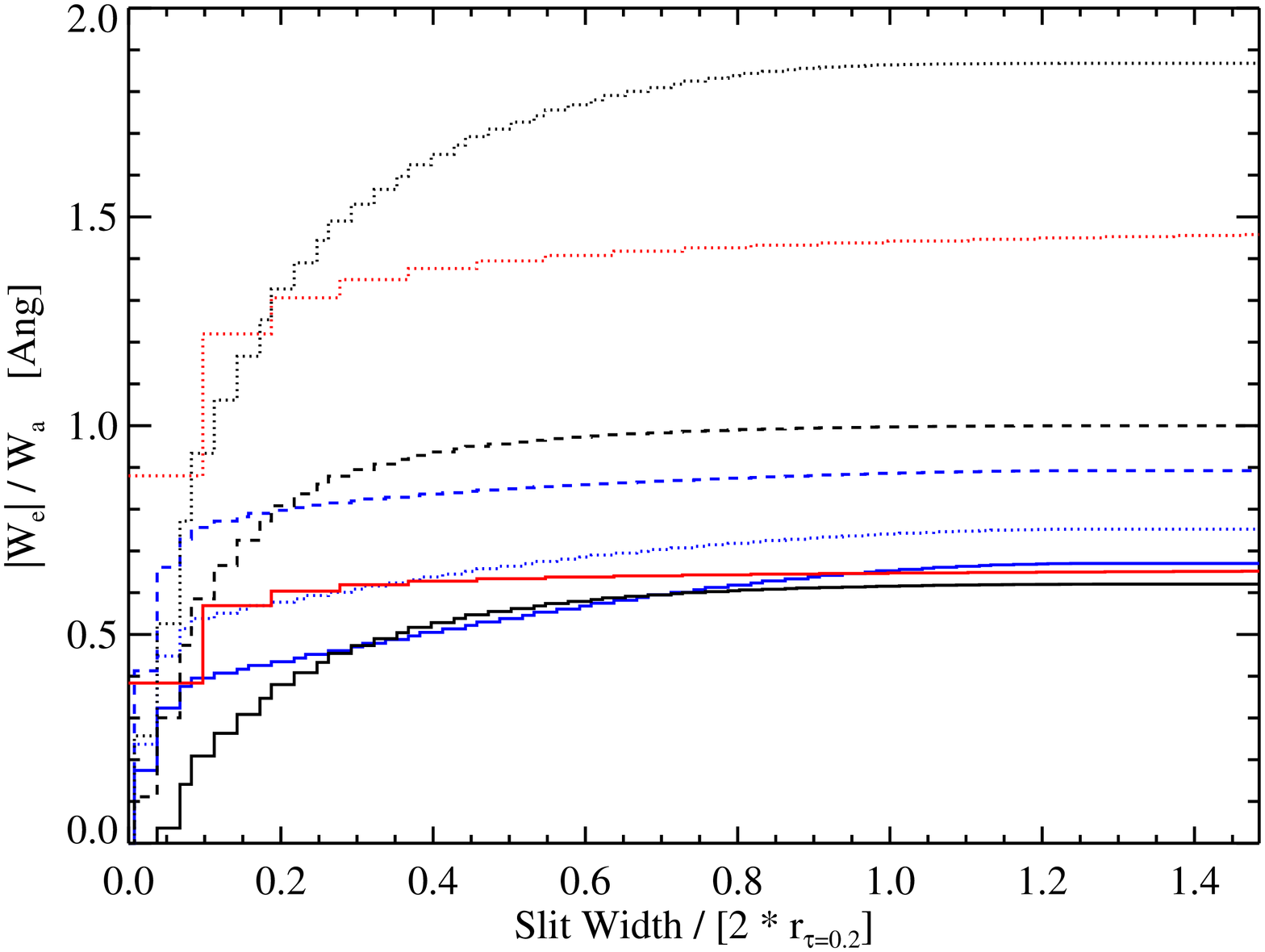}
\caption{
Absolute emission equivalent width ($|W_{\rm e}|$) relative to the observed
absorption equivalent width ($W_{\rm a}$) for a series of transitions:
(solid -- \mgiia; dotted -- \mgiib; dashed -- \feiib).  The
\ewe/\ewabs\ ratio is plotted as a function of slit width relative to
twice the radius \rtt, defined to be where the Sobolev optical depth 
$\tau_S = 0.2$.  The black curves correspond to the fiducial wind
model ($\S$~\ref{sec:fiducial}; $\mrtt \approx 15$\,kpc), the red curves are for the
LBG-partial covering scenario ($\S$~\ref{sec:Covering}; $\mrtt \approx
8$\,kpc), and the
radiation-driven wind ($\S$~\ref{sec:radiative}; $\mrtt = 40$kpc) has blue curves.
For all of the wind models, the $|W_{\rm e}|$/\ewabs\ ratio rises very steeply
with slit width and then plateaus at $\approx \mrtt$.  
Therefore, a slit that exceeds $\approx 0.5 \mrtt$ will admit nearly
all of the photons scattered to our sightline.
}
\label{fig:obs_slit}
\end{figure}

We explore the effects of slit-loss as follows.
We model the slit as a perfect, infinitely-long rectangle centered on the
galaxy.  We then tabulate the equivalent width of the line-emission
through slits with a range of widths, parameterized by \rtt: twice the radius
where the wind has a Sobolev optical depth of $\tau_S = 0.2$. 
The line-emission will be weak beyond
this radius because the photons have only a low probability of
being scattered.  For the fiducial wind model, 
the \mgiia\ transition has $\mrtt \approx 15$\,kpc 
(Figure~\ref{fig:fiducial_nvt}).  
The predicted equivalent width in emission \ewe\ relative to the absorption
equivalent width \ewabs\ is presented in
Figure~\ref{fig:obs_slit} for a series of transitions for the fiducial
wind model (black curves).
The \ewe/\ewabs\ curves rise very steeply with increasing slit width and then
plateau when the slit width reaches $\approx \mrtt$.  
For the fiducial model, the emission is concentrated toward the source; this
derives from the density and velocity profiles but is also a simple
consequence of geometric projection.
Figure~\ref{fig:obs_slit} reveals similar results for other wind
models. The curves rise so steeply that 
the effects of slit-loss are minor (order unity)
unless the slit-width is very small.   
Nevertheless, the results do motivate
extended aperture observations, e.g.\ integral field unit (IFU)
instrumentation, that would map the wind both spatially and spectrally.

Although several effects can reduce the line-emission
relative to the absorption of the outflow, our analysis
indicates that detectable line-emission should occur frequently. 
Furthermore, the line-emission could be suppressed so 
that it does not
exceed
the galaxy continuum yet still (partially) fills-in the absorption
profiles. Indeed, dust and anisotropic winds preferentially suppress
line-emission at $\mdv > 0 \mkms$
(Figures~\ref{fig:anisotropic},\ref{fig:dust}).
The remaining emission would still 
modify the observed absorption profiles (e.g.\
Figure~\ref{fig:noemiss}) and may complicate conclusions
regarding characteristics of the flow.  {\it An analysis of cool gas outflows
that entirely ignores line-emission may incorrectly conclude that the source is
partially covered, that the gas has a significantly lower peak optical depth,
and/or that a $\mdv \sim 0 \mkms$ component is
absent.}  
We now examine several quantitative effects of line-emission.

\begin{figure}
\includegraphics[height=3.5in,angle=90]{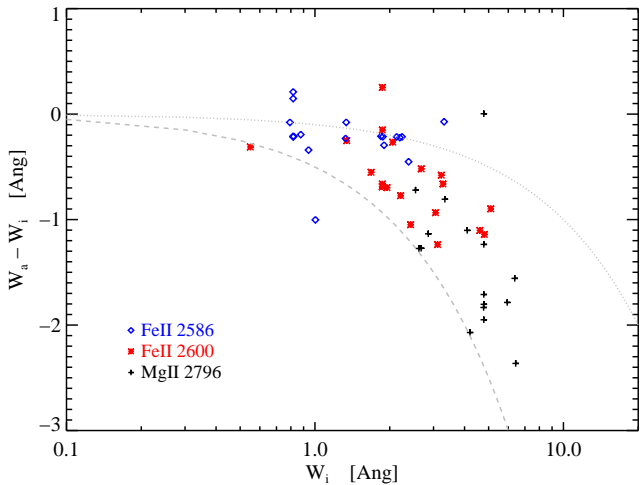}
\caption{
Difference between the `observed' absorption equivalent width \ewabs\
which includes the flux of scattered photons and the `intrinsic'
equivalent width $W_{\rm i}$ that ignores photon scattering.  The
dashed (dotted) curves trace a 50\%\ (10\%) reduction in \ewabs\
relative to $W_{\rm i}$.  One notes a reduction in \ewabs\ by $\approx
30-50\%$ for the \mgiia\ transition (the effect is generally larger
for \mgiib).  The effects of scattered photons are reduced for the
\ion{Fe}{2} transitions because a fraction (in fact a majority for
\feiia) of the absorbed photons fluoresce as \feiis\ emission at
longer wavelengths and do not `fill-in' the absorption profiles.
The figure shows results for all of the models presented in
Tables~\ref{tab:line_diag} and \ref{tab:plaw_diag}.
}
\label{fig:obs_ew}
\end{figure}

Figure~\ref{fig:obs_ew} demonstrates one observational
consequence: reduced measurements for the absorption
equivalent width \ewabs\ of the flow.  In the case of \ion{Mg}{2},
which has the most strongly affected transitions, \ewabs\ is reduced by
$30-50\%$ from the intrinsic equivalent width (the equivalent width
one would measure in the absence of scattered photons).  In turn, one may derive 
a systematically lower optical depth or velocity extent for the wind,
and therefore a lower total mass and kinetic energy.  
The effects are most pronounced for wind
scenarios where the peak optical depth occurs near $\mdv = 0 \mkms$.
Geometric projection limits the majority of scattered photon emission to
have $|\mdv| < 200 \mkms$; therefore,  the absorption profiles
are filled-in primarily at these velocities.  

\begin{figure}
\includegraphics[height=3.5in,angle=90]{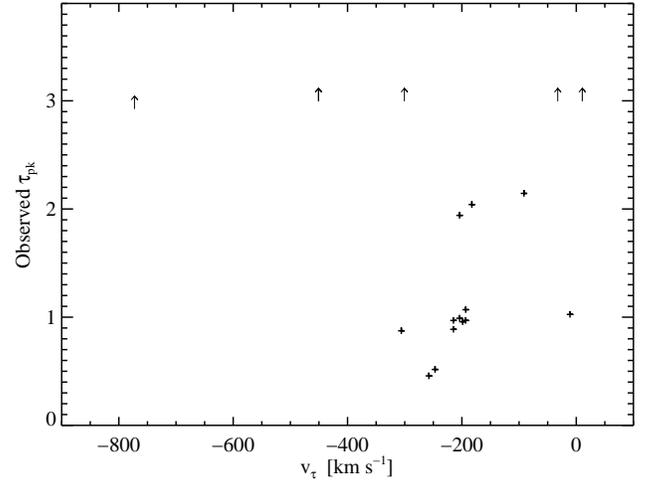}
\caption{
The observed peak optical depth $\tau_{\rm pk} \equiv -\ln [I_{\rm
  min}]$ for \mgiia\ profiles of the wind models studied in this paper
(Tables~\ref{tab:line_diag}, \ref{tab:plaw_diag}) with $I_{\rm min}$
the minimum normalized intensity of the absorption profile.  Cases
where \tpk\ exceeds 3 are presented as lower limits.  The \tpk\ values
are plotted against the velocity $v_\tau$ where $I(v_\tau) = I_{\rm min}$.  In all
of the models, the true peak optical depth $\mtpk^{\rm true} > 10$.
The much lower `observed' \tpk\ values occur because scattered \mgiia\
photons have filled-in the absorption profiles at velocities $\mdv
\gtrsim -400 \mkms$. These effects, therefore, are greatest for wind
models where the optical depth peaks near $\mdv \sim 0 \mkms$ because
the majority of scattered photons have this relative velocity.
Indeed, models with $v_\tau < -400 \mkms$ all show $\mtpk > 3$.
}
\label{fig:obs_peaktau}
\end{figure}

Another (related) consequence is the reduction of the peak depth of 
absorption.
In Figure~\ref{fig:obs_peaktau}, we plot the observed peak optical depth
$\tau_{\rm pk}$ for \mgiia\ 
versus the velocity where the profile has greatest depth 
for the various wind models (i.e.\
$\tau_{\rm pk}$ vs.\ $v_\tau$ from Tables~\ref{tab:line_diag} and
\ref{tab:plaw_diag}).   In every one of the models, the true peak
optical depth $\tau_{\rm pk}^{\rm true} > 10$.  For the
majority of cases with $v_\tau > -300\mkms$, one observes $\tau_{\rm pk} <
2$ and would infer the wind is not even optically thick!
This occurs because photons scattered by the wind have `filled-in' the
absorption at velocity $\mdv \approx 0 \mkms$.  In contrast, wind
models with $v_\tau < -400 \mkms$ all yield $\tau_{\rm pk} > 3$.  The
results are similar for the \mgiib\ profile.  In fact, one generally
measures a similar $\tau_{\rm pk}$ for each \ion{Mg}{2} line and 
may incorrectly conclude that the source is partially covered. 
In the case of \ion{Fe}{2}, one may even observe a (non-physical)
inversion in the apparent optical depths of \feiia\ and \feiib\ (e.g.\
Figure~\ref{fig:norm}). 

\begin{figure}
\includegraphics[height=3.5in,angle=90]{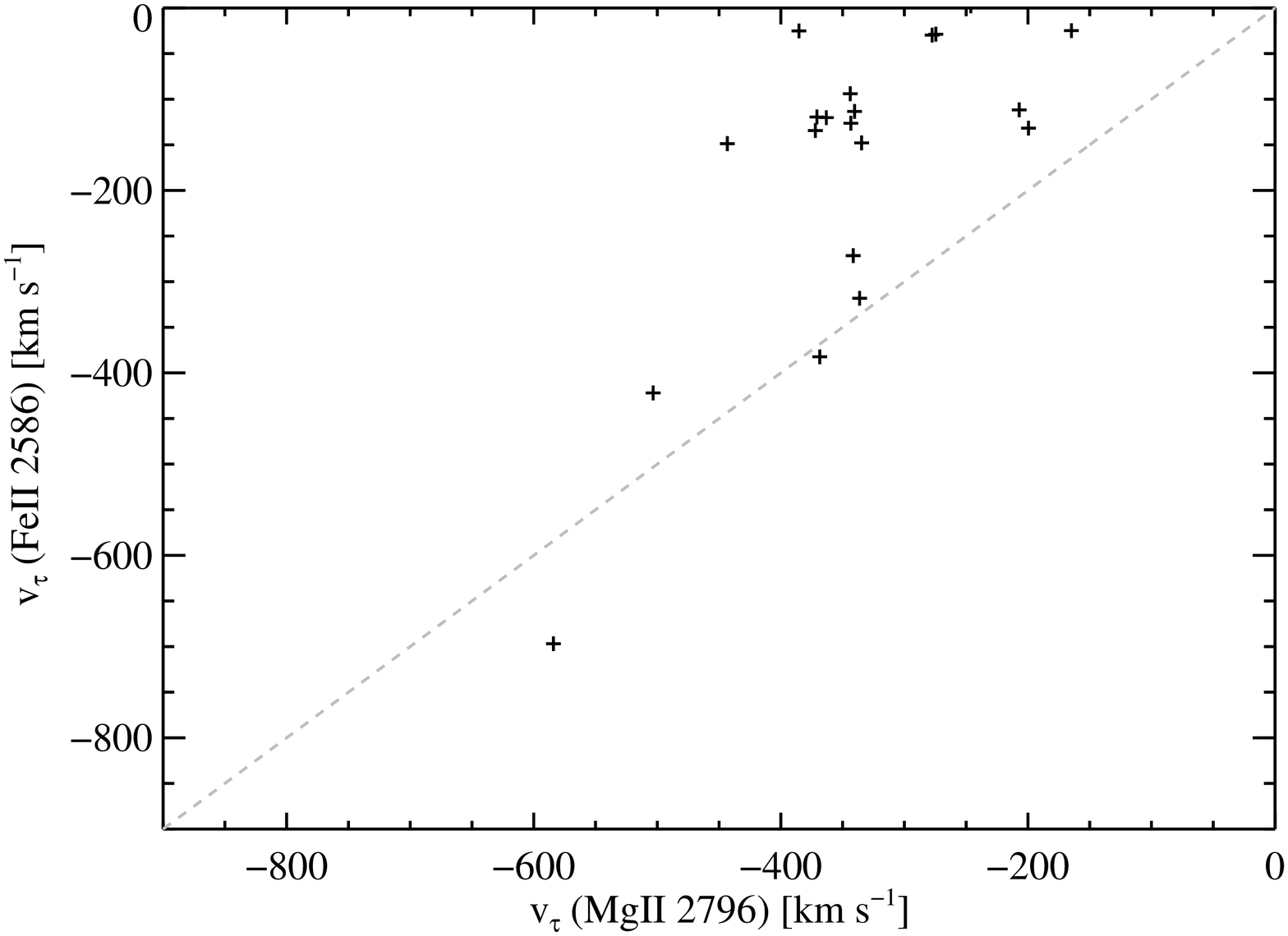}
\caption{
Comparison of the velocities for which the \feiia\ and \mgiia\ absorption
profiles have greatest depth.  Results for all of the models
studied in this paper are presented (Tables~\ref{tab:line_diag},
\ref{tab:plaw_diag}).  
It is evident that wind models with a peak optical depth near the
systemic velocity have a \mgiia\ absorption profile shifted blueward
by one to several hundred \kms.  Analysis of such profiles may lead to
the false conclusion 
that (i) the majority of mass in the wind is travelling at a higher
velocity; and (ii) there is no gas with $\mdv \sim 0 \mkms$.
The dashed curve traces the one-to-one line.
}
\label{fig:obs_vtau}
\end{figure}

These effects 
are reduced
for the \ion{Fe}{2} absorption profiles, especially for \feiia.  This
is because a significant fraction (even a majority) of the
\ion{Fe}{2} emission is florescent \ion{Fe}{2}* emission at longer
wavelengths which does not affect the absorption profiles. 
Therefore, the \ion{Fe}{2} absorption equivalent widths (\ewabs) more
closely follow the intrinsic values, one derives more accurate peak
optical depths, and the absorption kinematics more faithfully reflect
the motions of the flow.  Regarding the last point, one also
predicts a velocity offset between the \ion{Fe}{2} and \ion{Mg}{2}
absorption-line centroids (Figure~\ref{fig:obs_vtau}).
This affects the analysis of gas related to the ISM of the galaxy and
also material infalling at modest speeds.  
Figure~\ref{fig:obs_vtau} also emphasizes that
the \ion{Mg}{2} profiles may misrepresent the kinematics of the bulk of
the gas.  Analysis of these lines, without consideration of
line-emission, may lead to incorrect conclusions on the energetics
and mass flux of the wind.

\begin{figure}
\includegraphics[height=3.5in,angle=90]{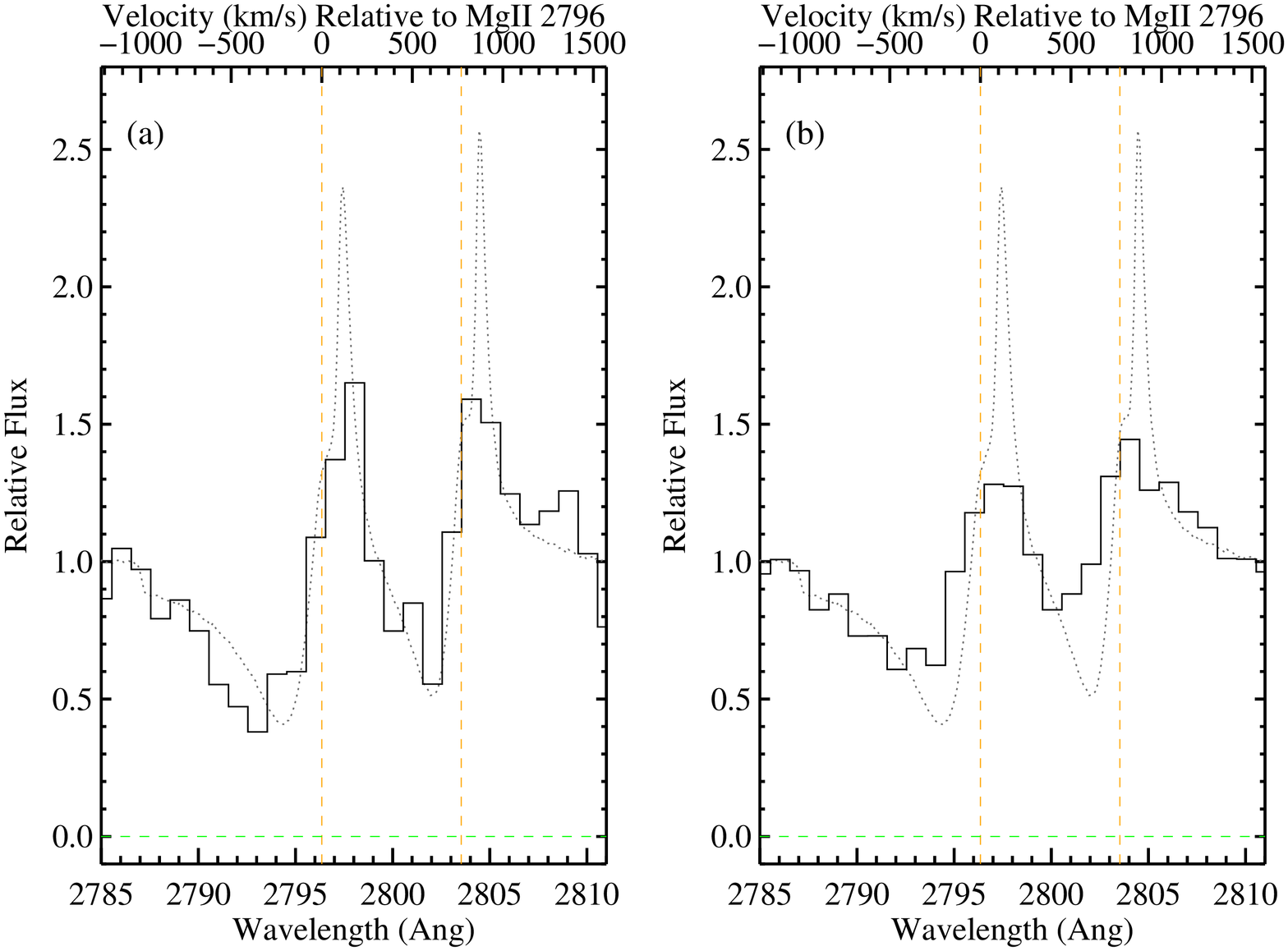}
\caption{
`Observations' of the \ion{Mg}{2} profiles for the ISM+wind model.  
(a) The true profile (dotted lines) has been convolved with a
Gaussian line-spread-function (FWHM~$= 250 \mkms$) and noise has been
added to give a S/N=7 per pixel.  Note the suppressed peak in 
line-emission; one may even infer the flux of \mgiia\ exceeds that of
\mgiib.  The \mgiia\ absorption is also shifted to
shorter wavelengths, i.e.\ to a greater velocity offset from systemic.  
(b) The solid curve shows a stack of 100 spectra of the ISM+wind
profile, each degraded to a S/N=2 per pixel, and offset from systemic
by a normal deviate with $\sigma = 100 \mkms$ to mimic uncertainty in
the redshift of the galaxies.  
This treatment is meant to illustrate the
implications of stacking galaxy spectra to study outflows
\citep[e.g.][S10]{wcp+09,rwk+10}.   The main difference from the
single galaxy observation shown in panel (a) is the smearing of
line-emission and absorption that reduces the height/depth of each.
The effects would likely be even more pronounced if one studied spectra with
a diversity of \ion{Mg}{2} profiles.    
}
\label{fig:obs_lris}
\end{figure}

We emphasize that all of these effects are heightened by the
relatively low spectral resolution and S/N characteristic of the data
commonly acquired for $z>0$ star-forming galaxies.  Figure~\ref{fig:obs_lris}a
shows one realization of the \mgiid\ doublet for the ISM+wind model
convolved with the line-spread-function of the Keck/LRIS spectrometer
(a Gaussian with FWHM=250\kms) and an assumed signal-to-noise of
S/N=7 per 1 \AA\ pixel.  Both the absorption
and emission are well detected, but it would be
difficult to resolve the issues discussed above (e.g.\ partial
covering, peak optical depth) with these data.  
One also notes several systematic effects of the lower spectral
resolution, e.g.\ reduced peak flux in the line-emission and a
systematic shift of \mgiia\ absorption to more negative velocity.
We have also modelled an observation of the ISM+wind model using
a stacked galaxy spectrum (Figure~\ref{fig:obs_lris}b).
Specifically, we
averaged 100 identical \ion{Mg}{2} profiles from the fiducial model
degraded to a S/N=2\,pix$^{-1}$ and shifted by a random velocity
offset with $\sigma = 100\mkms$.  
This treatment illustrates the effects of using stacked galaxy 
spectra to study outflows
\citep[e.g.][S10]{wcp+09,rwk+10}.   The main difference 
between this and 
the
single galaxy observation shown in panel (a) is the smearing of
line-emission and absorption that reduces the height/depth of each.
The effects would be even more pronounced if one studied spectra with
a diversity of \ion{Mg}{2} profiles.    
Special care is required, therefore, to interpret properties of the wind (and
ISM) from such spectral analysis.


\begin{figure}
\includegraphics[height=3.5in,angle=90]{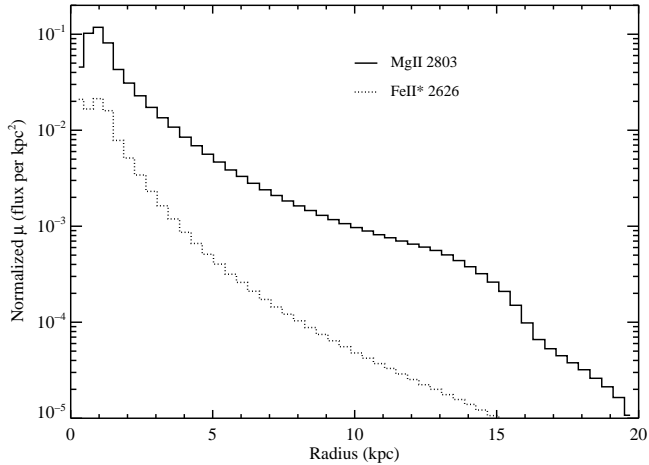}
\caption{
Surface brightness profiles of the fiducial wind model for the \mgiia\
(solid) and \feiie\ (dotted) transitions.  Both profiles peak at small
radii and decrease by several orders of magnitude before reaching the
outer edge of the wind.  The profiles are sufficiently shallow,
however, that the azimuthally integrated flux declines by only a
factor of $\approx 10$ from $r = 1$\,kpc to $r=10$\,kpc.  Therefore, a
sensitive IFU observation could map the emission (and, in principle,
the kinematics) from $r_{\rm inner}$ to approximately $r_{\rm outer}$.
}
\label{fig:obs_sb}
\end{figure}

The previous few paragraphs sounded a cautionary 
perspective on the implications for absorption-line analysis of
galactic-scale outflows in the presence of (expected) significant
line-emission.   While this is a necessary complication, 
we emphasize that such analysis remains one of the few observational
techniques at our disposal to study outflows.  Furthermore,
direct analysis of
the line-emission offers new and unique
constraints on the characteristics of the outflow.  And, when coupled
with the absorption-line data, the two sets of constraints may break
various degeneracies in the physical characteristics of the outflow.
We now consider a few examples.

The most obvious characteristics probed by the line-emission are
the size and morphology of the outflow \citep{rubin+10c}.  
Line emission is predicted to extend to radii where the Sobolev
optical depth exceeds a few tenths.  The principle
challenge is to achieve sufficient sensitivity to detect the
predicted,
low surface-brightness emission. As Figure~\ref{fig:obs_sb}
demonstrates, the surface brightness at the inner wind radius of our
fiducial wind exceeds that at the outer radius by several orders of magnitude.
Nevertheless, an instrument that sampled the entire wind (e.g.\ a
large format IFU or narrow band imager) may detect the emission in
azimuthally-averaged apertures. 
For the fiducial wind model, for example, the azimuthally integrated
flux falls by only a factor of 10 for projected radii of 1 to 15\,kpc.
Of the transitions considered in this paper, \ion{Mg}{2} emission is
preferred for this analysis because 
(i) it has the largest equivalent width in absorption and
(ii) there are fewer emission channels per absorption line than for
\ion{Fe}{2}.  The \ion{Mg}{2} lines (especially \mgiib) frequently have the
highest peak and integrated fluxes (Figure~\ref{fig:obs_sb}).  
On the other hand, the \ion{Mg}{2} transitions are more susceptible to
the effects of dust extinction and for some galaxies \feiis\ emission
could be dominant.  In either case, the study of spatially-extended
line-emission from photons scattered by a galactic-scale wind offers
a direct means to study the morphology and radial extent of these
phenomena.  The next generation of large-format optical and infrared
integral-field-units are well-suited to this scientific endeavor
(e.g.\ KCWI on Keck, MUSE on the VLT).

The kinematic measurements of the line-emission also offer insight into physical
characteristics of the wind.
In $\S$~\ref{sec:dust}, we emphasized that dust extinction
preferentially suppresses photons scattered off the backside of the
wind (e.g.\ $\mdv > 0 \mkms$) so that the line-centroid is shifted to
negative velocities.  Similarly, anisotropic models with reduced
emission off the backside yield emission lines that are centered
blueward of the galaxy's systemic redshift.  These effects are most
prominent in the \feiis\ emission.  The centroids of the \ion{Mg}{2}
lines, meanwhile, are sensitive to the optical depth
profile of the wind.   For example, a wind with flows exceeding
$\approx 600 \mkms$ will shift the \mgiia\ centroid to bluer
wavelengths (e.g.\ Figure~\ref{fig:norm}).
Although these effects are modest (tens of
\kms), they may be resolved by moderate resolution spectroscopy.

\begin{figure}
\includegraphics[height=3.5in,angle=90]{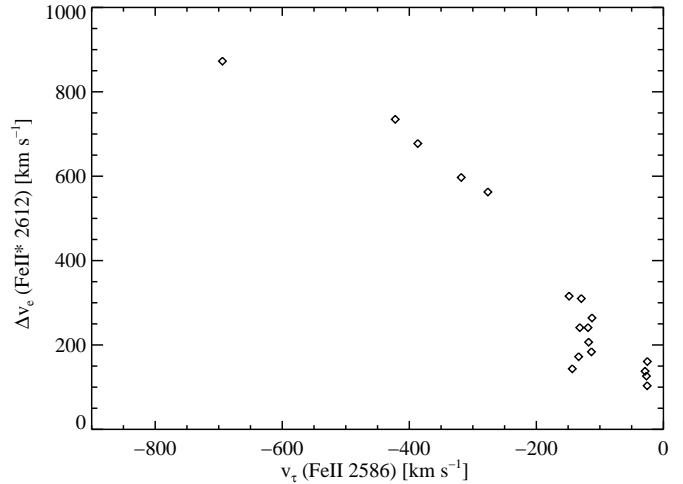}
\caption{
Plot of the 90\%\ velocity width ($\Delta v_e$) of \feiic\ emission versus the
velocity where the resonance lines achieve peak optical depth
$v_\tau$ in absorption.  The $\Delta v_e$ values rise steadily with
increasing offset of $v_\tau$ from systemic.  The width of the \feiic\
emission line, therefore, offers an independent diagnostic of the wind
speed.
}
\label{fig:obs_edelv}
\end{figure}

The emission-line velocity widths also reveal
characteristics of the wind.  In particular, the \feiis\ emission-line 
widths are
sensitive to the optical depth of the wind at large velocity offsets from
systemic.  This is illustrated in Figure~\ref{fig:obs_edelv} which plots
the 90\%\ velocity width $\Delta v_e$ of \feiic\ versus the velocity
where the wind optical depth is maximal ($v_\tau$).  We find that
$\Delta v_e$ increases with $v_\tau$ such that a large width for
\feiis\ emission requires a wind profile with large optical depths at
$\mdv < -200 \mkms$.  The broadening of \feiis\ emission occurs
because this emission is dominated by single scatterings 
which trace all components of the wind.  

Lastly, the line flux ratios of pairs of transitions 
are sensitive to characteristics of the wind.
The most obvious example is the relative emission of the \mgiid\
doublet.  Specifically, an outflow whose velocity exceeds the doublet
spacing ($\Delta v \approx 770 \mkms$) will convert \mgiia\ photons
into \mgiib\ emission resulting in a flux ratio that is inverted
relative to the intrinsic optical depth profiles (e.g.\
Figure~\ref{fig:norm}).  Another important example is the relative
flux of \feiis\ emission relative to \ion{Mg}{2}.  
A dusty medium with velocity near systemic (e.g.\ an
ISM component) may significantly suppress \ion{Mg}{2} emission yet
have a significant \feiis\ line flux because the latter is not
resonantly trapped.  

\section{Summary}
\label{sec:summary}

In this paper, we have explored the predicted absorption and
emission-line profiles of a set of simple galactic-scale outflow
models.  This analysis implemented Monte Carlo radiative transfer
techniques to propogate resonant photons through an expanding medium,
allowing for their conversion to non-resonant photons (e.g.\ \feiis).
Our work focused on the \mgiid\ and \ion{Fe}{2} UV1
multiplet of rest-frame UV transitions, but the results apply to
most other lines used to probe cool gas.  Our primary
findings are summarized as follows:

\begin{enumerate}

\item Isotropic, dust-free wind models conserve photon flux.
  Therefore, the blue-shifted absorption-line profiles commonly
  observed in star-forming galaxies are predicted to be accompanied by
  emission-lines with similar equivalent width.  This holds even for
  non-extreme anisotropic and dust-extincted scenarios.

\item The line-emission occurs preferentially at the systemic velocity
  of the galaxy and `fills-in' the absorption profiles at velocity
  offsets $|\mdv| < 200 \mkms$ from systemic.  For transitions that
  are only coupled to the ground-state (e.g.\ \ion{Mg}{2}, \lya,
  \ion{Na}{1}),  this implies much lower, absorption-line equivalent widths (by up to
  $50\%$) and observed absorption profiles that are significantly offset in velocity
  from the intrinsic optical depth profile.

\item Analysis of cool gas outflows that entirely ignores this
  line-emission may incorrectly conclude that the source is partially
  covered, that the gas has a significantly lower peak optical depth, and/or that 
  gas with velocities near systemic (e.g.\ from the ISM or even
  an infalling component) is absent.

\item Resonance transitions that are strongly coupled to
  non-resonant lines (e.g.\ \ion{Fe}{2}, \ion{Si}{2}) 
  produce emission dominated by the
  optically-thin, fine-structure transitions.  As such, these resonance
  absorption lines offer the best
  characterization of the opacity of the wind and also of gas 
  with velocities near systemic.  

\item  Dust extinction modestly affects 
  models where resonance photons are trapped for only a few scatterings.  Models
  with high opacity at small radii and at systemic velocity (e.g.\ with an
  optically thick, ISM component)  can effectively extinguish
  resonantly trapped emission (\ion{Mg}{2}) but have weaker
  effect on non-resonant lines (\feiis).

\item We examined two scenarios designed to mimic the wind model
  proposed by \cite{steidel+10} for $z \sim 3$ Lyman break galaxies.  Our
  implementation of this model genreates substantial line-emission
  from scattered photons that greatly modifies the predicted
  line-profiles so that this model does not reproduce the observed
  line-profiles. 


\item Significant line-emission is a generic prediction of simple wind
  models, even in the presence of dust, anisotropic flows, and when
  viewed through finite apertures.  We have explored the 2D emission
  maps (Figure~\ref{fig:fiducial_ifu_mgii}) and surface brightness 
  profiles (Figure~\ref{fig:obs_sb}) of
  the winds.  Sensitive, spatially-extended observations will map the
  morphology and radial extent of the outflows.  These data afford the
  best opportunity to estimate the energetics and mass-flux of
  galactic-scale outflows.

\item The kinematics and flux ratios of the emission lines 
  constrain the 
  speed, opacity, dust extinction, and morphology of the
  wind.  When combined with absorption-line analysis, one may develop
  yet tighter constraints on these characteristics.

\end{enumerate}

\acknowledgments

We acknowledge valuable conversations with D. Koo, J. Hennawi,
A. Coil, and A. Wolfe. 
J.X.P and K.R. are partially supported
by an NSF CAREER grant (AST--0548180), and 
by NSF grant AST-0908910.


\begin{thebibliography}{74}
\expandafter\ifx\csname natexlab\endcsname\relax\def\natexlab#1{#1}\fi

\bibitem[{{Aguirre} {et~al.}(2001){Aguirre}, {Hernquist}, {Schaye}, {Weinberg},
  {Katz}, \& {Gardner}}]{ahs+01}
{Aguirre}, A., {Hernquist}, L., {Schaye}, J., {Weinberg}, D.~H., {Katz}, N., \&
  {Gardner}, J. 2001, \apj, 560, 599

\bibitem[{{Alton} {et~al.}(1999){Alton}, {Davies}, \& {Bianchi}}]{adb99}
{Alton}, P.~B., {Davies}, J.~I., \& {Bianchi}, S. 1999, \aap, 343, 51

\bibitem[{{Charlot} \& {Fall}(2000)}]{cf00}
{Charlot}, S., \& {Fall}, S.~M. 2000, \apj, 539, 718

\bibitem[{{Chen} {et~al.}(2010){Chen}, {Tremonti}, {Heckman}, {Kauffmann},
  {Weiner}, {Brinchmann}, \& {Wang}}]{cth+10}
{Chen}, Y., {Tremonti}, C.~A., {Heckman}, T.~M., {Kauffmann}, G., {Weiner},
  B.~J., {Brinchmann}, J., \& {Wang}, J. 2010, \aj, 140, 445

\bibitem[{{Chevalier} \& {Clegg}(1985)}]{cc85}
{Chevalier}, R.~A., \& {Clegg}, A.~W. 1985, \nat, 317, 44

\bibitem[{{Dijkstra} {et~al.}(2006){Dijkstra}, {Haiman}, \&
  {Spaans}}]{Dijkstra_2006}
{Dijkstra}, M., {Haiman}, Z., \& {Spaans}, M. 2006, \apj, 649, 14

\bibitem[{{France} {et~al.}(2010){France}, {Nell}, {Green}, \&
  {Leitherer}}]{france10}
{France}, K., {Nell}, N., {Green}, J.~C., \& {Leitherer}, C. 2010, \apjl, 722,
  L80

\bibitem[{{Fujita} {et~al.}(2009){Fujita}, {Martin}, {Mac Low}, {New}, \&
  {Weaver}}]{fmm+09}
{Fujita}, A., {Martin}, C.~L., {Mac Low}, M., {New}, K.~C.~B., \& {Weaver}, R.
  2009, \apj, 698, 693

\bibitem[{{Hamann} {et~al.}(2010){Hamann}, {Kanekar}, {Prochaska}, {Murphy},
  {Ellison}, {Malec}, {Milutinovic}, \& {Ubachs}}]{hkp+10}
{Hamann}, F., {Kanekar}, N., {Prochaska}, J.~X., {Murphy}, M.~T., {Ellison},
  S., {Malec}, A.~L., {Milutinovic}, N., \& {Ubachs}, W. 2010, ArXiv e-prints

\bibitem[{{Harrington}(1973)}]{Harrington_1973}
{Harrington}, J.~P. 1973, \mnras, 162, 43

\bibitem[{{Hartigan} {et~al.}(1999){Hartigan}, {Morse}, {Tumlinson}, {Raymond},
  \& {Heathcote}}]{hmt+99}
{Hartigan}, P., {Morse}, J.~A., {Tumlinson}, J., {Raymond}, J., \& {Heathcote},
  S. 1999, \apj, 512, 901

\bibitem[{{Heckman} {et~al.}(1990){Heckman}, {Armus}, \& {Miley}}]{ham90}
{Heckman}, T.~M., {Armus}, L., \& {Miley}, G.~K. 1990, \apjs, 74, 833

\bibitem[{{Hughes} {et~al.}(1990){Hughes}, {Robson}, \& {Gear}}]{hrg90}
{Hughes}, D.~H., {Robson}, E.~I., \& {Gear}, W.~K. 1990, \mnras, 244, 759

\bibitem[{{Jeffery} \& {Branch}(1990)}]{Jeffery_Branch}
{Jeffery}, D.~J., \& {Branch}, D. 1990, in Supernovae, Jerusalem Winter School
  for Theoretical Physics, ed. {J.~C.~Wheeler, T.~Piran, \& S.~Weinberg},
  149--+

\bibitem[{{Kasen} {et~al.}(2002){Kasen}, {Branch}, {Baron}, \&
  {Jeffery}}]{Kasen_2002}
{Kasen}, D., {Branch}, D., {Baron}, E., \& {Jeffery}, D. 2002, \apj, 565, 380

\bibitem[{{Kasen} {et~al.}(2006){Kasen}, {Thomas}, \& {Nugent}}]{Kasen_2006}
{Kasen}, D., {Thomas}, R.~C., \& {Nugent}, P. 2006, \apj, 651, 366

\bibitem[{{Kasen} \& {Woosley}(2007)}]{kw07}
{Kasen}, D., \& {Woosley}, S.~E. 2007, \apj, 656, 661

\bibitem[{{Kasen} \& {Woosley}(2009)}]{kw09}
---. 2009, \apj, 703, 2205

\bibitem[{{Kasen} {et~al.}(2010)}]{Kasen_lyman}
{Kasen}, D., {et~al.} 2010, ApJ in preparation

\bibitem[{{Kere{\v s}} {et~al.}(2009){Kere{\v s}}, {Katz}, {Dav{\'e}},
  {Fardal}, \& {Weinberg}}]{kkd+09}
{Kere{\v s}}, D., {Katz}, N., {Dav{\'e}}, R., {Fardal}, M., \& {Weinberg},
  D.~H. 2009, \mnras, 396, 2332

\bibitem[{{Kinney} {et~al.}(1993){Kinney}, {Bohlin}, {Calzetti}, {Panagia}, \&
  {Wyse}}]{kbc+93}
{Kinney}, A.~L., {Bohlin}, R.~C., {Calzetti}, D., {Panagia}, N., \& {Wyse},
  R.~F.~G. 1993, \apjs, 86, 5

\bibitem[{{Kobulnicky} \& {Martin}(2010)}]{km10}
{Kobulnicky}, H.~A., \& {Martin}, C.~L. 2010, \apj, 718, 724

\bibitem[{{Kurucz}(2005)}]{kurucz05}
{Kurucz}, R.~L. 2005, Memorie della Societa Astronomica Italiana Supplementi,
  8, 14

\bibitem[{{Laursen} {et~al.}(2009){Laursen}, {Razoumov}, \&
  {Sommer-Larsen}}]{Laursen_2009}
{Laursen}, P., {Razoumov}, A.~O., \& {Sommer-Larsen}, J. 2009, \apj, 696, 853

\bibitem[{{Lehnert} {et~al.}(1999){Lehnert}, {Heckman}, \& {Weaver}}]{lhw99}
{Lehnert}, M.~D., {Heckman}, T.~M., \& {Weaver}, K.~A. 1999, \apj, 523, 575

\bibitem[{{Lowenthal} {et~al.}(1997){Lowenthal}, {Koo}, {Guzman}, {Gallego},
  {Phillips}, {Faber}, {Vogt}, {Illingworth}, \& {Gronwall}}]{lkg+97}
{Lowenthal}, J.~D., {et~al.} 1997, \apj, 481, 673

\bibitem[{{Martin}(1999)}]{martin99}
{Martin}, C.~L. 1999, \apj, 513, 156

\bibitem[{{Martin}(2005)}]{martin05}
---. 2005, \apj, 621, 227

\bibitem[{{Martin} \& {Bouch{\'e}}(2009)}]{mb09}
{Martin}, C.~L., \& {Bouch{\'e}}, N. 2009, \apj, 703, 1394

\bibitem[{{M{\'e}nard} {et~al.}(2008){M{\'e}nard}, {Nestor}, {Turnshek},
  {Quider}, {Richards}, {Chelouche}, \& {Rao}}]{mnt+08}
{M{\'e}nard}, B., {Nestor}, D., {Turnshek}, D., {Quider}, A., {Richards}, G.,
  {Chelouche}, D., \& {Rao}, S. 2008, \mnras, 385, 1053

\bibitem[{{M{\'e}nard} {et~al.}(2010){M{\'e}nard}, {Scranton}, {Fukugita}, \&
  {Richards}}]{msf+10}
{M{\'e}nard}, B., {Scranton}, R., {Fukugita}, M., \& {Richards}, G. 2010,
  \mnras, 405, 1025

\bibitem[{{Morton}(2003)}]{morton03}
{Morton}, D.~C. 2003, \apjs, 149, 205

\bibitem[{{Murray} {et~al.}(2010){Murray}, {M{\'e}nard}, \& {Thompson}}]{mmt10}
{Murray}, N., {M{\'e}nard}, B., \& {Thompson}, T.~A. 2010, ArXiv e-prints

\bibitem[{{Murray} {et~al.}(2005){Murray}, {Quataert}, \& {Thompson}}]{mqt05}
{Murray}, N., {Quataert}, E., \& {Thompson}, T.~A. 2005, \apj, 618, 569

\bibitem[{{Neufeld}(1990)}]{Neufeld_1990}
{Neufeld}, D.~A. 1990, \apj, 350, 216

\bibitem[{{Nulsen} {et~al.}(1998){Nulsen}, {Barcons}, \& {Fabian}}]{nbf98}
{Nulsen}, P.~E.~J., {Barcons}, X., \& {Fabian}, A.~C. 1998, \mnras, 301, 168

\bibitem[{{Oppenheimer} \& {Dav{\'e}}(2006)}]{od06}
{Oppenheimer}, B.~D., \& {Dav{\'e}}, R. 2006, \mnras, 373, 1265

\bibitem[{{Pettini} {et~al.}(1998){Pettini}, {Kellogg}, {Steidel}, {Dickinson},
  {Adelberger}, \& {Giavalisco}}]{pks+98}
{Pettini}, M., {Kellogg}, M., {Steidel}, C.~C., {Dickinson}, M., {Adelberger},
  K.~L., \& {Giavalisco}, M. 1998, \apj, 508, 539

\bibitem[{{Pettini} {et~al.}(2002){Pettini}, {Rix}, {Steidel}, {Adelberger},
  {Hunt}, \& {Shapley}}]{prs+02}
{Pettini}, M., {Rix}, S.~A., {Steidel}, C.~C., {Adelberger}, K.~L., {Hunt},
  M.~P., \& {Shapley}, A.~E. 2002, \apj, 569, 742

\bibitem[{{Pettini} {et~al.}(2000){Pettini}, {Steidel}, {Adelberger},
  {Dickinson}, \& {Giavalisco}}]{psa+00}
{Pettini}, M., {Steidel}, C.~C., {Adelberger}, K.~L., {Dickinson}, M., \&
  {Giavalisco}, M. 2000, \apj, 528, 96

\bibitem[{{Phillips}(1993)}]{phillips93}
{Phillips}, A.~C. 1993, \aj, 105, 486

\bibitem[{{Prochaska} {et~al.}(2006){Prochaska}, {Chen}, \& {Bloom}}]{pcb06}
{Prochaska}, J.~X., {Chen}, H.-W., \& {Bloom}, J.~S. 2006, \apj, 648, 95

\bibitem[{{Prochaska} {et~al.}(2007){Prochaska}, {Chen}, {Dessauges-Zavadsky},
  \& {Bloom}}]{pcd+07}
{Prochaska}, J.~X., {Chen}, H.-W., {Dessauges-Zavadsky}, M., \& {Bloom}, J.~S.
  2007, \apj, 666, 267

\bibitem[{{Prochaska} \& {Wolfe}(2001)}]{pw01}
{Prochaska}, J.~X., \& {Wolfe}, A.~M. 2001, \apjl, 560, L33

\bibitem[{{Radovich} {et~al.}(2001){Radovich}, {Kahanp{\"a}{\"a}}, \&
  {Lemke}}]{rkl01}
{Radovich}, M., {Kahanp{\"a}{\"a}}, J., \& {Lemke}, D. 2001, \aap, 377, 73

\bibitem[{{Rubin} {et~al.}(2010{\natexlab{a}}){Rubin}, {Prochaska},
  {M{\'e}nard}, {Murray}, {Kasen}, {Koo}, \& {Phillips}}]{rubin+10c}
{Rubin}, K.~H.~R., {Prochaska}, J.~X., {M{\'e}nard}, B., {Murray}, N., {Kasen},
  D., {Koo}, D.~C., \& {Phillips}, A.~C. 2010{\natexlab{a}}, ArXiv e-prints

\bibitem[{{Rubin} {et~al.}(2010{\natexlab{b}}){Rubin}, {Weiner}, {Koo},
  {Martin}, {Prochaska}, {Coil}, \& {Newman}}]{rwk+10}
{Rubin}, K.~H.~R., {Weiner}, B.~J., {Koo}, D.~C., {Martin}, C.~L., {Prochaska},
  J.~X., {Coil}, A.~L., \& {Newman}, J.~A. 2010{\natexlab{b}}, \apj, 719, 1503

\bibitem[{{Rupke} {et~al.}(2005{\natexlab{a}}){Rupke}, {Veilleux}, \&
  {Sanders}}]{rvs05a}
{Rupke}, D.~S., {Veilleux}, S., \& {Sanders}, D.~B. 2005{\natexlab{a}}, \apjs,
  160, 87

\bibitem[{{Rupke} {et~al.}(2005{\natexlab{b}}){Rupke}, {Veilleux}, \&
  {Sanders}}]{rvs05b}
---. 2005{\natexlab{b}}, \apjs, 160, 115

\bibitem[{{Sato} {et~al.}(2009){Sato}, {Martin}, {Noeske}, {Koo}, \&
  {Lotz}}]{smn+09}
{Sato}, T., {Martin}, C.~L., {Noeske}, K.~G., {Koo}, D.~C., \& {Lotz}, J.~M.
  2009, \apj, 696, 214

\bibitem[{{Savage} \& {Sembach}(1996)}]{ss96}
{Savage}, B.~D., \& {Sembach}, K.~R. 1996, \araa, 34, 279

\bibitem[{{Scannapieco} {et~al.}(2006){Scannapieco}, {Pichon}, {Aracil},
  {Petitjean}, {Thacker}, {Pogosyan}, {Bergeron}, \& {Couchman}}]{spa+06}
{Scannapieco}, E., {Pichon}, C., {Aracil}, B., {Petitjean}, P., {Thacker},
  R.~J., {Pogosyan}, D., {Bergeron}, J., \& {Couchman}, H.~M.~P. 2006, \mnras,
  365, 615

\bibitem[{{Schaye}(2001)}]{schaye01a}
{Schaye}, J. 2001, \apjl, 559, L1

\bibitem[{{Shapley} {et~al.}(2003){Shapley}, {Steidel}, {Pettini}, \&
  {Adelberger}}]{shapley03}
{Shapley}, A.~E., {Steidel}, C.~C., {Pettini}, M., \& {Adelberger}, K.~L. 2003,
  \apj, 588, 65

\bibitem[{{Silva} \& {Viegas}(2002)}]{silva02}
{Silva}, A.~I., \& {Viegas}, S.~M. 2002, \mnras, 329, 135

\bibitem[{{Simcoe} {et~al.}(2002){Simcoe}, {Sargent}, \& {Rauch}}]{ssr02}
{Simcoe}, R.~A., {Sargent}, W.~L.~W., \& {Rauch}, M. 2002, \apj, 578, 737

\bibitem[{{Snow} {et~al.}(1994){Snow}, {Lamers}, {Lindholm}, \&
  {Odell}}]{sll+94}
{Snow}, T.~P., {Lamers}, H.~J.~G.~L.~M., {Lindholm}, D.~M., \& {Odell}, A.~P.
  1994, \apjs, 95, 163

\bibitem[{{Sobolev}(1960)}]{sobolev60}
{Sobolev}, V.~V. 1960, {Moving envelopes of stars}

\bibitem[{{Socrates} {et~al.}(2008){Socrates}, {Davis}, \&
  {Ramirez-Ruiz}}]{sdr08}
{Socrates}, A., {Davis}, S.~W., \& {Ramirez-Ruiz}, E. 2008, \apj, 687, 202

\bibitem[{{Somerville} {et~al.}(2001){Somerville}, {Primack}, \&
  {Faber}}]{spf01}
{Somerville}, R.~S., {Primack}, J.~R., \& {Faber}, S.~M. 2001, \mnras, 320, 504

\bibitem[{{Steidel} {et~al.}(2010){Steidel}, {Erb}, {Shapley}, {Pettini},
  {Reddy}, {Bogosavljevi{\'c}}, {Rudie}, \& {Rakic}}]{steidel+10}
{Steidel}, C.~C., {Erb}, D.~K., {Shapley}, A.~E., {Pettini}, M., {Reddy}, N.,
  {Bogosavljevi{\'c}}, M., {Rudie}, G.~C., \& {Rakic}, O. 2010, \apj, 717, 289

\bibitem[{{Steidel} {et~al.}(1996){Steidel}, {Giavalisco}, {Pettini},
  {Dickinson}, \& {Adelberger}}]{sgp+96}
{Steidel}, C.~C., {Giavalisco}, M., {Pettini}, M., {Dickinson}, M., \&
  {Adelberger}, K.~L. 1996, \apjl, 462, L17+

\bibitem[{{Strickland} \& {Heckman}(2009)}]{sh09}
{Strickland}, D.~K., \& {Heckman}, T.~M. 2009, \apj, 697, 2030

\bibitem[{{Strickland} {et~al.}(2004){Strickland}, {Heckman}, {Colbert},
  {Hoopes}, \& {Weaver}}]{shc+04}
{Strickland}, D.~K., {Heckman}, T.~M., {Colbert}, E.~J.~M., {Hoopes}, C.~G., \&
  {Weaver}, K.~A. 2004, \apj, 606, 829

\bibitem[{{Swinbank} {et~al.}(2005){Swinbank}, {Smail}, {Bower}, {Borys},
  {Chapman}, {Blain}, {Ivison}, {Howat}, {Keel}, \& {Bunker}}]{ssb+05}
{Swinbank}, A.~M., {et~al.} 2005, \mnras, 359, 401

\bibitem[{{Tasitsiomi}(2006)}]{Tomi_2006}
{Tasitsiomi}, A. 2006, \apj, 645, 792

\bibitem[{{Tremonti} {et~al.}(2007){Tremonti}, {Moustakas}, \&
  {Diamond-Stanic}}]{tmd07}
{Tremonti}, C.~A., {Moustakas}, J., \& {Diamond-Stanic}, A.~M. 2007, \apjl,
  663, L77

\bibitem[{{Veilleux} {et~al.}(2003){Veilleux}, {Shopbell}, {Rupke},
  {Bland-Hawthorn}, \& {Cecil}}]{vsr+03}
{Veilleux}, S., {Shopbell}, P.~L., {Rupke}, D.~S., {Bland-Hawthorn}, J., \&
  {Cecil}, G. 2003, \aj, 126, 2185

\bibitem[{{Verhamme} {et~al.}(2006){Verhamme}, {Schaerer}, \&
  {Maselli}}]{Verhame_2006}
{Verhamme}, A., {Schaerer}, D., \& {Maselli}, A. 2006, \aap, 460, 397

\bibitem[{{Walter} {et~al.}(2002){Walter}, {Weiss}, \& {Scoville}}]{wws02}
{Walter}, F., {Weiss}, A., \& {Scoville}, N. 2002, \apjl, 580, L21

\bibitem[{{Weiner} {et~al.}(2009){Weiner}, {Coil}, {Prochaska}, {Newman},
  {Cooper}, {Bundy}, {Conselice}, {Dutton}, {Faber}, {Koo}, {Lotz}, {Rieke}, \&
  {Rubin}}]{wcp+09}
{Weiner}, B.~J., {et~al.} 2009, \apj, 692, 187

\bibitem[{{Westmoquette} {et~al.}(2008){Westmoquette}, {Smith}, \&
  {Gallagher}}]{wsg08}
{Westmoquette}, M.~S., {Smith}, L.~J., \& {Gallagher}, J.~S. 2008, \mnras, 383,
  864

\bibitem[{{York} {et~al.}(2006){York}, {Khare}, {Vanden Berk}, {Kulkarni},
  {Crotts}, {Lauroesch}, {Richards}, {Schneider}, {Welty}, {Alsayyad}, {Kumar},
  {Lundgren}, {Shanidze}, {Smith}, {Vanlandingham}, {Baugher}, {Hall},
  {Jenkins}, {Menard}, {Rao}, {Tumlinson}, {Turnshek}, {Yip}, \&
  {Brinkmann}}]{ykv+06}
{York}, D.~G., {et~al.} 2006, \mnras, 367, 945

\bibitem[{{Zheng} \& {Miralda-Escud{\'e}}(2002)}]{Zheng_2002}
{Zheng}, Z., \& {Miralda-Escud{\'e}}, J. 2002, \apj, 578, 33

\end{thebibliography}




\end{document}